\documentclass[dvipsnames]{article}
\usepackage{fullpage}
\usepackage[square,numbers]{natbib}
\usepackage{color}
\usepackage{amsfonts}
\usepackage{amsmath}
\usepackage{xcolor}
\usepackage{hyperref}
\usepackage{graphicx}
\usepackage{booktabs}
\usepackage{notation}
\usepackage{pandoc}
\usepackage{comment}
\usepackage[inline, shortlabels]{enumitem}
\setlist{itemjoin ={,\enspace},itemjoin* = { and\enspace}}

\newcommand{\fontbold}{\bf}

\begin{document}
\title{Overview of the TREC 2022 Fair Ranking Track}
\author{
  Michael D. Ekstrand\\
  \texttt{michaelekstrand@boisestate.edu}
  \and
  Graham McDonald \\
  \texttt{graham.mcdonald@glasgow.ac.uk}
  \and
  Amifa Raj\\
  \texttt{amifaraj@u.boisestate.edu}
  \and
  Isaac Johnson\\
  \texttt{isaac@wikimedia.org}
}

\maketitle

\setcounter{tocdepth}{1}
\tableofcontents

\section{Introduction}
The TREC Fair Ranking Track aims to provide a platform for participants to develop and evaluate novel retrieval algorithms that can provide a fair exposure to a mixture of demographics or attributes, such as ethnicity, that are represented by relevant documents in response to a search query. For example, particular demographics or attributes can be represented by the documents' topical content or authors.

The 2022 Fair Ranking Track adopted a resource allocation task. The task focused on supporting Wikipedia editors who are looking to improve the encyclopedia's coverage of topics under the purview of a WikiProject.\footnote{\url{https://en.wikipedia.org/wiki/WikiProject}} WikiProject coordinators and/or Wikipedia editors search for Wikipedia documents that are in need of editing to improve the quality of the article. The 2022 Fair Ranking track aimed to ensure that documents that are about, or somehow represent, certain protected characteristics receive a fair exposure to the Wikipedia editors, so that the documents have an fair opportunity of being improved and, therefore, be well-represented in Wikipedia. The under-representation of particular protected characteristics in Wikipedia can result in systematic biases that can have a negative human, social, and economic impact, particularly for disadvantaged or protected societal groups~\cite{pedreshi2008discrimination,redi2020taxonomy}.

\section{Task Definition}\label{sec:task-definition}
The 2022 Fair Ranking Track used an \emph{ad hoc} retrieval protocol. Participants were provided with a corpus of documents (a subset of the English language Wikipedia) and a set of queries. A query was of the form of a short list of search terms that represent a WikiProject. Each document in the corpus was relevant to zero to many WikiProjects and associated with zero to many fairness categories.   

There were two tasks in the 2022 Fair Ranking Track. In each of the tasks, for a given query, participants were to produce document rankings that are: 

\begin{enumerate}
\item Relevant to a particular WikiProject.
\item Provide a fair exposure to articles that are associated to particular protected attributes.
\end{enumerate}

The tasks shared a topic set, the corpus, the basic problem structure and the fairness objective. However, they differed in their target user persona, system output (static ranking vs.\ sequences of rankings) and evaluation metrics. The common problem setup was as follows:

\begin{itemize}
    \item \textbf{Queries} were provided by the organizers and derived from the topics of existing or hypothetical WikiProjects.
    \item \textbf{Documents} were Wikipedia articles that may or may not be relevant to any particular WikiProject that is represented by a query.
    \item \textbf{Rankings} were ranked lists of articles for editors to consider working on.
    \item \textbf{Fairness} of exposure should be achieved with respect to the protected attributes associated with the documents. Documents can be associated to many different fairness attributes. The official track evaluation focused on intersectional fairness and, as such, evaluated how fairly systems rank documents with respect to all of the fairness categories. However, individual teams could choose whether to optimise their systems with respect to all, a subset of, or individual fairness categories.
\end{itemize}

\subsection{Task 1: WikiProject Coordinators}\label{subsec:task1}
The first task focused on WikiProject coordinators as users of the search system; their goal is to search for relevant articles and produce a ranked list of articles needing work that other editors can then consult when looking for work to do.\\

\noindent \textbf{Output}: The output for this task was a \textbf{single ranking per query}, consisting of \textbf{500 articles}.\\

Evaluation was a multi-objective assessment of rankings by the following two criteria:

\begin{itemize}
    \item Relevance to a WikiProject topic. We will provide relevance assessments for the articles derived from existing Wikipedia data; Ranking relevance will be computed with nDCG, using binary relevance and logarithmic decay.
    \item Fairness with respect to the exposure of different fairness categories associated to the articles returned in response to a query.
\end{itemize}

Section \ref{sec:eval-task1} contains details on the evaluation metrics.

\subsection{Task 2: Wikipedia Editors}\label{subsec:task2}

The second task focused on individual Wikipedia editors looking for work associated with a project.
The conceptual model is that rather than maintaining a fixed work list as in Task 1, a WikiProject coordinator would create a saved search, and when an editor looks for work they re-run the search.
This means that different editors may receive different rankings for the same query, and differences in these rankings may be leveraged for providing fairness.\\

\noindent \textbf{Output}: The output of this task is \textbf{100 rankings per query}, each consisting of \textbf{20 articles}. \\

Evaluation was a multi-objective assessment of rankings by the following three criteria:

\begin{itemize}
    \item Relevance to a WikiProject topic. We will provide relevance assessments for articles derived from existing Wikipedia data. Ranking relevance will be computed with nDCG.
    \item Work needed on the article (articles needing more work preferred). We provide the output of an article quality assessment tool for each article in the corpus; for the purposes of this track, we assume lower-quality articles need more work. 
    \item Fairness with respect to the exposure of different fairness categories associated to the articles returned in response to a query.
\end{itemize}

The goal of this task was \textit{not} to be fair to work-needed levels; rather, we consider work-needed and topical relevance to be two components of a multi-objective notion of relevance, so that between two documents with the same topical relevance, the one with more work needed is more relevant to the query in the context of looking for articles to improve.

This task used \emph{expected exposure} to compare the exposure article subjects receive in result rankings to the \emph{ideal} (or \emph{target}) \emph{exposure} they would receive based on their relevance and work-needed \cite{diaz2020evaluating}.
This addresses fundamental limits in the ability to provide fair exposure in a single ranking by examining the exposure over multiple rankings.

For each query, participants provided 100 rankings, which we considered to be samples from the distribution realized by a stochastic ranking policy (given a query $\query$, a distribution $\pi_\query$ over truncated permutations of the documents).
Note that this is how we interpret the queries, but it did not mean that a stochastic policy is how the system should have been implemented --- other implementation designs were certainly possible.
The objective was to provide equitable exposure to documents of comparable relevance and work-needed, aggregated by protected attribute. Section~\ref{sec:eval-task2} has details on the evaluation metrics.

\section{Data}
This section provides details of the format of the test collection, topics and ground truth. Further details about data generation and limitations can be found in Section~\ref{sec:limitations}.

\subsection{Obtaining the Data}

The corpus and query data set is distributed via Globus, and can be obtained in two ways. First, it can be obtained via Globus, from our repository at \url{https://boi.st/TREC2022Globus}. From this site, you can log in using your institution's Globus account or your own Google account, and synchronize it to your local Globus install or download it with Globus Connect Personal.\footnote{\url{https://www.globus.org/globus-connect-personal}}  This method has robust support for restarting downloads and dealing with intermittent connections. Second, it can be downloaded directly via HTTP from:\\ \url{https://data.boisestate.edu/library/Ekstrand-2021/TRECFairRanking2022/}. 

The runs and evaluation qrels will be made available in the ordinary TREC archives.

\subsection{Corpus}\label{subsec:corpus}

The corpus consisted of articles from English Wikipedia.
We removed all redirect articles, but left the wikitext (markup Wikipedia uses to describe formatting) intact.
This was provided as a JSON file, with one record per line, and compressed with gzip (\texttt{trec\_corpus.json.gz}). Each record contains the following fields:

\begin{description}
    \item[id] The unique numeric Wikipedia article identifier.
    \item[title] The article title.
    \item[url] The article URL, to comply with Wikipedia licensing attribution requirements.
\end{description}

\noindent The three available formats of the corpus are as follows:
\begin{description}
    \item[text:] The full article text, with Wiki markup (\texttt{text} file only)
    \item[plain:] The full article text, without Wiki markup (\texttt{plain} file only)
    \item[html:] The full article text, rendered into HTML (\texttt{html} file only)
\end{description}

The contents of this corpus were prepared in accordance with, and licensed under, the CC BY-SA 3.0 license.\footnote{\url{https://creativecommons.org/licenses/by-sa/3.0/}} The raw Wikipedia dump files used to produce this corpus are available in the \texttt{source} directory; this is primarily for archival purposes, because Wikipedia does not publish dumps indefinitely.

\subsection{Queries}\label{subsec:topics}

The queries are in the \texttt{2022} directory, in the file \texttt{train\_topics\_meta.jsonl}.
Each of the queries map to a single Wikiproject. The queries are constructed from extracted keywords from articles that are relevant to a Wikiproject. The following fields are provided:

\begin{description}
    \item[id] A query identifier (int)
    \item[title] The Wikiproject title (string)
    \item[keywords] A collection of search keywords forming the query text (list of str). We cleaned and parsed the Wiki articles and then used KeyBert \cite{grootendorst2020keybert} to extract the most representative words of those articles. For each Wikiproject, we aggregated the extracted keywords from relevant articles and, after some manual filtering, used those as query texts for that particular Wikiproject.
    \item[homepage] The URL for the Wikiproject. This is provided for attribution and not expected to be used by your system as it will not be present in the evaluation data (string)
    \item[rel\_docs] A list of the page IDs of relevant pages (list of int)
\end{description}

The keywords are the primary query text. The scope is there to provide some additional context and potentially support techniques for refining system queries.

In addition to query relevance, for Task 2: Wikipedia Editors (Section~\ref{subsec:task2}), participants will also be expected to return relevant documents that need more editing work done more highly than relevant documents that need less work done.

\subsection{Fairness Categories}\label{subsec:fairnesscat}

Fairness ground truth labels for the following fairness categories are also in the \texttt{2022} directory, in the \texttt{trec\_2022\_articles\_discrete.json.gz} file. While we provide the raw values for each fairness category with the data, for most categories we also map the raw values to a reduced, fixed set of categories that will be used to judging systems.\footnote{For more information, see: \url{https://public.paws.wmcloud.org/User:Isaac_(WMF)/TREC/TREC_2022_Data.ipynb}}

\begin{description}
    \item[Geographic location (article topic)] The geographical location associated with the article topic. Both the associated countries---e.g., United Kingdom---and sub-continental regions---e.g., Northern Europe---are provided but systems will be evaluated using sub-continental regions (and not countries). An article can have 0 to many regions associated with it.
    \item[Geographic location (article sources)] The geographic location associated with the article based on the article's sources. Same categories as article geographic location above.
    \item[Gender (biographies only)] The gender of the individual about which the biography pertains. Gender has been reduced to four distinct categories: Man, Woman, Non-binary, and Unknown (missing data or not a biography).
    \item[Age of the topic] How old the subject of the article is. For example, the birth date of a person in a biographical article, the date that an event occurred for articles that are about an event, or the creation date of a piece of art or music when the article is about the piece of art or music. The raw years are mapped to four distinct categories: Unknown, Pre-1900s, 20th century, and 21st century.
    \item[Occupation (biographies only)] The occupation of the subject of an article. An article have 0 (unknown) to many occupations associated with it. There are 32 distinct occupation categories included in the data.
    \item[Alphabetical] Editors often work through articles in alphabetical order and this can result in articles about subjects / topics that start with letters that appear earlier in the alphabet getting more exposure to the editors. Therefore, it is important that articles from later in the alphabet also get a fair exposure to the editors. The first letter is mapped to four discrete categories: a-d, e-k, l-r, and s-.
    \item[Age of the article] The length of time the article has existed. The date is mapped to one of four discrete categories: 2001-2006, 2007-2011, 2012-2016, and 2017-2022.
    \item[Popularity (\# pageviews)] Number of times the page was viewed in February 2022. The number of pageviews are normalized and mapped to four discrete categories: Low, Medium-Low, Medium-High, and High.
    \item[Replication of articles in other languages] The number of other language Wikipedias that the article is replicated in. This can range from English-only to all 300+ languages of Wikipedia but is mapped to three discrete categories: English only, 2-4 languages, and 5+ languages.
\end{description}

For the purposes of multidimensional fairness, we treated the dimensions as independent, and took the outer product of the fairness categories as a combined fairness space.  The resulting space had over 11M dimensions, requiring care in implementing target alignments and metrics.

\subsection{Metadata}\label{subsec:metadata}
We provide a simple Wikimedia quality score (a float between 0 and 1 where 0 is no content on the page and 1 is high quality) for optimizing for work-needed in Task 2. Work-needed can be operationalized as the reverse---i.e. 1 minus this quality score.  The discretized quality scores will be used as work-needed for final system evaluation.

This data is provided together in a metadata file (\texttt{trec\_metadata.json.gz}), in which each line is the metadata for one article represented as a JSON record with the following keys:

\begin{description}
    \item[page\_id] Unique page identifier (int)
    \item[quality\_score] Continuous measure of article quality with 0 representing low quality and 1 representing high quality (float in range $[0,1]$)
    \item[quality\_score\_disc] Discrete quality score in which the quality score is mapped to six ordinal categories from low to high: Stub, Start, C, B, GA, FA (string)
    \item[Group Alignments] The group alignments associated to an article as described in Section~\ref{subsec:fairnesscat}.
\end{description}

\subsection{Output}\label{subsec:output}

For \textbf{Task 1}, participants outputted results in rank order in a tab-separated file with two columns:

\begin{description}
    \item[id] The query ID for the topic
    \item[page\_id] ID for the recommended article
\end{description}

\noindent For \textbf{Task 2}, this file had 3 columns, to account for repeated rankings per query:

\begin{description}
    \item[id] Query ID
    \item[rep\_number] Repeat Number (1-100)
    \item[page\_id] ID for the recommended article
\end{description}

\section{Evaluation Metrics}

Each task was evaluated with its own metric designed for that task setting. The goal of these metrics was to measure the extent to which a system (1) exposed relevant documents, and (2) exposed those documents in a way that is fair to article topic groups, defined by the mentioned fairness constraints of the article's subject.

This faces a problem in that Wikipedia itself has well-documented biases: if we target the current group distribution within Wikipedia, we will reward systems that simply reproduce Wikipedia's existing biases instead of promoting social equity. However, if we simply target equal exposure for groups, we would ignore potential real disparities in topical relevance. Due to the biases in Wikipedia's coverage, and the inability to retrieve documents that don't exist to fill in coverage gaps, there is not good empirical data on what the distribution for any particular topic \textit{should} be if systemic biases did not exist in either Wikipedia or society (the ``world as it could and should be'' \citep{mitchell:fairness}). Therefore, in this track we adopted a compromise: we \textbf{averaged} the empirical distribution of groups among relevant documents with the world population (for location) or equality (for gender) to derive the target group distribution.

Code to implement the metrics is found at \url{https://github.com/fair-trec/trec2022-fair-public}.

\subsection{Preliminaries}

The tasks were to retrieve documents $\doc$ from a corpus $\Corpus$ that are relevant to a query $\query$.  $\relvec{\query} \in [0,1]^{|\Corpus|}$ is a vector of relevance judgements for query $q$.
We denote a ranked list by $\ranking$; $\ranking_i$ is the document at position $i$ (starting from 1), and $\rankInv_\doc$ is the rank of document $\doc$.
For Task 1, each system returned a single ranked list; for Task 2, it returned a sequence of rankings $\rankSeq$.

We represented the group alignment of a document $\doc$ with an \textit{alignment vector} $\avec{\doc} \in [0,1]^{|\groups|}$. $\dal{\doc}{\group}$ is document $\doc$'s alignment with group $\group$.  $\amat \in [0,1]^{|\Corpus| \times |\groups|}$ is the alignment matrix for all documents. $\avec{\world}$ denotes the distribution of the world.\footnote{Obtained from \url{https://en.wikipedia.org/wiki/List_of_continents_and_continental_subregions_by_population}}

We considered fairness with respect to two group sets, $\groups_\geo$ and $\groups_\gender$.
We operationalized this intersectional objective by letting $\groups = \groups_\geo \times \groups_\gender$, the Cartesian product of the two group sets.
Further, alignment under either group set may be unknown; we represented this case by treating ``unknown'' as its own group ($\group_?$) in each set.
In the product set, a document's alignment may be unknown for either or both groups.

In all metrics, we use \textbf{log discounting} to compute attention weights:

$$\weight_i = \frac{1}{\log_2 \max (i, 2)}$$

Task 2 also considered the work each document needs, represented by $\workDoc \in \{1,2,3,4\}$.

\subsection{Task 1: WikiProject Coordinators (Single Rankings)}
\label{sec:eval-task1}

For the single-ranking Task 1, we adopted attention-weighted rank fairness (AWRF), first described by \citet{sapiezynski2019quantifying} and named by \citet{Raj2020-om}. AWRF computes a vector $\dlist$ of the cumulated exposure a list gives to each group, and a target vector $\dtarget$; we then compared these with the Jenson-Shannon divergence:

\begin{align}
\dlist' & = \sum_i \weight_i \avec{\ranking_i} & \text{cumulated attention} \nonumber\\
\dlist & = \frac{\dlist'}{\|\dlist'\|_1} & \text{normalize to a distribution} \nonumber \\
\dtarget & = \frac{1}{2}\left( \amat^\transpose \relvec{\query} + \avec{\world} \right) \nonumber \\
\AWRF(\ranking) & = 1 - \Djs{\dlist}{\dtarget}
\end{align}

For Task 1, we ignored documents that are fully unknown for the purposes of computing $\dlist$ and $\dtarget$; they do not contribute exposure to any group.

The resulting metric is in the range $[0,1]$, with 1 representing a maximally-fair ranking (the distance from the target distribution is minimized). We combined it with an ordinary nDCG metric for utility:

\begin{align}
    \nDCG(\ranking) = \frac{\sum_i \weight_i \rel{\query}{\doc}}{\mathrm{ideal}} \\
    \MOne(\ranking) = \AWRF(\ranking) \times \nDCG(\ranking)
\end{align}

To score well on the final metric $\MOne$, a run must be \textbf{both} accurate and fair.

\subsection{Task 2: Wikipedia Editors (Multiple Rankings)}
\label{sec:eval-task2}

For Task 2, we used Expected Exposure \citep{diaz2020evaluating} to compare the exposure each group receives in the sequence of rankings to the exposure it would receive in a sequence of rankings drawn from an \textit{ideal policy} with the following properties:

\begin{itemize}
    \item Relevant documents come before irrelevant documents
    \item Relevant documents are sorted in nonincreasing order of work needed
    \item Within each work-needed bin of relevant documents, group exposure is fairly distributed according to the average of the distribution of relevant documents and the distribution of global population (the same average target as before).
\end{itemize}

We have encountered some confusion about whether this task is requiring fairness towards work-needed; as we have designed the metric, work-needed is considered to be a part of (graded) relevance: a document is more relevant if it is relevant to the topic and needs significant work.  In the Expected Exposure framework, this combined relevance is used to derive the target policies.

To apply expected exposure, we first define the exposure $\docExp$ a document $\doc$ receives in sequence $\rankSeq$:

\begin{equation}
    \docExp = \frac{1}{|\rankSeq|} \sum_{\ranking \in \rankSeq} w_{\rankInv_\doc}
\end{equation}

This forms an exposure vector $\expVec \in \mathbb{R}^{|\Corpus|}$.  It is aggregated into a group exposure vector $\groupExp$, including ``unknown'' as a group:

\begin{equation}
    \groupExp = \amat^\transpose \expVec
\end{equation}

Our implementation rearranges the mean and aggregate operations, but the result is mathematically equivalent.

We then compare these system exposures with the target exposures $\tgtExpVec$ for each query.  This starts with the per-document ideal exposure; if $m_\work$ is the number of relevant documents with work-needed level $\work \in \{1, 2, 3, 4\}$, then according to \citet{diaz2020evaluating} the ideal exposure for document $\doc$ is computed as:

\begin{equation}
    \targetExposure_\doc = \frac{1}{m_{\workDoc}} \sum_{i = m_{>\workDoc} + 1}^{m_{\ge \workDoc}} \weight_i
\end{equation}

We use this to compute the non-averaged target distribution $\tilde\groupExp^*$:

\begin{equation}
    \tilde\groupExp^* = \amat^\transpose \tgtExpVec
\end{equation}

Since we include ``unknown'' as a group, we have a challenge with computing the target distribution by averaging the empirical distribution of relevant documents and the global population --- global population does not provide any information on the proportion of relevant articles for which the fairness attributes are relevant.  Our solution, therefore, is to average the distribution of \emph{known-group} documents with the world population, and re-normalize so the final distribution is a probability distribution, but derive the proportion of known- to unknown-group documents entirely from the empirical distribution of relevant documents.  Extended to handle partially-unknown documents, this procedure proceeds as follows:

\begin{itemize}
    \item Average the distribution of fully-known documents (both gender and location are known) with the global intersectional population (global population by location and equality by gender).
    \item Average the distribution of documents with unknown location but known gender with the equality gender distribution.
    \item Average the distribution of documents with unknown gender but known location with the world population.
\end{itemize}

The result is the target group exposure $\groupExp^*$.  We use this to measure the \textbf{expected exposure loss}:

\begin{align}
    \MTwo(\rankSeq_\query) & = \| \groupExp - \groupExp^* \|_2 \\
    & = \groupExp \cdot \groupExp - 2\groupExp \cdot \groupExp^* + \groupExp^* \cdot \groupExp^* \nonumber \\
    \operatorname{EE-D}(\rankSeq_\query) & = \groupExp^* \cdot \groupExp^* \\
    \operatorname{EE-R}(\rankSeq_\query) & = \groupExp \cdot \groupExp^*
\end{align}

Lower $\MTwo$ is better.  It decomposes into two submetrics, the \textbf{expected exposure disparity} (EE-D) that measures overall inequality in exposure independent of relevance, for which lower is better; and the \textbf{expected exposure relevance} (EE-L) that measures exposure/relevance alignment, for which higher is better \citep{diaz2020evaluating}.
\section{Results}\label{sec:results}
\begin{table}[tb]
    \centering
    \begin{tabular}{lrrrl}
\toprule
{} &   nDCG &   AWRF &  Score &          95\% CI \\
\midrule
\textbf{tmt5           } & 0.7242 & 0.4988 & 0.3626 &  (0.326, 0.397) \\
\textbf{UoGRelvOnlyT1  } & 0.6044 & 0.5246 & 0.3254 &  (0.284, 0.372) \\
\textbf{UoGTrT1ColPRF  } & 0.6044 & 0.5246 & 0.3254 &  (0.283, 0.369) \\
\textbf{UoGTrExpE2     } & 0.5977 & 0.5243 & 0.3230 &  (0.280, 0.368) \\
\textbf{0mt5           } & 0.6216 & 0.4778 & 0.2990 &  (0.267, 0.332) \\
\textbf{0mt5\_p         } & 0.5841 & 0.5015 & 0.2949 &  (0.262, 0.326) \\
\textbf{tmt5\_p         } & 0.5728 & 0.5121 & 0.2946 &  (0.260, 0.327) \\
\textbf{FRT\_constraint } & 0.5749 & 0.4793 & 0.2782 &  (0.245, 0.312) \\
\textbf{bm25\_p         } & 0.5434 & 0.5026 & 0.2773 &  (0.241, 0.312) \\
\textbf{UoGTrQE        } & 0.5368 & 0.4983 & 0.2734 &  (0.240, 0.309) \\
\textbf{UoGTrExpE1     } & 0.5176 & 0.5122 & 0.2716 &  (0.238, 0.308) \\
\textbf{UDInfo\_F\_bm25  } & 0.5666 & 0.4719 & 0.2708 &  (0.236, 0.302) \\
\textbf{ans\_bm25       } & 0.5661 & 0.4719 & 0.2706 &  (0.237, 0.303) \\
\textbf{UDInfo\_F\_mlp2  } & 0.5655 & 0.4718 & 0.2703 &  (0.235, 0.302) \\
\textbf{FRT\_attention  } & 0.5893 & 0.4484 & 0.2702 &  (0.231, 0.311) \\
\textbf{UDInfo\_F\_lgbm2 } & 0.5645 & 0.4719 & 0.2698 &  (0.235, 0.302) \\
\textbf{UDInfo\_F\_mlp4  } & 0.5638 & 0.4719 & 0.2695 &  (0.234, 0.301) \\
\textbf{UDInfo\_F\_lgbm4 } & 0.5631 & 0.4723 & 0.2693 &  (0.235, 0.302) \\
\textbf{FRT\_diversity  } & 0.5305 & 0.4909 & 0.2641 &  (0.229, 0.299) \\
\textbf{rmit\_cidda\_ir\_5} & 0.5417 & 0.4525 & 0.2485 &  (0.215, 0.282) \\
\textbf{rmit\_cidda\_ir\_1} & 0.5438 & 0.4416 & 0.2433 &  (0.210, 0.277) \\
\textbf{rmit\_cidda\_ir\_4} & 0.5388 & 0.4435 & 0.2431 &  (0.209, 0.278) \\
\textbf{rmit\_cidda\_ir\_7} & 0.5382 & 0.4443 & 0.2426 &  (0.209, 0.276) \\
\textbf{rmit\_cidda\_ir\_3} & 0.5365 & 0.4447 & 0.2420 &  (0.208, 0.275) \\
\textbf{rmit\_cidda\_ir\_6} & 0.5343 & 0.4457 & 0.2418 &  (0.208, 0.276) \\
\textbf{rmit\_cidda\_ir\_8} & 0.5322 & 0.4469 & 0.2415 &  (0.208, 0.276) \\
\textbf{rmit\_cidda\_ir\_2} & 0.5197 & 0.4443 & 0.2345 &  (0.201, 0.269) \\
\bottomrule
\end{tabular}

    \caption{Task 1 runs. Higher score is better (for all metrics). CI is 95\% bootstrapped CI of score.}
    \label{tab:task1:runs}
\end{table}

\begin{figure}[tb]
    \centering
    \includegraphics[width=0.5\textwidth]{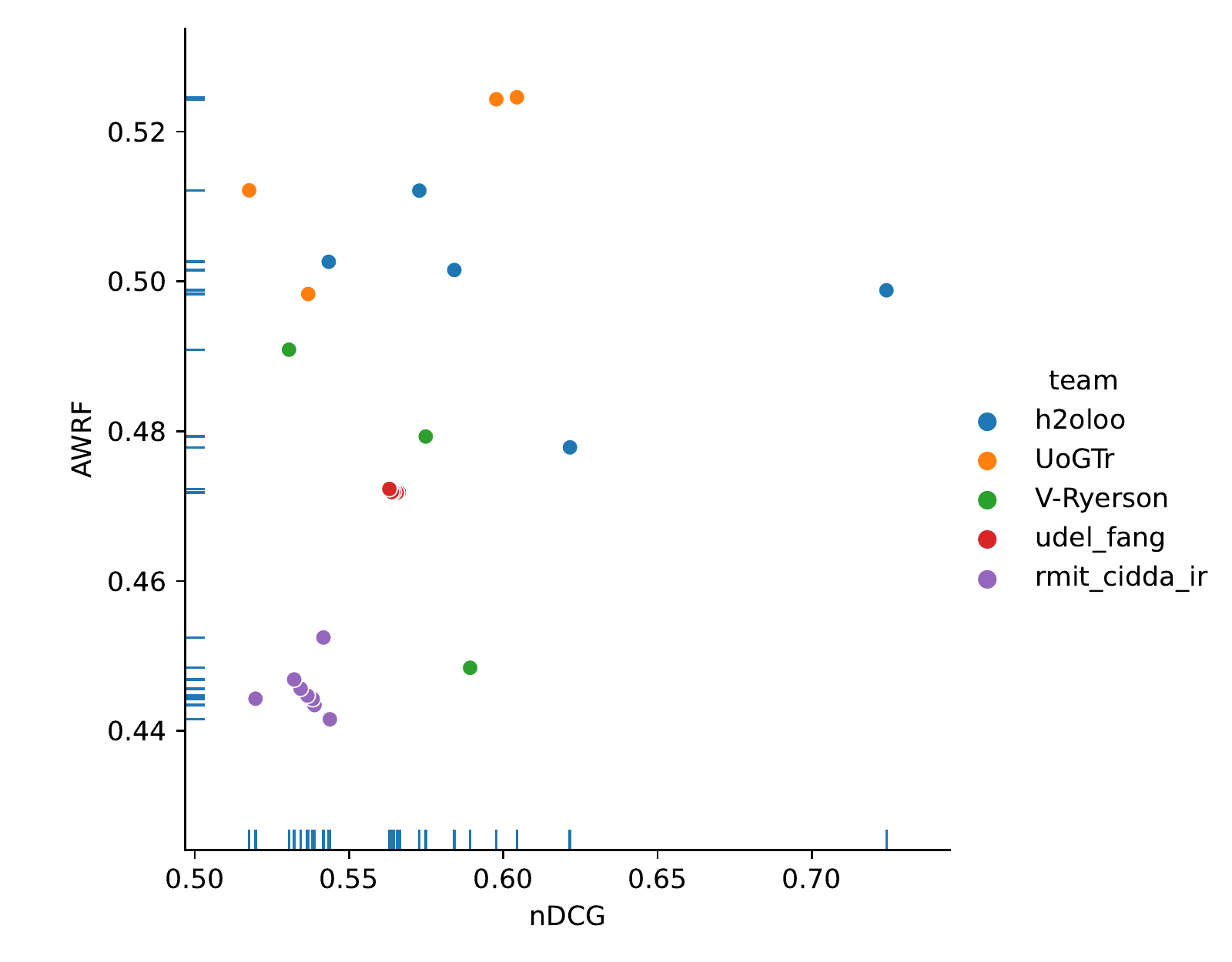}
    \caption{Task 1 submissions by individual component metrics (NDCG and AWRF).  Higher values are better for both metrics.}
    \label{fig:task1:ndcg-awrf}
\end{figure}

\begin{table}[tbp]
    \centering
    \begin{tabular}{lrrrrrrrrr}
 & Overall & age & alpha & gender & langs & occ & pop & src-geo & sub-geo \\
tmt5 & \fontbold 0.3626 & \fontbold 0.6860 & \fontbold 0.7190 & \fontbold 0.6795 & \fontbold 0.6786 & \fontbold 0.6825 & \fontbold 0.6313 & \fontbold 0.6453 & \fontbold 0.6450 \\
UoGRelvOnlyT1 & 0.3254 & 0.5843 & 0.5916 & 0.5266 & 0.5896 & 0.5267 & 0.5802 & 0.5548 & 0.5389 \\
UoGTrT1ColPRF & 0.3254 & 0.5843 & 0.5916 & 0.5266 & 0.5896 & 0.5267 & 0.5802 & 0.5548 & 0.5389 \\
UoGTrExpE2 & 0.3230 & 0.5797 & 0.5869 & 0.5453 & 0.5849 & 0.5443 & 0.5765 & 0.5482 & 0.5344 \\
0mt5 & 0.2990 & 0.5833 & 0.6164 & 0.5834 & 0.5817 & 0.5845 & 0.5333 & 0.5529 & 0.5496 \\
0mt5\_p & 0.2949 & 0.5552 & 0.5801 & 0.5519 & 0.5530 & 0.5514 & 0.5112 & 0.5315 & 0.5352 \\
tmt5\_p & 0.2946 & 0.5481 & 0.5696 & 0.5437 & 0.5463 & 0.5421 & 0.5077 & 0.5271 & 0.5303 \\
FRT\_constraint & 0.2782 & 0.5511 & 0.5712 & 0.5330 & 0.5494 & 0.5358 & 0.5220 & 0.5144 & 0.5059 \\
bm25\_p & 0.2773 & 0.5178 & 0.5395 & 0.5156 & 0.5165 & 0.5147 & 0.4768 & 0.4969 & 0.5005 \\
UoGTrQE & 0.2734 & 0.5216 & 0.5291 & 0.4813 & 0.5225 & 0.4807 & 0.4975 & 0.4828 & 0.4906 \\
UoGTrExpE1 & 0.2716 & 0.5073 & 0.4890 & 0.4643 & 0.5094 & 0.4641 & 0.5027 & 0.4778 & 0.4657 \\
UDInfo\_F\_bm25 & 0.2708 & 0.5282 & 0.5606 & 0.5320 & 0.5289 & 0.5335 & 0.4810 & 0.5026 & 0.4983 \\
ans\_bm25 & 0.2706 & 0.5275 & 0.5601 & 0.5315 & 0.5282 & 0.5331 & 0.4801 & 0.5021 & 0.4978 \\
UDInfo\_F\_mlp2 & 0.2703 & 0.5268 & 0.5597 & 0.5311 & 0.5269 & 0.5326 & 0.4803 & 0.5015 & 0.4976 \\
FRT\_attention & 0.2702 & 0.5278 & 0.5841 & 0.5100 & 0.5281 & 0.5162 & 0.5213 & 0.5160 & 0.4881 \\
UDInfo\_F\_lgbm2 & 0.2698 & 0.5261 & 0.5587 & 0.5304 & 0.5272 & 0.5320 & 0.4795 & 0.5006 & 0.4972 \\
UDInfo\_F\_mlp4 & 0.2695 & 0.5251 & 0.5581 & 0.5294 & 0.5250 & 0.5309 & 0.4789 & 0.4999 & 0.4965 \\
UDInfo\_F\_lgbm4 & 0.2693 & 0.5249 & 0.5574 & 0.5290 & 0.5263 & 0.5307 & 0.4791 & 0.4991 & 0.4962 \\
FRT\_diversity & 0.2641 & 0.5195 & 0.5270 & 0.5020 & 0.5169 & 0.4998 & 0.4835 & 0.4861 & 0.4828 \\
rmit\_cidda\_ir\_5 & 0.2485 & 0.4788 & 0.5366 & 0.5021 & 0.4932 & 0.5038 & 0.4405 & 0.4738 & 0.4617 \\
rmit\_cidda\_ir\_1 & 0.2433 & 0.4383 & 0.5377 & 0.5064 & 0.4850 & 0.5101 & 0.4223 & 0.4774 & 0.4698 \\
rmit\_cidda\_ir\_4 & 0.2431 & 0.4444 & 0.5319 & 0.5008 & 0.4780 & 0.5052 & 0.4278 & 0.4700 & 0.4570 \\
rmit\_cidda\_ir\_7 & 0.2426 & 0.4321 & 0.5317 & 0.5012 & 0.4858 & 0.5052 & 0.4232 & 0.4718 & 0.4638 \\
rmit\_cidda\_ir\_3 & 0.2420 & 0.4302 & 0.5297 & 0.4994 & 0.4846 & 0.5037 & 0.4208 & 0.4707 & 0.4636 \\
rmit\_cidda\_ir\_6 & 0.2418 & 0.4312 & 0.5278 & 0.4969 & 0.4845 & 0.5015 & 0.4218 & 0.4686 & 0.4613 \\
rmit\_cidda\_ir\_8 & 0.2415 & 0.4315 & 0.5259 & 0.4949 & 0.4844 & 0.4996 & 0.4219 & 0.4671 & 0.4600 \\
rmit\_cidda\_ir\_2 & 0.2345 & 0.4122 & 0.5134 & 0.4829 & 0.4805 & 0.4875 & 0.4169 & 0.4552 & 0.4473 \\
\end{tabular}

    \caption{Task 1 $\MOne$ on individual fairness dimensions.}
    \label{tab:task1:single}
\end{table}

\begin{figure}[tpb]
    \includegraphics[width=\textwidth]{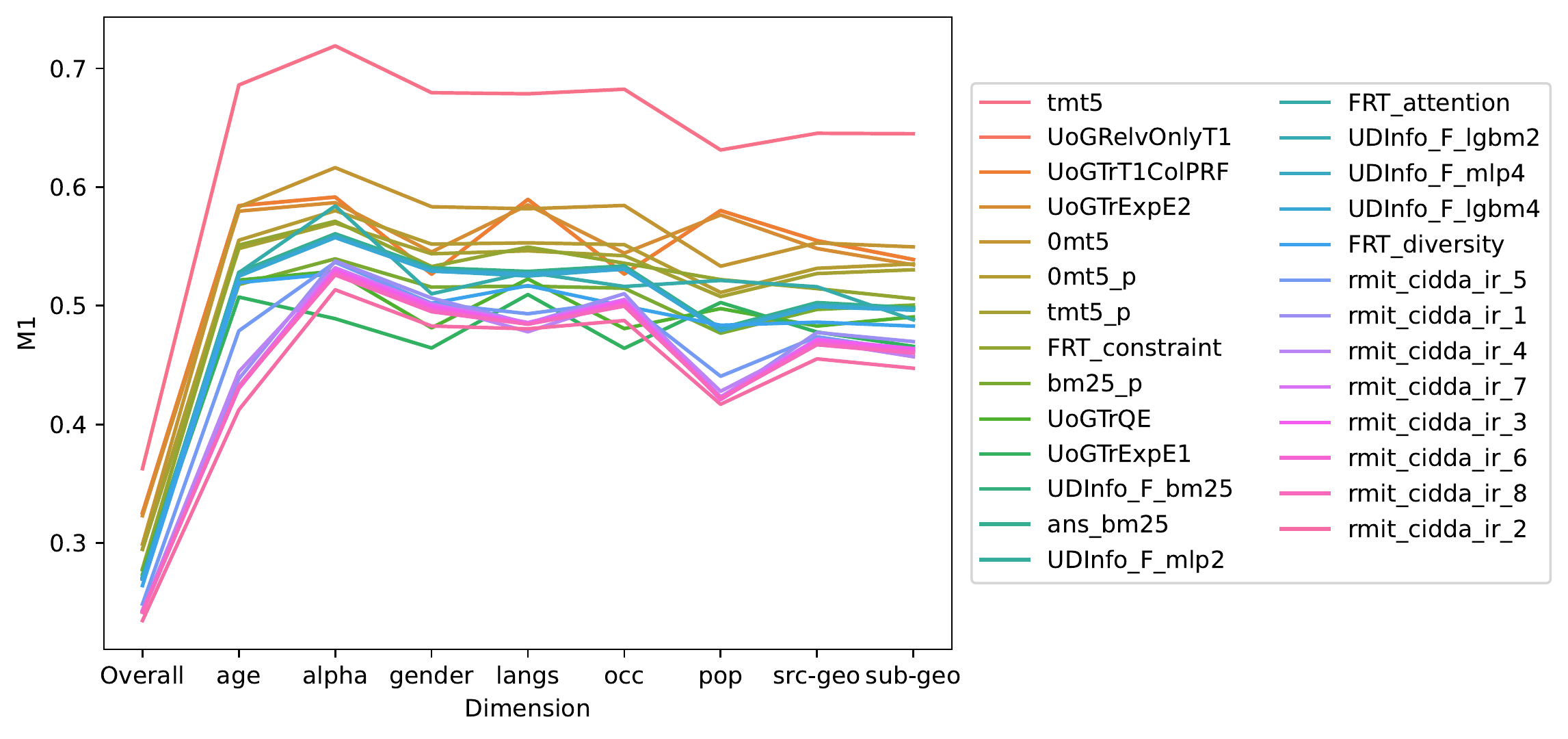}
    \caption{Task 1 $\MOne$ on individual fairness dimensions.}
    \label{fig:task1:single}
\end{figure}

\begin{table}[tbp]
    \centering
    \begin{tabular}{lrrrr}
 & Overall & 2021 & Internal & Demographic \\
tmt5 & \fontbold 0.3626 & \fontbold 0.6034 & \fontbold 0.5699 & \fontbold 0.5389 \\
UoGRelvOnlyT1 & 0.3254 & 0.4757 & 0.5154 & 0.4338 \\
UoGTrT1ColPRF & 0.3254 & 0.4757 & 0.5154 & 0.4338 \\
UoGTrExpE2 & 0.3230 & 0.4874 & 0.5160 & 0.4395 \\
0mt5 & 0.2990 & 0.5143 & 0.4787 & 0.4563 \\
0mt5\_p & 0.2949 & 0.5019 & 0.4702 & 0.4450 \\
tmt5\_p & 0.2946 & 0.4987 & 0.4722 & 0.4428 \\
FRT\_constraint & 0.2782 & 0.4691 & 0.4704 & 0.4152 \\
bm25\_p & 0.2773 & 0.4716 & 0.4408 & 0.4181 \\
UoGTrQE & 0.2734 & 0.4411 & 0.4564 & 0.3866 \\
UoGTrExpE1 & 0.2716 & 0.4208 & 0.4277 & 0.3828 \\
UDInfo\_F\_bm25 & 0.2708 & 0.4666 & 0.4294 & 0.4145 \\
ans\_bm25 & 0.2706 & 0.4660 & 0.4285 & 0.4141 \\
UDInfo\_F\_mlp2 & 0.2703 & 0.4660 & 0.4284 & 0.4138 \\
FRT\_attention & 0.2702 & 0.4362 & 0.4420 & 0.3877 \\
UDInfo\_F\_lgbm2 & 0.2698 & 0.4656 & 0.4276 & 0.4133 \\
UDInfo\_F\_mlp4 & 0.2695 & 0.4649 & 0.4273 & 0.4127 \\
UDInfo\_F\_lgbm4 & 0.2693 & 0.4645 & 0.4270 & 0.4123 \\
FRT\_diversity & 0.2641 & 0.4515 & 0.4382 & 0.4012 \\
rmit\_cidda\_ir\_5 & 0.2485 & 0.4300 & 0.3834 & 0.3818 \\
rmit\_cidda\_ir\_1 & 0.2433 & 0.4375 & 0.3568 & 0.3899 \\
rmit\_cidda\_ir\_4 & 0.2431 & 0.4270 & 0.3603 & 0.3824 \\
rmit\_cidda\_ir\_7 & 0.2426 & 0.4327 & 0.3547 & 0.3868 \\
rmit\_cidda\_ir\_3 & 0.2420 & 0.4320 & 0.3527 & 0.3866 \\
rmit\_cidda\_ir\_6 & 0.2418 & 0.4297 & 0.3531 & 0.3851 \\
rmit\_cidda\_ir\_8 & 0.2415 & 0.4284 & 0.3533 & 0.3840 \\
rmit\_cidda\_ir\_2 & 0.2345 & 0.4160 & 0.3394 & 0.3735 \\
\end{tabular}

    \caption{Task 1 $\MOne$ on subsets of the fairness dimensions.}
    \label{tab:task1:subsets}
\end{table}

\begin{figure}[tpb]
    \includegraphics[width=\textwidth]{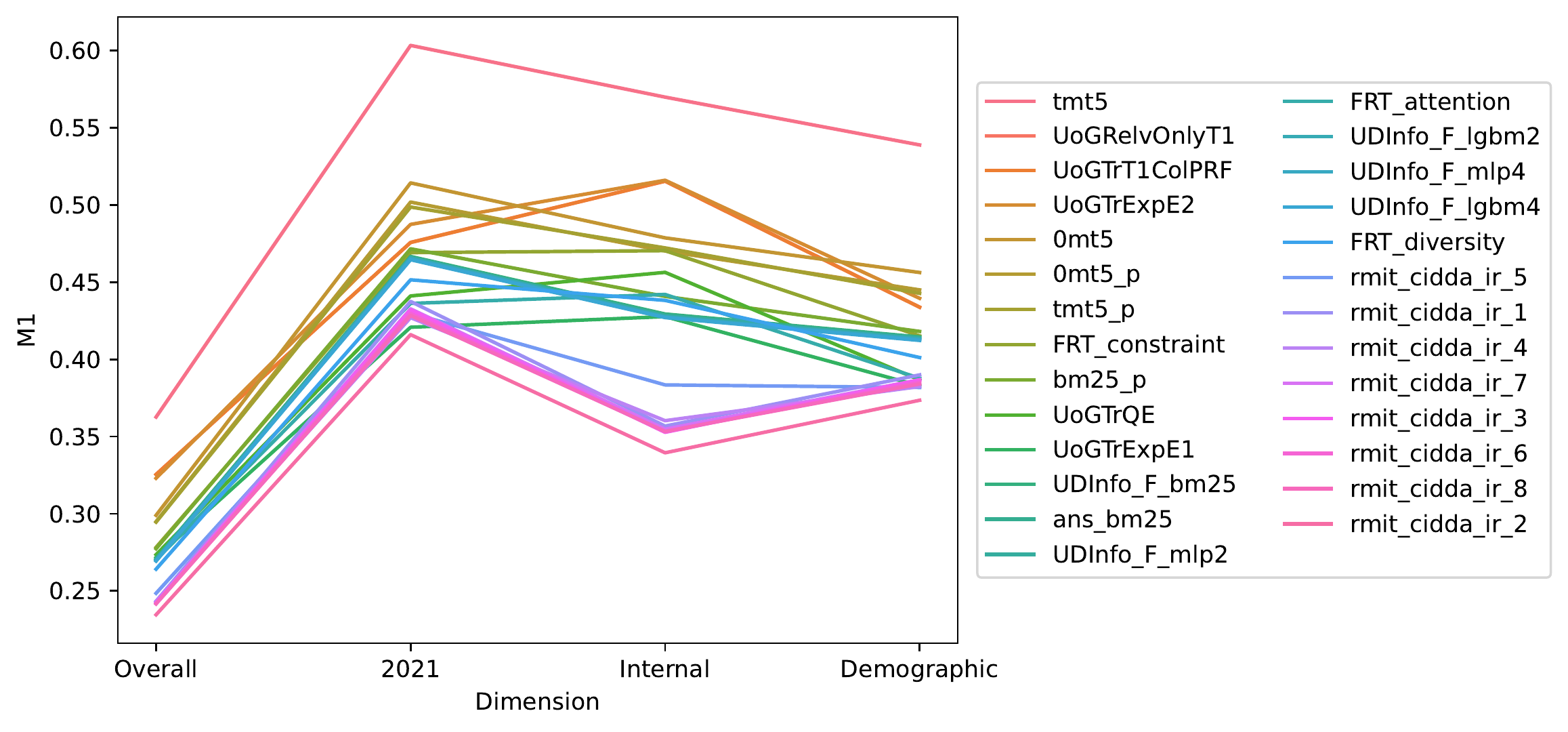}
    \caption{Task 1 $\MOne$ on subsets of the fairness dimensions.}
    \label{fig:task1:subsets}
\end{figure}

This year 5 different teams submitted a total of 24 runs. All 5 teams participated in Task 1: Single Rankings (27 runs total), while 2 groups participated in Task 2: Multiple Rankings (11 runs total).

\subsection{Task 1: WikiProject Coordinators (Single Rankings)}

Approaches for Task 1 included:
\begin{itemize}
    \item Relevance ranking by ColBERT-PRF.
    \item Relevance ranking by ColBERT-E2E and a heuristic approach that re-ranks to match target exposure using diversification.
    \item Query rewriting strategy to expand query.
    \item BM25 ranking from pyserini and pre-traned BERT for semantic score and re-ranked to fit target distribution at each ranking position by using greedy diversification, providing higher attention to protected group, and ensuring fairness in each position.
    \item Relevance ranking using BM25 from pyserini and LambdaMART lerning-to-rank model and a multi-layer-perception to re-rank.
    \item BM25 ranking from pyserini and weighted Reciprocal Ranking Fusion to diversify the ranking.
    \item Relevance-only approaches.
\end{itemize}

Table~\ref{tab:task1:runs} shows the submitted systems ranked by the official Task 1 metric $\MOne$ and its component parts nDCG and AWRF. Figure~\ref{fig:task1:ndcg-awrf} plots the runs with the component metrics on the $x$ and $y$ axes. Unlike last year, we see less clustering of approaches from individual teams: two teams approaches had similar performance, while others are more scattered throughout the space.

We also computed fairness on individual dimension (Table~\ref{tab:task1:single} and Figure~\ref{fig:task1:single}), and on the three subsets identified in Section~\ref{subsec:fairnesscat} (Table~\ref{tab:task1:subsets} and Figure~\ref{fig:task1:subsets}).
For Task 1 the best-performing system overall also performed best on each individual category and subset; however, ordering of other systems changed between subsets or categories.

\subsection{Task 2: Wikipedia Editors (Multiple Rankings)}\label{sub:t2results}

Approaches for Task 2 included:
\begin{itemize}
    \item Multi armed bandit strategies to select rankings from a pool of rankings considering each fairness category and observing the exposure and fairness-relevance relationship. 
    \item Epsilon-greedy with weighted ranking, epsilon-decay strategy, and randomisation are used in ranking selection process.
    \item Relevance-only ranking.
\end{itemize}

\begin{figure}[tb]
    \centering
    \includegraphics[width=0.5\textwidth]{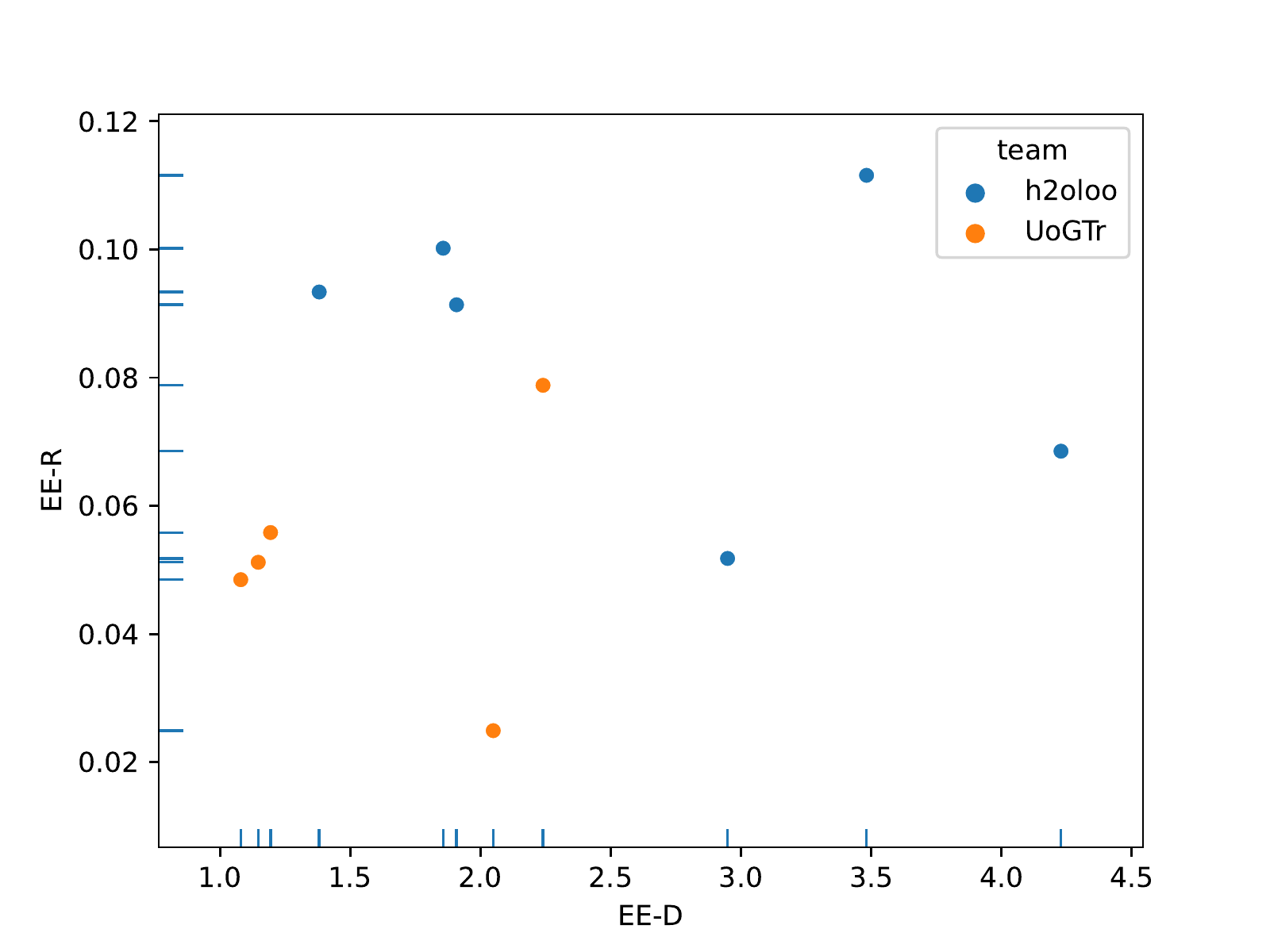}
    \caption{Task 2 submissions by expected exposure subcomponents. Lower EE-D is better; higher EE-R is better.}
    \label{fig:task2:eed-eer}
\end{figure}

Table~\ref{tab:task2:runs} shows the submitted systems ranked by the official Task 2 metric EE-L and its component parts EE-D and EE-R. Figure~\ref{fig:task2:eed-eer} plots the runs with the component metrics on the $x$ and $y$ axes. Overall, the submitted systems generally performed better for one of the component metrics than they did for the other.

\begin{figure}[tpb]
    \includegraphics[width=\textwidth]{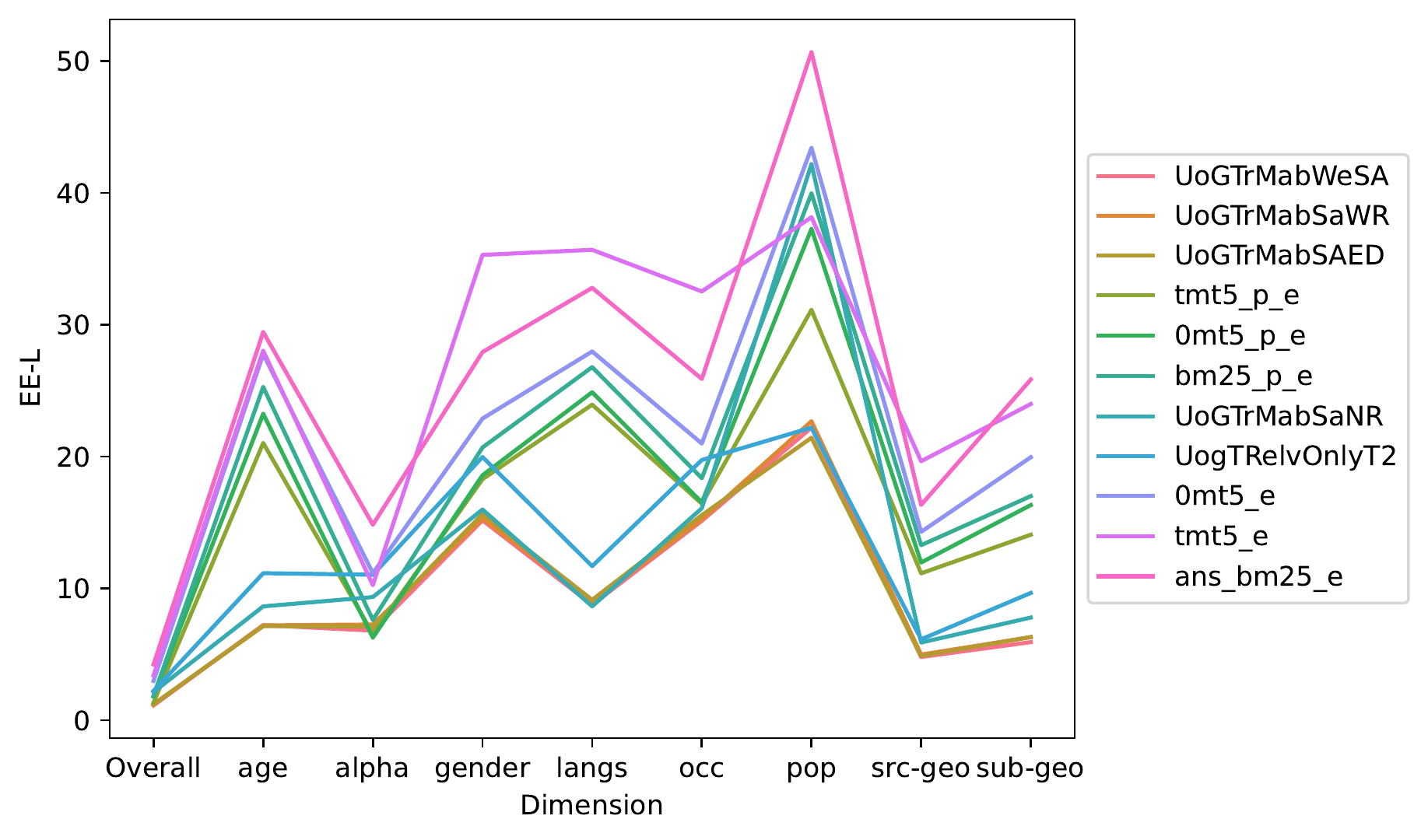}
    \caption{Task 2 EE-L on subsets of the fairness dimensions.}
    \label{fig:task2:single}
\end{figure}

\begin{figure}[tpb]
    \includegraphics[width=\textwidth]{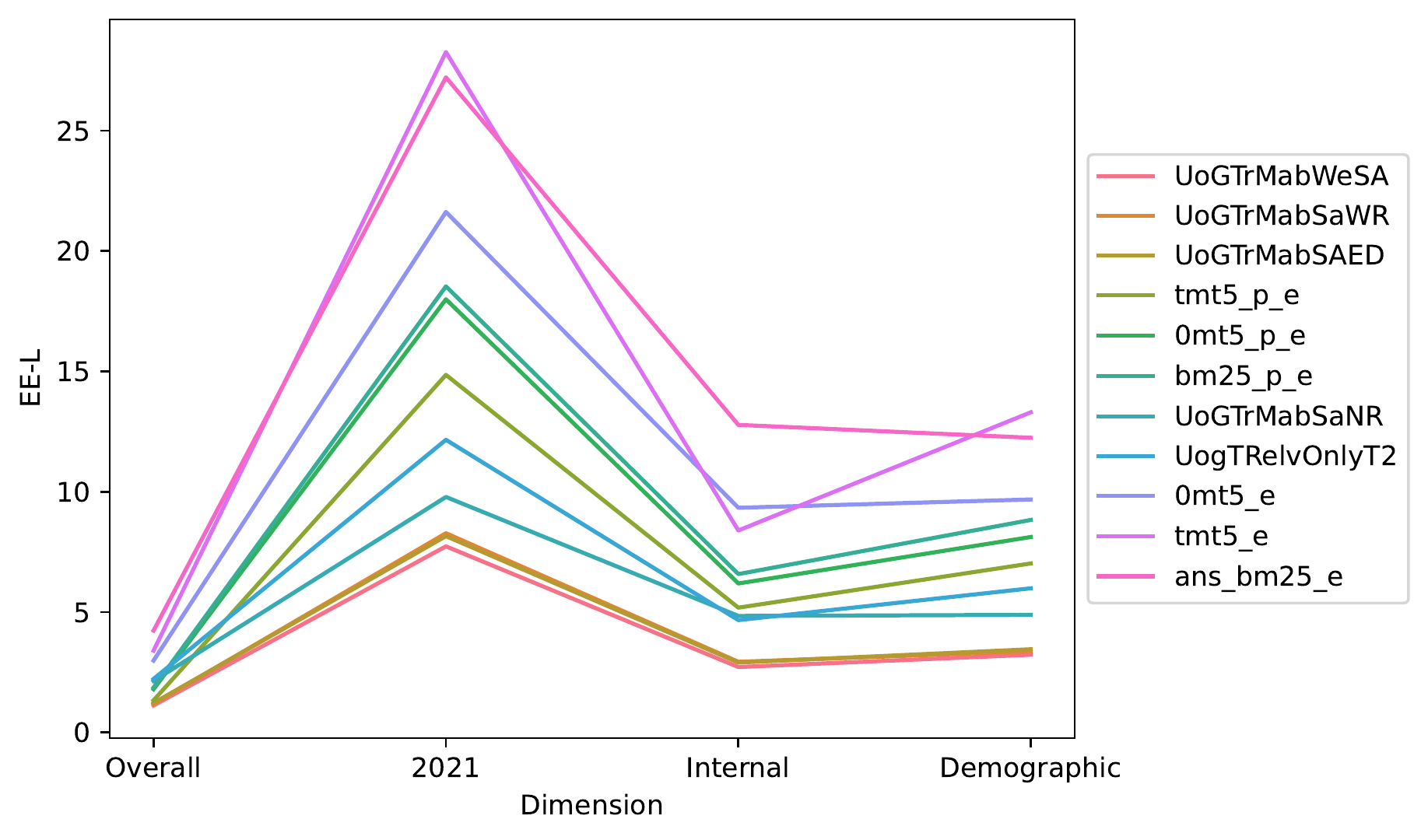}
    \caption{Task 2 EE-L on subsets of the fairness dimensions.}
    \label{fig:task2:subsets}
\end{figure}

We also computed fairness on individual dimension (Table~\ref{tab:task2:single} and Figure~\ref{fig:task2:single}), and on the three subsets identified in Section~\ref{subsec:fairnesscat} (Table~\ref{tab:task2:subsets} and Figure~\ref{fig:task2:subsets}).  Unlike Task 1, we see more difference in fairness between different single attributes and subsets.

\begin{table}[tb]
    \centering
    \begin{tabular}{lrrrl}
 & EE-R & EE-D & EE-L & EE-L 95\% CI \\
\textbf{UoGTrMabWeSA} & 0.0485 & 1.0791 & 1.1231 & (0.9790, 1.3096) \\
\textbf{UoGTrMabSaWR} & 0.0512 & 1.1462 & 1.1847 & (1.0258, 1.3739) \\
\textbf{UoGTrMabSAED} & 0.0558 & 1.1935 & 1.2228 & (1.0507, 1.4185) \\
\textbf{tmt5\_p\_e} & 0.0934 & 1.3803 & 1.3345 & (1.1409, 1.5793) \\
\textbf{0mt5\_p\_e} & 0.1002 & 1.8563 & 1.7968 & (1.5600, 2.0779) \\
\textbf{bm25\_p\_e} & 0.0914 & 1.9077 & 1.8659 & (1.5343, 2.2897) \\
\textbf{UoGTrMabSaNR} & 0.0249 & 2.0486 & 2.1397 & (1.8286, 2.6452) \\
\textbf{UogTRelvOnlyT2} & 0.0788 & 2.2398 & 2.2231 & (1.8645, 2.7786) \\
\textbf{0mt5\_e} & 0.0518 & 2.9483 & 2.9856 & (2.5022, 3.7271) \\
\textbf{tmt5\_e} & 0.1116 & 3.4819 & 3.3997 & (3.0171, 3.8588) \\
\textbf{ans\_bm25\_e} & 0.0685 & 4.2286 & 4.2324 & (2.7795, 8.3580) \\
\end{tabular}

    \caption{Task 2 runs. Lower EE-L is better. Confidence intervals are bootstrapped.}
    \label{tab:task2:runs}
\end{table}

\begin{table}[tbp]
    \centering
    \begin{tabular}{lrrrrrrrrr}
 & Overall & age & alpha & gender & langs & occ & pop & src-geo & sub-geo \\
UoGTrMabWeSA & \fontbold 1.123 & 7.215 & 6.798 & \fontbold 15.163 & 8.756 & \fontbold 15.127 & 22.197 & \fontbold 4.798 & \fontbold 5.925 \\
UoGTrMabSaWR & 1.185 & 7.194 & 7.259 & 15.291 & 8.978 & 15.226 & 22.674 & 4.979 & 6.303 \\
UoGTrMabSAED & 1.223 & \fontbold 7.136 & 7.111 & 15.636 & 9.127 & 15.551 & \fontbold 21.423 & 4.884 & 6.314 \\
tmt5\_p\_e & 1.334 & 21.024 & 6.554 & 18.282 & 23.936 & 16.420 & 31.123 & 11.142 & 14.070 \\
0mt5\_p\_e & 1.797 & 23.243 & \fontbold 6.266 & 18.579 & 24.876 & 16.516 & 37.273 & 11.962 & 16.320 \\
bm25\_p\_e & 1.866 & 25.273 & 7.607 & 20.687 & 26.782 & 18.361 & 39.964 & 13.278 & 16.995 \\
UoGTrMabSaNR & 2.140 & 8.637 & 9.343 & 15.981 & \fontbold 8.640 & 16.078 & 42.181 & 5.894 & 7.790 \\
UogTRelvOnlyT2 & 2.223 & 11.152 & 11.037 & 19.964 & 11.686 & 19.738 & 22.193 & 6.136 & 9.654 \\
0mt5\_e & 2.986 & 27.765 & 11.149 & 22.881 & 27.967 & 20.988 & 43.410 & 14.288 & 19.946 \\
tmt5\_e & 3.400 & 28.021 & 10.255 & 35.293 & 35.676 & 32.521 & 38.150 & 19.642 & 23.982 \\
ans\_bm25\_e & 4.232 & 29.440 & 14.830 & 27.922 & 32.793 & 25.886 & 50.670 & 16.335 & 25.838 \\
\end{tabular}

    \caption{Task 2 EE-L on individual fairness dimensions.}
    \label{tab:task2:single}
\end{table}

\begin{table}[tbp]
    \centering
    \begin{tabular}{lrrrr}
 & Overall & 2021 & Internal & Demographic \\
UoGTrMabWeSA & \fontbold 1.123 & \fontbold 7.732 & \fontbold 2.719 & \fontbold 3.230 \\
UoGTrMabSaWR & 1.185 & 8.276 & 2.936 & 3.373 \\
UoGTrMabSAED & 1.223 & 8.156 & 2.904 & 3.458 \\
tmt5\_p\_e & 1.334 & 14.856 & 5.185 & 7.021 \\
0mt5\_p\_e & 1.797 & 17.995 & 6.190 & 8.120 \\
bm25\_p\_e & 1.866 & 18.532 & 6.580 & 8.829 \\
UoGTrMabSaNR & 2.140 & 9.783 & 4.842 & 4.885 \\
UogTRelvOnlyT2 & 2.223 & 12.157 & 4.667 & 5.989 \\
0mt5\_e & 2.986 & 21.622 & 9.334 & 9.679 \\
tmt5\_e & 3.400 & 28.262 & 8.391 & 13.300 \\
ans\_bm25\_e & 4.232 & 27.213 & 12.777 & 12.242 \\
\end{tabular}

    \caption{Task 2 EE-L on subsets of the fairness dimensions.}
    \label{tab:task2:subsets}
\end{table}

\label{sec:limitations}
\section{Limitations}

The data and metrics in this task address a few specific types of unfairness, and do so partially. This is fundamentally true of any fairness intervention, and does not in any way diminish the value of the effort --- it is impossible for any data set, task definition, or metric to fully capture fairness in a universal way, and all data and analyses have limitations.

Some of the limitations of the data and task include:

\begin{itemize}
    \item \textbf{Fairness criteria}
    \begin{itemize}
        \item \textbf{Gender}: For each Wikipedia article, we ascertain whether it is a biography, and, if so, which gender identity can be associated with the person it is about.\footnote{Code: \url{https://github.com/geohci/miscellaneous-wikimedia/blob/master/wikidata-properties-spark/wikidata_gender_information.ipynb}} This data is directly determined via Wikidata based on the instance-of property indicating the article is about a human (P31:Q5 in Wikidata terms) and then collecting the value associated with the sex-or-gender property (P21). Coverage here is quite high at 99.98\% of biographies on Wikipedia having associated gender data on Wikidata.
        
        Assigning gender identities to people is not a process without errors, biases, and ethical concerns. Applying the taxonomy developed by Pinney et al.~\cite{pinney2023much} to this work yields the following summary: the primary referent of the gender data is the subject; we do use a gender variable and it's binary+other; gender determination is done via annotators (see details below); the gender data is used to measure bias and the goal is to audit system behavior.
        The process for assigning gender (annotation) is subject to some community-defined technical limitations\footnote{\url{https://www.wikidata.org/wiki/Property_talk:P21\#Documentation}} and the Wikidata policy on living people\footnote{\url{https://www.wikidata.org/wiki/Wikidata:Living_people}}. While a separate project, English Wikipedia's policies on gender identity\footnote{\url{https://en.wikipedia.org/wiki/Wikipedia:Manual_of_Style/Gender_identity}} likely inform how many editors handle gender; in particular, this policy explicitly favors the most recent reliably-sourced \textit{self-identification} for gender, so misgendering a biography subject is a violation of Wikipedia policy; there may be erroneous data, but such data seems to be a violation of policy instead of a policy decision. Wikidata:WikiProject LGBT has documented some clear limitations of gender data on Wikidata and a list of further discussions and considerations.\footnote{\url{https://www.wikidata.org/wiki/Wikidata:WikiProject_LGBT/gender}} Since we are using gender data to calculate aggregate statistics, we judged these limitations to be less problematic than it would be if we were making decisions about individuals. 
        
        In our analysis (see Appendix A), we handle nonbinary gender identities by using four gender categories: unknown, male, female, and third.
        
        We advise great care when working with the gender data, particularly outside the immediate context of the TREC task (either its original instance or using the data to evaluate comparable systems).
        
        \item \textbf{Geography}: For each Wikipedia article, we also ascertained which, if any, countries and continents are relevant to the content.\footnote{Code: \url{https://github.com/geohci/wiki-region-groundtruth/blob/main/wiki-region-data.ipynb}} This was determined by directly looking up several community-maintained Wikidata structured data statements about the article. These properties were checked for the presence of countries, which were then mapped to continents via the United Nation's geoscheme.\footnote{\url{https://en.wikipedia.org/wiki/United_Nations_geoscheme}} While this data must meet Wikidata's verifiability guidelines,\footnote{\url{https://www.wikidata.org/wiki/Wikidata:Verifiability}} it does suffer from varying levels of incompleteness. For example, only 74\% of people on Wikidata have a country of citizenship property.\footnote{\url{https://humaniki.wmcloud.org/gender-by-country}} Furthermore, structured data is itself limited---e.g., country of citizenship does not appropriately capture people who are considered stateless though these people may have many strong ties to a country. It is not easy to evaluate whether this data is missing at random or biased against certain regions of the world. Care should be taken when interpreting the absence of associated continents in the data. Further details can be found in the code repository.\footnote{\url{https://github.com/geohci/wiki-region-groundtruth}}
        
        We also identify the associated countries and continents with the sources in the article. Each source is mapped to a country based on the URL or publisher associated with it. These mappings are built via a mixture of Wikidata, country extraction from whois records, and heuristics related to the top-level domain of the URL.\footnote{See code for more details: \url{https://github.com/geohci/geo-provenance}} This is the only inferred attribute used that is not maintained by Wikimedians and thus is much more likely to contain errors. Because this data was also incomplete, we had the assessors annotate an additional 15,000 items to help add to the data and better understand its quality. The feedback from assessors was that it was a difficult task---i.e. inferring country information for a generic website or publisher is often not easy and can be quite ambiguous at times. While most items were only assessed once, 101 publishers received multiple assessments with 85 (84\%) of these in agreement and 32 URLs received multiple assessments with 26 (81\%) of these in agreement. We also had a few items for which we had already inferred regions that we had the assessors check: only 6 publishers were checked but 5 (83\%) were in agreement and 90 URLs were checked with 82 (91\%) in agreement. While this leaves uncertainty about the publisher data, it does suggest that the URL data is reasonable quality because a few of those in disagreement appear to be assessor errors.
        
        \item \textbf{Age}: We calculate the associated age of the subject of the article in a similar manner to article topic geography (extracting dates from several pre-determined properties on Wikidata).\footnote{See the code for more details: \url{https://gitlab.wikimedia.org/isaacj/miscellaneous-wikimedia/-/blob/master/wikidata-properties-spark/article-age.ipynb}} While geography is a multi-label feature, age is mapped to a single value via the median of the associated years and that median value is bucketed as pre-1900s, 20th century, or 21st century (and beyond). Like geography, it is not clear how many articles should have associated categories---e.g., many articles like those for plant species, do not clearly map to any specific time period---but it is safe to assume all biographies should have associated year data and we see 91.5\% coverage suggesting relatively complete data.
        
        \item \textbf{Occupation}: For each Wikipedia biography, we also ascertained which occupations could be associated with the person it is about. This data is directly determined via Wikidata by collecting the values associated with the occupation property (P106). For each occupation value, we then mapped it to one of 32 higher-order occupations based on the occupation ontology (using P279, the sub-class of, property for each occupation)\footnote{For more details, see the code: \url{https://gitlab.wikimedia.org/isaacj/miscellaneous-wikimedia/-/blob/master/wikidata-properties-spark/wikidata_occupation_taxonomy.ipynb}}. The 32 higher-order occupations were hand-selected to give sufficient detail while remaining a manageable number of categories. On English Wikipedia, 92.1\% of biographies have at least one associated occupation value that could be mapped to the 32 higher-order occupations.
        
        \item \textbf{Popularity}: For each article, we calculated how many pageviews it received in February 2022. These pageview counts are based on webrequest logs\footnote{\url{https://wikitech.wikimedia.org/wiki/Analytics/Data_Lake/Traffic/Webrequest}} and filter out views from user-agents that explicitly identify themselves as spiders\footnote{See: \url{https://meta.wikimedia.org/wiki/Research:Page_view}} and actors (shared user-agent and IP address) that seem to be automated in that they view more than 800 pages per hour\footnote{\url{https://wikitech.wikimedia.org/wiki/Analytics/Data_Lake/Traffic/BotDetection}}. These heuristics are not perfect however and traffic can be easily miscategorized if, for example, automated requests come from many different IPs or devices or actual users share a proxy that gives them the same IP and user-agent. The raw counts of pageviews were then converted into relative values between 0 and 1 by square-root transforming the value and normalizing to the 99th percentile of pageviews. Finally, these values were bucketed as [0 - 0.125), [0.125 - 0.250), [0.25 - 0.5), [0.5 - 1].
        
        \item \textbf{Sitelinks}: The Wikimedia editor community maintains article sitelinks, or interlanguage links---i.e. explicit connection of articles about the same subject across language editions---via Wikidata. Almost all Wikipedia articles (99.92\% for English)\footnote{\url{https://wikidata-analytics.wmcloud.org/app/WD_percentUsageDashboard}} have a corresponding Wikidata item and editors work to merge Wikidata items that are about the same subject so the sitelinks are aligned. Though there is no empirical data, it is generally accepted that most articles are appropriately linked to their corresponding other-language equivalents, especially in languages with shared scripts where simple approaches such as searching for an article title is often sufficient to identify matches.
        
        \item \textbf{Other}: several fairness criteria are relatively straightforward and thus do not have many attached limitations. Specifically, the first letter of the article title (Alphabetical) and age of the article.
    \end{itemize}
    \item \textbf{Relevance Criteria}
    \begin{itemize}
        \item \textbf{WikiProject Relevance}: For the training queries, relevance was obtained from page lists for existing WikiProjects. While WikiProjects have broad coverage of English Wikipedia and we selected for WikiProjects that had tagged new articles in the recent months in the training data as a proxy for activity, it is certain that almost all WikiProjects are incomplete in tagging relevant content (itself a strong motivation for this task). While it is not easy to measure just how incomplete they are, it should not be assumed that content that has not been tagged as relevant to a WikiProject in the training data is indeed irrelevant.\footnote{Current Wikiproject tags were extracted from the database tables maintained by the PageAssessments extension: \url{https://www.mediawiki.org/wiki/Extension:PageAssessments}}

        \item \textbf{Work-needed}: Our proxy for work-needed is a coarse proxy. It is based on just a few simple features (page length, sections, images, categories, links, and references) and does not reflect the nuances of the work needed to craft a top-quality Wikipedia article.\footnote{For further details, see: \url{https://meta.wikimedia.org/wiki/Research:Prioritization_of_Wikipedia_Articles/Language-Agnostic_Quality\#V2}} A fully-fledged system for supporting Wikiprojects would also include a more nuanced approach to understanding the work needed for each article and how to appropriately allocate this work.
    \end{itemize}
    \item \textbf{Task Definition}
    \begin{itemize}
        \item \textbf{Existing Article Bias}: The task is limited to topics for which English Wikipedia already has articles. These tasks are not able to counteract biases in the processes by which articles come to exist (or are deleted~\cite{tripodi2021ms})---recommending articles that should exist but don't is an interesting area for future study.
        
        \item \textbf{Fairness constructs}: we focus on several fairness constructs in this challenge as metrics for which there is high data coverage and a clear mechanism for which "unfair" coverage might arise. That does not mean these are the most important constructs, but others---e.g., religion, sexuality, culture, race---generally are either more challenging to model or map to fairness goals~\cite{redi2020taxonomy}.
    \end{itemize}
\end{itemize}
\bibliography{proposal}
\appendix
\hypertarget{3e5dade8}{}
\hypertarget{page-alignments}{%
\section{Page Alignments}\label{page-alignments}}

This notebook computes the \emph{page alignments} from the Wikipedia
metadata. These are then used by the task-specific alignment notebooks
to compute target distributions and page alignment subsets for retrieved
pages.

\textbf{Warning:} this notebook takes quite a bit of memory to run.

\hypertarget{d06661fd}{}
\hypertarget{setup}{%
\subsection{Setup}\label{setup}}

We begin by loading necessary libraries:

\hypertarget{0a93bc62}{}
\begin{Shaded}
\begin{Highlighting}[]
\ImportTok{import}\NormalTok{ sys}
\ImportTok{from}\NormalTok{ pathlib }\ImportTok{import}\NormalTok{ Path}
\ImportTok{import}\NormalTok{ pandas }\ImportTok{as}\NormalTok{ pd}
\ImportTok{import}\NormalTok{ xarray }\ImportTok{as}\NormalTok{ xr}
\ImportTok{import}\NormalTok{ numpy }\ImportTok{as}\NormalTok{ np}
\ImportTok{import}\NormalTok{ matplotlib.pyplot }\ImportTok{as}\NormalTok{ plt}
\ImportTok{import}\NormalTok{ seaborn }\ImportTok{as}\NormalTok{ sns}
\ImportTok{import}\NormalTok{ gzip}
\ImportTok{import}\NormalTok{ json}
\ImportTok{from}\NormalTok{ natural.size }\ImportTok{import}\NormalTok{ binarysize}
\end{Highlighting}
\end{Shaded}

\leavevmode\vadjust pre{\hypertarget{2c1fa7a7}{}}%
Set up progress bar and logging support:

\hypertarget{3f00e55b}{}
\begin{Shaded}
\begin{Highlighting}[]
\ImportTok{from}\NormalTok{ tqdm.auto }\ImportTok{import}\NormalTok{ tqdm}
\NormalTok{tqdm.pandas(leave}\OperatorTok{=}\VariableTok{False}\NormalTok{)}
\end{Highlighting}
\end{Shaded}

\hypertarget{23c385e0}{}
\begin{Shaded}
\begin{Highlighting}[]
\ImportTok{import}\NormalTok{ sys, logging}
\NormalTok{logging.basicConfig(level}\OperatorTok{=}\NormalTok{logging.INFO, stream}\OperatorTok{=}\NormalTok{sys.stderr)}
\NormalTok{log }\OperatorTok{=}\NormalTok{ logging.getLogger(}\StringTok{\textquotesingle{}PageAlignments\textquotesingle{}}\NormalTok{)}
\end{Highlighting}
\end{Shaded}

\leavevmode\vadjust pre{\hypertarget{5377d31a-3a38-4339-ab55-63a0e0284605}{}}%
And set up an output directory:

\hypertarget{8ea49a9e-355e-4f88-a81a-cef60ccd66a5}{}
\begin{Shaded}
\begin{Highlighting}[]
\ImportTok{from}\NormalTok{ wptrec.save }\ImportTok{import}\NormalTok{ OutRepo}
\NormalTok{output }\OperatorTok{=}\NormalTok{ OutRepo(}\StringTok{\textquotesingle{}data/metric{-}tables\textquotesingle{}}\NormalTok{)}
\end{Highlighting}
\end{Shaded}

\hypertarget{9114cb11-feac-4fa5-912f-ebb4e9190ef3}{}
\hypertarget{loading-data}{%
\subsection{Loading Data}\label{loading-data}}

Now we need to load the data.

\hypertarget{3c1c85d3-1cf4-48a3-9e4a-218ddda6a675}{}
\hypertarget{static-data}{%
\subsubsection{Static Data}\label{static-data}}

We need a set of subregions that are folded into
\href{https://en.wikipedia.org/wiki/United_Nations_geoscheme_for_Oceania}{Oceania}:

\hypertarget{76fc27f7-8d41-43d9-8576-0b11fb09082a}{}
\begin{Shaded}
\begin{Highlighting}[]
\NormalTok{oc\_regions }\OperatorTok{=}\NormalTok{ [}
    \StringTok{\textquotesingle{}Australia and New Zealand\textquotesingle{}}\NormalTok{,}
    \StringTok{\textquotesingle{}Melanesia\textquotesingle{}}\NormalTok{,}
    \StringTok{\textquotesingle{}Micronesia\textquotesingle{}}\NormalTok{,}
    \StringTok{\textquotesingle{}Polynesia\textquotesingle{}}\NormalTok{,}
\NormalTok{]}
\end{Highlighting}
\end{Shaded}

\leavevmode\vadjust pre{\hypertarget{808051bb-4112-4d6d-b9c4-3c81acb82fb9}{}}%
And finally a name for unknown:

\hypertarget{bf79fdb6-c8d4-4833-bae6-13416ffbef2f}{}
\begin{Shaded}
\begin{Highlighting}[]
\NormalTok{UNKNOWN }\OperatorTok{=} \StringTok{\textquotesingle{}@UNKNOWN\textquotesingle{}}
\end{Highlighting}
\end{Shaded}

\leavevmode\vadjust pre{\hypertarget{768ec207-ccde-4adb-bf8c-86181420a15a}{}}%
Now all our background data is set up.

\hypertarget{3ddf9824}{}
\hypertarget{page-data}{%
\subsubsection{Page Data}\label{page-data}}

Finally, we load the page metadata. This is a little manual to manage
memory usage. Two memory usage tricks:

\begin{itemize}
\tightlist
\item
  Only import the things we need
\item
  Use \texttt{sys.intern} for strings representing categoricals to
  decrease memory use
\end{itemize}

Bonus is that, through careful logic, we get a progress bar.

\hypertarget{fbc0362f-2993-48d5-b8b4-4d2a156e1cd1}{}
\begin{Shaded}
\begin{Highlighting}[]
\CommentTok{\# META\_FILE\_TAG = \textquotesingle{}discrete\textquotesingle{}}
\NormalTok{META\_FILE\_TAG }\OperatorTok{=} \StringTok{\textquotesingle{}discrete\_assessed\textquotesingle{}}
\end{Highlighting}
\end{Shaded}

\hypertarget{2f7f76a8-5868-4516-890b-d09c888e24e9}{}
\begin{Shaded}
\begin{Highlighting}[]
\NormalTok{page\_path }\OperatorTok{=}\NormalTok{ Path(}\SpecialStringTok{f\textquotesingle{}data/trec\_2022\_articles\_}\SpecialCharTok{\{}\NormalTok{META\_FILE\_TAG}\SpecialCharTok{\}}\SpecialStringTok{.json.gz\textquotesingle{}}\NormalTok{)}
\NormalTok{page\_file\_size }\OperatorTok{=}\NormalTok{ page\_path.stat().st\_size}
\NormalTok{binarysize(page\_file\_size)}
\end{Highlighting}
\end{Shaded}

\begin{verbatim}
'238.76 MiB'
\end{verbatim}

\hypertarget{28667620-8853-4a24-872d-a9144be7c4f5}{}
\hypertarget{definitions}{%
\paragraph{Definitions}\label{definitions}}

Let's define the different attributes we need to extract:

\hypertarget{d0801581-d3b5-484b-8b5f-f838c861e8ef}{}
\begin{Shaded}
\begin{Highlighting}[]
\NormalTok{SUB\_GEO\_ATTR }\OperatorTok{=} \StringTok{\textquotesingle{}page\_subcont\_regions\textquotesingle{}}
\NormalTok{SRC\_GEO\_ATTR }\OperatorTok{=} \StringTok{\textquotesingle{}source\_subcont\_regions\textquotesingle{}}
\NormalTok{GENDER\_ATTR }\OperatorTok{=} \StringTok{\textquotesingle{}gender\textquotesingle{}}
\NormalTok{OCC\_ATTR }\OperatorTok{=} \StringTok{\textquotesingle{}occupations\textquotesingle{}}
\NormalTok{BASIC\_ATTRS }\OperatorTok{=}\NormalTok{ [}
    \StringTok{\textquotesingle{}page\_id\textquotesingle{}}\NormalTok{,}
    \StringTok{\textquotesingle{}first\_letter\_category\textquotesingle{}}\NormalTok{,}
    \StringTok{\textquotesingle{}creation\_date\_category\textquotesingle{}}\NormalTok{,}
    \StringTok{\textquotesingle{}relative\_pageviews\_category\textquotesingle{}}\NormalTok{,}
    \StringTok{\textquotesingle{}num\_sitelinks\_category\textquotesingle{}}\NormalTok{,}
\NormalTok{]}
\end{Highlighting}
\end{Shaded}

\hypertarget{2002989c-2606-432b-b94c-2e558fab6e9c}{}
\hypertarget{read-data}{%
\paragraph{Read Data}\label{read-data}}

Now, we're going to process by creating lists we can reassemble with
\texttt{pd.DataFrame.from\_records}. We'll fill these with tuples and
dictionaries as appropriate.

\hypertarget{cc6e2d99-a10d-4d55-bb8d-0cac56cd8af7}{}
\begin{Shaded}
\begin{Highlighting}[]
\NormalTok{qual\_recs }\OperatorTok{=}\NormalTok{ []}
\NormalTok{sub\_geo\_recs }\OperatorTok{=}\NormalTok{ []}
\NormalTok{src\_geo\_recs }\OperatorTok{=}\NormalTok{ []}
\NormalTok{gender\_recs }\OperatorTok{=}\NormalTok{ []}
\NormalTok{occ\_recs }\OperatorTok{=}\NormalTok{ []}
\NormalTok{att\_recs }\OperatorTok{=}\NormalTok{ []}
\NormalTok{seen\_pages }\OperatorTok{=} \BuiltInTok{set}\NormalTok{()}
\end{Highlighting}
\end{Shaded}

\leavevmode\vadjust pre{\hypertarget{b4a9b41f-6dd4-44b6-9e7e-c0510a2e4b25}{}}%
And we're off.

\hypertarget{50069a6d-dbd8-4fb7-a1c5-e31ef3d5586b}{}
\begin{Shaded}
\begin{Highlighting}[]
\ControlFlowTok{with}\NormalTok{ tqdm(total}\OperatorTok{=}\NormalTok{page\_file\_size, desc}\OperatorTok{=}\StringTok{\textquotesingle{}compressed input\textquotesingle{}}\NormalTok{, unit}\OperatorTok{=}\StringTok{\textquotesingle{}B\textquotesingle{}}\NormalTok{, unit\_scale}\OperatorTok{=}\VariableTok{True}\NormalTok{) }\ImportTok{as}\NormalTok{ fpb:}
    \ControlFlowTok{with} \BuiltInTok{open}\NormalTok{(page\_path, }\StringTok{\textquotesingle{}rb\textquotesingle{}}\NormalTok{) }\ImportTok{as}\NormalTok{ gzf, gzip.GzipFile(fileobj}\OperatorTok{=}\NormalTok{gzf, mode}\OperatorTok{=}\StringTok{\textquotesingle{}r\textquotesingle{}}\NormalTok{) }\ImportTok{as}\NormalTok{ decoded:}
        \ControlFlowTok{for}\NormalTok{ line }\KeywordTok{in}\NormalTok{ decoded:}
\NormalTok{            line }\OperatorTok{=}\NormalTok{ json.loads(line)}
\NormalTok{            page }\OperatorTok{=}\NormalTok{ line[}\StringTok{\textquotesingle{}page\_id\textquotesingle{}}\NormalTok{]}
            \ControlFlowTok{if}\NormalTok{ page }\KeywordTok{in}\NormalTok{ seen\_pages:}
                \ControlFlowTok{continue}
            \ControlFlowTok{else}\NormalTok{:}
\NormalTok{                seen\_pages.add(page)}
            
            \CommentTok{\# page quality}
\NormalTok{            qual\_recs.append((page, line[}\StringTok{\textquotesingle{}qual\_cat\textquotesingle{}}\NormalTok{]))}
            
            \CommentTok{\# page geography}
            \ControlFlowTok{for}\NormalTok{ geo }\KeywordTok{in}\NormalTok{ line[SUB\_GEO\_ATTR]:}
\NormalTok{                sub\_geo\_recs.append((page, sys.}\BuiltInTok{intern}\NormalTok{(geo)))}
            
            \CommentTok{\# src geography}
\NormalTok{            psg }\OperatorTok{=}\NormalTok{ \{}\StringTok{\textquotesingle{}page\_id\textquotesingle{}}\NormalTok{: page\}}
            \ControlFlowTok{for}\NormalTok{ g, v }\KeywordTok{in}\NormalTok{ line[SRC\_GEO\_ATTR].items():}
                \ControlFlowTok{if}\NormalTok{ g }\OperatorTok{==} \StringTok{\textquotesingle{}UNK\textquotesingle{}}\NormalTok{:}
\NormalTok{                    g }\OperatorTok{=}\NormalTok{ UNKNOWN}
\NormalTok{                psg[sys.}\BuiltInTok{intern}\NormalTok{(g)] }\OperatorTok{=}\NormalTok{ v}
\NormalTok{            src\_geo\_recs.append(psg)}
            
            \CommentTok{\# genders}
            \ControlFlowTok{for}\NormalTok{ g }\KeywordTok{in}\NormalTok{ line[GENDER\_ATTR]:}
\NormalTok{                gender\_recs.append((page, sys.}\BuiltInTok{intern}\NormalTok{(g)))}
            
            \CommentTok{\# occupations}
            \ControlFlowTok{for}\NormalTok{ occ }\KeywordTok{in}\NormalTok{ line[OCC\_ATTR]:}
\NormalTok{                occ\_recs.append((page, sys.}\BuiltInTok{intern}\NormalTok{(occ)))}
            
            \CommentTok{\# other attributes}
\NormalTok{            att\_recs.append(}\BuiltInTok{tuple}\NormalTok{((sys.}\BuiltInTok{intern}\NormalTok{(line[a]) }\ControlFlowTok{if} \BuiltInTok{isinstance}\NormalTok{(line[a], }\BuiltInTok{str}\NormalTok{) }\ControlFlowTok{else}\NormalTok{ line[a])}
                                  \ControlFlowTok{for}\NormalTok{ a }\KeywordTok{in}\NormalTok{ BASIC\_ATTRS))}
            
\NormalTok{            fpb.update(gzf.tell() }\OperatorTok{{-}}\NormalTok{ fpb.n)  }\CommentTok{\# update the progress bar}
\end{Highlighting}
\end{Shaded}

\begin{Shaded}
\begin{Highlighting}[]
\FunctionTok{\{}\DataTypeTok{"model\_id"}\FunctionTok{:}\StringTok{"7a8ed81f35ca4fa0b50c58c43638f3e0"}\FunctionTok{,}\DataTypeTok{"version\_major"}\FunctionTok{:}\DecValTok{2}\FunctionTok{,}\DataTypeTok{"version\_minor"}\FunctionTok{:}\DecValTok{0}\FunctionTok{\}}
\end{Highlighting}
\end{Shaded}

\hypertarget{8f28ebe0-1dff-49a7-b099-2e469d60cb9f}{}
\hypertarget{reassemble-dfs}{%
\paragraph{Reassemble DFs}\label{reassemble-dfs}}

Now we will assemble these records into data frames.

\hypertarget{1026cef3-3dcf-4f9c-8f57-0442ba62e1ff}{}
\begin{Shaded}
\begin{Highlighting}[]
\NormalTok{quality }\OperatorTok{=}\NormalTok{ pd.DataFrame.from\_records(qual\_recs, columns}\OperatorTok{=}\NormalTok{[}\StringTok{\textquotesingle{}page\_id\textquotesingle{}}\NormalTok{, }\StringTok{\textquotesingle{}quality\textquotesingle{}}\NormalTok{])}
\end{Highlighting}
\end{Shaded}

\hypertarget{8df5bc49-4ed7-4be5-b1e5-7920bf353b10}{}
\begin{Shaded}
\begin{Highlighting}[]
\NormalTok{sub\_geo }\OperatorTok{=}\NormalTok{ pd.DataFrame.from\_records(sub\_geo\_recs, columns}\OperatorTok{=}\NormalTok{[}\StringTok{\textquotesingle{}page\_id\textquotesingle{}}\NormalTok{, }\StringTok{\textquotesingle{}sub\_geo\textquotesingle{}}\NormalTok{])}
\NormalTok{sub\_geo.info()}
\end{Highlighting}
\end{Shaded}

\begin{verbatim}
<class 'pandas.core.frame.DataFrame'>
RangeIndex: 3773443 entries, 0 to 3773442
Data columns (total 2 columns):
 #   Column   Dtype 
---  ------   ----- 
 0   page_id  int64 
 1   sub_geo  object
dtypes: int64(1), object(1)
memory usage: 57.6+ MB
\end{verbatim}

\hypertarget{6419221c-d3dd-4d9e-ba40-762a6dee0b17}{}
\begin{Shaded}
\begin{Highlighting}[]
\NormalTok{src\_geo }\OperatorTok{=}\NormalTok{ pd.DataFrame.from\_records(src\_geo\_recs)}
\NormalTok{src\_geo.info()}
\end{Highlighting}
\end{Shaded}

\begin{verbatim}
<class 'pandas.core.frame.DataFrame'>
RangeIndex: 6460210 entries, 0 to 6460209
Data columns (total 25 columns):
 #   Column                     Dtype  
---  ------                     -----  
 0   page_id                    int64  
 1   Northern America           float64
 2   @UNKNOWN                   float64
 3   Northern Europe            float64
 4   Western Asia               float64
 5   Western Europe             float64
 6   Western Africa             float64
 7   Southern Europe            float64
 8   Australia and New Zealand  float64
 9   Central America            float64
 10  Eastern Asia               float64
 11  South America              float64
 12  Eastern Europe             float64
 13  Northern Africa            float64
 14  Eastern Africa             float64
 15  Southern Asia              float64
 16  Polynesia                  float64
 17  South-eastern Asia         float64
 18  Central Asia               float64
 19  Caribbean                  float64
 20  Southern Africa            float64
 21  Middle Africa              float64
 22  Antarctica                 float64
 23  Melanesia                  float64
 24  Micronesia                 float64
dtypes: float64(24), int64(1)
memory usage: 1.2 GB
\end{verbatim}

\hypertarget{2bf62b4b-c366-4dda-bfe5-728157e2f341}{}
\begin{Shaded}
\begin{Highlighting}[]
\NormalTok{gender }\OperatorTok{=}\NormalTok{ pd.DataFrame.from\_records(gender\_recs, columns}\OperatorTok{=}\NormalTok{[}\StringTok{\textquotesingle{}page\_id\textquotesingle{}}\NormalTok{, }\StringTok{\textquotesingle{}gender\textquotesingle{}}\NormalTok{])}
\NormalTok{gender.info()}
\end{Highlighting}
\end{Shaded}

\begin{verbatim}
<class 'pandas.core.frame.DataFrame'>
RangeIndex: 1850219 entries, 0 to 1850218
Data columns (total 2 columns):
 #   Column   Dtype 
---  ------   ----- 
 0   page_id  int64 
 1   gender   object
dtypes: int64(1), object(1)
memory usage: 28.2+ MB
\end{verbatim}

\hypertarget{0f8c37ab-9a56-462e-b0a2-971463b48b3f}{}
\begin{Shaded}
\begin{Highlighting}[]
\NormalTok{occupations }\OperatorTok{=}\NormalTok{ pd.DataFrame.from\_records(occ\_recs, columns}\OperatorTok{=}\NormalTok{[}\StringTok{\textquotesingle{}page\_id\textquotesingle{}}\NormalTok{, }\StringTok{\textquotesingle{}occ\textquotesingle{}}\NormalTok{])}
\NormalTok{occupations.info()}
\end{Highlighting}
\end{Shaded}

\begin{verbatim}
<class 'pandas.core.frame.DataFrame'>
RangeIndex: 2445899 entries, 0 to 2445898
Data columns (total 2 columns):
 #   Column   Dtype 
---  ------   ----- 
 0   page_id  int64 
 1   occ      object
dtypes: int64(1), object(1)
memory usage: 37.3+ MB
\end{verbatim}

\hypertarget{52f7dff1-f946-453c-aa9d-64b16073137e}{}
\begin{Shaded}
\begin{Highlighting}[]
\NormalTok{cat\_attrs }\OperatorTok{=}\NormalTok{ pd.DataFrame.from\_records(att\_recs, columns}\OperatorTok{=}\NormalTok{BASIC\_ATTRS)}
\NormalTok{cat\_attrs.info()}
\end{Highlighting}
\end{Shaded}

\begin{verbatim}
<class 'pandas.core.frame.DataFrame'>
RangeIndex: 6460210 entries, 0 to 6460209
Data columns (total 5 columns):
 #   Column                       Dtype 
---  ------                       ----- 
 0   page_id                      int64 
 1   first_letter_category        object
 2   creation_date_category       object
 3   relative_pageviews_category  object
 4   num_sitelinks_category       object
dtypes: int64(1), object(4)
memory usage: 246.4+ MB
\end{verbatim}

\hypertarget{14ed691f-f5bc-4447-abbc-35bae4de32c5}{}
\begin{Shaded}
\begin{Highlighting}[]
\NormalTok{all\_pages }\OperatorTok{=}\NormalTok{ np.array(}\BuiltInTok{list}\NormalTok{(seen\_pages))}
\NormalTok{all\_pages }\OperatorTok{=}\NormalTok{ np.sort(all\_pages)}
\NormalTok{all\_pages }\OperatorTok{=}\NormalTok{ pd.Series(all\_pages)}
\end{Highlighting}
\end{Shaded}

\hypertarget{cc2d4340-c566-4875-8236-7a80421072eb}{}
\begin{Shaded}
\begin{Highlighting}[]
\KeywordTok{del}\NormalTok{ src\_geo\_recs, sub\_geo\_recs}
\KeywordTok{del}\NormalTok{ gender\_recs, occ\_recs}
\KeywordTok{del}\NormalTok{ seen\_pages}
\end{Highlighting}
\end{Shaded}

\hypertarget{b1e154b6-1216-41fa-94fc-0c6f686cf4a1}{}
\begin{Shaded}
\begin{Highlighting}[]
\OperatorTok{\%}\NormalTok{reset }\OperatorTok{{-}}\NormalTok{f out}
\end{Highlighting}
\end{Shaded}

\begin{verbatim}
Flushing output cache (1 entries)
\end{verbatim}

\hypertarget{9978065d-a54b-49be-a912-e9012a088dca}{}
\begin{Shaded}
\begin{Highlighting}[]
\ImportTok{import}\NormalTok{ gc}
\NormalTok{gc.collect()}
\end{Highlighting}
\end{Shaded}

\begin{verbatim}
0
\end{verbatim}

\hypertarget{de0d0123-f424-44e5-bec6-9b8e3a3be5cd}{}
\hypertarget{helper-functions}{%
\subsection{Helper Functions}\label{helper-functions}}

These functions will help with further computations.

\hypertarget{e3f26b88-d363-4bae-b6e8-fab1568c1c9c}{}
\hypertarget{normalize-distribution}{%
\subsubsection{Normalize Distribution}\label{normalize-distribution}}

We are going to compute a number of data frames that are alignment
vectors, such that each row is to be a multinomial distribution. This
function normalizes such a frame.

\hypertarget{337588d2-15ae-4dfc-bcdb-f1ae9f796083}{}
\begin{Shaded}
\begin{Highlighting}[]
\KeywordTok{def}\NormalTok{ norm\_align\_matrix(df):}
\NormalTok{    df }\OperatorTok{=}\NormalTok{ df.fillna(}\DecValTok{0}\NormalTok{)}
\NormalTok{    sums }\OperatorTok{=}\NormalTok{ df.}\BuiltInTok{sum}\NormalTok{(axis}\OperatorTok{=}\StringTok{\textquotesingle{}columns\textquotesingle{}}\NormalTok{)}
    \ControlFlowTok{return}\NormalTok{ df.div(sums, axis}\OperatorTok{=}\StringTok{\textquotesingle{}rows\textquotesingle{}}\NormalTok{)}
\end{Highlighting}
\end{Shaded}

\hypertarget{9a89ae2f}{}
\hypertarget{page-alignments}{%
\subsection{Page Alignments}\label{page-alignments}}

All of our metrics require page "alignments": the protected-group
membership of each page.

\hypertarget{4ee8291d-dc02-4664-9456-b76a7c2f0536}{}
\hypertarget{quality}{%
\subsubsection{Quality}\label{quality}}

Quality isn't an alignment, but we're going to save it here:

\hypertarget{075d03bb-2bc7-43a6-b6d0-b5fe67f63b4c}{}
\begin{Shaded}
\begin{Highlighting}[]
\NormalTok{output.save\_table(quality, }\StringTok{\textquotesingle{}page{-}quality\textquotesingle{}}\NormalTok{, parquet}\OperatorTok{=}\VariableTok{True}\NormalTok{)}
\end{Highlighting}
\end{Shaded}

\begin{verbatim}
INFO:wptrec.save:saving CSV to data\metric-tables\page-quality.csv.gz
INFO:wptrec.save:data\metric-tables\page-quality.csv.gz: 35.62 MiB
INFO:wptrec.save:saving Parquet to data\metric-tables\page-quality.parquet
INFO:wptrec.save:data\metric-tables\page-quality.parquet: 9.14 MiB
\end{verbatim}

\hypertarget{c665c07b}{}
\hypertarget{page-geography}{%
\subsubsection{Page Geography}\label{page-geography}}

Let's start with the straight page geography alignment for the public
evaluation of the training queries. We've already loaded it above.

We need to do a little cleanup on this data:

\begin{itemize}
\tightlist
\item
  Align pages with no known geography with '@UNKNOWN' (to sort before
  known categories)
\item
  Replace Oceania subregions with Oceania
\end{itemize}

\hypertarget{d7ad9c07-3d61-4507-8ef4-c815ce86bb96}{}
\begin{Shaded}
\begin{Highlighting}[]
\NormalTok{sub\_geo.head()}
\end{Highlighting}
\end{Shaded}

\begin{verbatim}
   page_id           sub_geo
0      303  Northern America
1      307  Northern America
2      316  Northern America
3      324  Northern America
4      330   Southern Europe
\end{verbatim}

\leavevmode\vadjust pre{\hypertarget{7fe022cb-48c1-47ad-afc3-b7ebbbf50657}{}}%
Let's start by turning this into a wide frame:

\hypertarget{0fd86087-e9c0-46ac-b9d2-3c32b9c374bf}{}
\begin{Shaded}
\begin{Highlighting}[]
\NormalTok{sub\_geo\_align }\OperatorTok{=}\NormalTok{ sub\_geo.assign(x}\OperatorTok{=}\DecValTok{1}\NormalTok{).pivot(index}\OperatorTok{=}\StringTok{\textquotesingle{}page\_id\textquotesingle{}}\NormalTok{, columns}\OperatorTok{=}\StringTok{\textquotesingle{}sub\_geo\textquotesingle{}}\NormalTok{, values}\OperatorTok{=}\StringTok{\textquotesingle{}x\textquotesingle{}}\NormalTok{)}
\NormalTok{sub\_geo\_align.fillna(}\DecValTok{0}\NormalTok{, inplace}\OperatorTok{=}\VariableTok{True}\NormalTok{)}
\NormalTok{sub\_geo\_align.head()}
\end{Highlighting}
\end{Shaded}

\begin{verbatim}
sub_geo  Antarctica  Australia and New Zealand  Caribbean  Central America  \
page_id                                                                      
303             0.0                        0.0        0.0              0.0   
307             0.0                        0.0        0.0              0.0   
316             0.0                        0.0        0.0              0.0   
324             0.0                        0.0        0.0              0.0   
330             0.0                        0.0        0.0              0.0   

sub_geo  Central Asia  Eastern Africa  Eastern Asia  Eastern Europe  \
page_id                                                               
303               0.0             0.0           0.0             0.0   
307               0.0             0.0           0.0             0.0   
316               0.0             0.0           0.0             0.0   
324               0.0             0.0           0.0             0.0   
330               0.0             0.0           0.0             0.0   

sub_geo  Melanesia  Micronesia  ...  Northern Europe  Polynesia  \
page_id                         ...                               
303            0.0         0.0  ...              0.0        0.0   
307            0.0         0.0  ...              0.0        0.0   
316            0.0         0.0  ...              0.0        0.0   
324            0.0         0.0  ...              0.0        0.0   
330            0.0         0.0  ...              0.0        0.0   

sub_geo  South America  South-eastern Asia  Southern Africa  Southern Asia  \
page_id                                                                      
303                0.0                 0.0              0.0            0.0   
307                0.0                 0.0              0.0            0.0   
316                0.0                 0.0              0.0            0.0   
324                0.0                 0.0              0.0            0.0   
330                0.0                 0.0              0.0            0.0   

sub_geo  Southern Europe  Western Africa  Western Asia  Western Europe  
page_id                                                                 
303                  0.0             0.0           0.0             0.0  
307                  0.0             0.0           0.0             0.0  
316                  0.0             0.0           0.0             0.0  
324                  0.0             0.0           0.0             0.0  
330                  1.0             0.0           0.0             0.0  

[5 rows x 23 columns]
\end{verbatim}

\leavevmode\vadjust pre{\hypertarget{157c9a96-f1a7-4e32-849f-0c2181b1f51e}{}}%
Now we need to collapse Oceania into one column.

\hypertarget{beff97bf-279d-410d-89a8-5bb0b07a7c3a}{}
\begin{Shaded}
\begin{Highlighting}[]
\NormalTok{ocean }\OperatorTok{=}\NormalTok{ sub\_geo\_align.loc[:, oc\_regions].}\BuiltInTok{sum}\NormalTok{(axis}\OperatorTok{=}\StringTok{\textquotesingle{}columns\textquotesingle{}}\NormalTok{)}
\NormalTok{sub\_geo\_align }\OperatorTok{=}\NormalTok{ sub\_geo\_align.drop(columns}\OperatorTok{=}\NormalTok{oc\_regions)}
\NormalTok{sub\_geo\_align[}\StringTok{\textquotesingle{}Oceania\textquotesingle{}}\NormalTok{] }\OperatorTok{=}\NormalTok{ ocean}
\end{Highlighting}
\end{Shaded}

\leavevmode\vadjust pre{\hypertarget{877d4ad5-3bd3-4e96-9bda-382a5c584b7a}{}}%
Next we need to add the Unknown column and expand this.

Sum the items to find total amounts, and then create a series for
unknown:

\hypertarget{3bab9730-dc73-43c0-b586-c66b5c0969e3}{}
\begin{Shaded}
\begin{Highlighting}[]
\NormalTok{sub\_geo\_sums }\OperatorTok{=}\NormalTok{ sub\_geo\_align.}\BuiltInTok{sum}\NormalTok{(axis}\OperatorTok{=}\StringTok{\textquotesingle{}columns\textquotesingle{}}\NormalTok{)}
\NormalTok{sub\_geo\_unknown }\OperatorTok{=} \OperatorTok{\textasciitilde{}}\NormalTok{(sub\_geo\_sums }\OperatorTok{\textgreater{}} \DecValTok{0}\NormalTok{)}
\NormalTok{sub\_geo\_unknown }\OperatorTok{=}\NormalTok{ sub\_geo\_unknown.astype(}\StringTok{\textquotesingle{}f8\textquotesingle{}}\NormalTok{)}
\NormalTok{sub\_geo\_unknown }\OperatorTok{=}\NormalTok{ sub\_geo\_unknown.reindex(all\_pages, fill\_value}\OperatorTok{=}\DecValTok{1}\NormalTok{)}
\end{Highlighting}
\end{Shaded}

\leavevmode\vadjust pre{\hypertarget{bddec02e-fc45-427b-bdb0-5b20c0ef5a2f}{}}%
Now let's join this with the original frame:

\hypertarget{387afe95-b66b-4125-acc1-20375d20a888}{}
\begin{Shaded}
\begin{Highlighting}[]
\NormalTok{sub\_geo\_align }\OperatorTok{=}\NormalTok{ sub\_geo\_unknown.to\_frame(UNKNOWN).join(sub\_geo\_align, how}\OperatorTok{=}\StringTok{\textquotesingle{}left\textquotesingle{}}\NormalTok{)}
\NormalTok{sub\_geo\_align }\OperatorTok{=}\NormalTok{ norm\_align\_matrix(sub\_geo\_align)}
\NormalTok{sub\_geo\_align.head()}
\end{Highlighting}
\end{Shaded}

\begin{verbatim}
     @UNKNOWN  Antarctica  Caribbean  Central America  Central Asia  \
12        1.0         0.0        0.0              0.0           0.0   
25        1.0         0.0        0.0              0.0           0.0   
39        1.0         0.0        0.0              0.0           0.0   
290       1.0         0.0        0.0              0.0           0.0   
303       0.0         0.0        0.0              0.0           0.0   

     Eastern Africa  Eastern Asia  Eastern Europe  Middle Africa  \
12              0.0           0.0             0.0            0.0   
25              0.0           0.0             0.0            0.0   
39              0.0           0.0             0.0            0.0   
290             0.0           0.0             0.0            0.0   
303             0.0           0.0             0.0            0.0   

     Northern Africa  ...  Northern Europe  South America  South-eastern Asia  \
12               0.0  ...              0.0            0.0                 0.0   
25               0.0  ...              0.0            0.0                 0.0   
39               0.0  ...              0.0            0.0                 0.0   
290              0.0  ...              0.0            0.0                 0.0   
303              0.0  ...              0.0            0.0                 0.0   

     Southern Africa  Southern Asia  Southern Europe  Western Africa  \
12               0.0            0.0              0.0             0.0   
25               0.0            0.0              0.0             0.0   
39               0.0            0.0              0.0             0.0   
290              0.0            0.0              0.0             0.0   
303              0.0            0.0              0.0             0.0   

     Western Asia  Western Europe  Oceania  
12            0.0             0.0      0.0  
25            0.0             0.0      0.0  
39            0.0             0.0      0.0  
290           0.0             0.0      0.0  
303           0.0             0.0      0.0  

[5 rows x 21 columns]
\end{verbatim}

\hypertarget{65422f09-8269-4d4b-b5d7-95b6fc33ae5d}{}
\begin{Shaded}
\begin{Highlighting}[]
\NormalTok{sub\_geo\_align.sort\_index(axis}\OperatorTok{=}\StringTok{\textquotesingle{}columns\textquotesingle{}}\NormalTok{, inplace}\OperatorTok{=}\VariableTok{True}\NormalTok{)}
\NormalTok{sub\_geo\_align.info()}
\end{Highlighting}
\end{Shaded}

\begin{verbatim}
<class 'pandas.core.frame.DataFrame'>
Int64Index: 6460210 entries, 12 to 70194530
Data columns (total 21 columns):
 #   Column              Dtype  
---  ------              -----  
 0   @UNKNOWN            float64
 1   Antarctica          float64
 2   Caribbean           float64
 3   Central America     float64
 4   Central Asia        float64
 5   Eastern Africa      float64
 6   Eastern Asia        float64
 7   Eastern Europe      float64
 8   Middle Africa       float64
 9   Northern Africa     float64
 10  Northern America    float64
 11  Northern Europe     float64
 12  Oceania             float64
 13  South America       float64
 14  South-eastern Asia  float64
 15  Southern Africa     float64
 16  Southern Asia       float64
 17  Southern Europe     float64
 18  Western Africa      float64
 19  Western Asia        float64
 20  Western Europe      float64
dtypes: float64(21)
memory usage: 1.3 GB
\end{verbatim}

\leavevmode\vadjust pre{\hypertarget{035a69ba}{}}%
And convert this to an xarray for multidimensional usage:

\hypertarget{608b86cf}{}
\begin{Shaded}
\begin{Highlighting}[]
\NormalTok{sub\_geo\_xr }\OperatorTok{=}\NormalTok{ xr.DataArray(sub\_geo\_align, dims}\OperatorTok{=}\NormalTok{[}\StringTok{\textquotesingle{}page\textquotesingle{}}\NormalTok{, }\StringTok{\textquotesingle{}sub\_geo\textquotesingle{}}\NormalTok{])}
\NormalTok{sub\_geo\_xr}
\end{Highlighting}
\end{Shaded}

\begin{verbatim}
<xarray.DataArray (page: 6460210, sub_geo: 21)>
array([[1., 0., 0., ..., 0., 0., 0.],
       [1., 0., 0., ..., 0., 0., 0.],
       [1., 0., 0., ..., 0., 0., 0.],
       ...,
       [1., 0., 0., ..., 0., 0., 0.],
       [1., 0., 0., ..., 0., 0., 0.],
       [1., 0., 0., ..., 0., 0., 0.]])
Coordinates:
  * page     (page) int64 12 25 39 290 ... 70194480 70194481 70194489 70194530
  * sub_geo  (sub_geo) object '@UNKNOWN' 'Antarctica' ... 'Western Europe'
\end{verbatim}

\hypertarget{fe14d543}{}
\begin{Shaded}
\begin{Highlighting}[]
\NormalTok{binarysize(sub\_geo\_xr.nbytes)}
\end{Highlighting}
\end{Shaded}

\begin{verbatim}
'1.90 GiB'
\end{verbatim}

\hypertarget{d821286d-acb0-47cb-bf46-ee9ed43527a5}{}
\begin{Shaded}
\begin{Highlighting}[]
\NormalTok{output.save\_table(sub\_geo\_align, }\StringTok{\textquotesingle{}page{-}sub{-}geo{-}align\textquotesingle{}}\NormalTok{, parquet}\OperatorTok{=}\VariableTok{True}\NormalTok{)}
\end{Highlighting}
\end{Shaded}

\begin{verbatim}
INFO:wptrec.save:saving CSV to data\metric-tables\page-sub-geo-align.csv.gz
INFO:wptrec.save:data\metric-tables\page-sub-geo-align.csv.gz: 23.97 MiB
INFO:wptrec.save:saving Parquet to data\metric-tables\page-sub-geo-align.parquet
INFO:wptrec.save:data\metric-tables\page-sub-geo-align.parquet: 13.20 MiB
\end{verbatim}

\hypertarget{3ed855ea-3296-4add-a213-aba02955ae1a}{}
\hypertarget{page-source-geography}{%
\subsubsection{Page Source Geography}\label{page-source-geography}}

We now need to do a similar setup for page source geography, which comes
to us as a multinomial distribution already.

\hypertarget{0c88b853-9197-4883-8121-f5e58dbc7a27}{}
\begin{Shaded}
\begin{Highlighting}[]
\NormalTok{src\_geo.head()}
\end{Highlighting}
\end{Shaded}

\begin{verbatim}
   page_id  Northern America  @UNKNOWN  Northern Europe  Western Asia  \
0       12              50.0      42.0             40.0           2.0   
1       25              42.0     152.0             16.0           NaN   
2       39              24.0      25.0              6.0           NaN   
3      290              15.0      13.0              3.0           NaN   
4      303             202.0      23.0              9.0           NaN   

   Western Europe  Western Africa  Southern Europe  Australia and New Zealand  \
0             NaN             NaN              NaN                        NaN   
1             3.0             2.0              NaN                        NaN   
2             5.0             NaN              NaN                        NaN   
3             1.0             NaN              NaN                        NaN   
4             4.0             NaN              NaN                        NaN   

   Central America  ...  Southern Asia  Polynesia  South-eastern Asia  \
0              NaN  ...            NaN        NaN                 NaN   
1              NaN  ...            NaN        NaN                 NaN   
2              NaN  ...            NaN        NaN                 NaN   
3              NaN  ...            NaN        NaN                 NaN   
4              NaN  ...            NaN        NaN                 NaN   

   Central Asia  Caribbean  Southern Africa  Middle Africa  Antarctica  \
0           NaN        NaN              NaN            NaN         NaN   
1           NaN        NaN              NaN            NaN         NaN   
2           NaN        NaN              NaN            NaN         NaN   
3           NaN        NaN              NaN            NaN         NaN   
4           NaN        NaN              NaN            NaN         NaN   

   Melanesia  Micronesia  
0        NaN         NaN  
1        NaN         NaN  
2        NaN         NaN  
3        NaN         NaN  
4        NaN         NaN  

[5 rows x 25 columns]
\end{verbatim}

\leavevmode\vadjust pre{\hypertarget{e95d6351-850f-4508-9693-a60a3fc11457}{}}%
Set up the index:

\hypertarget{a986a9da-2bfb-4189-8db5-e67d7298bfb0}{}
\begin{Shaded}
\begin{Highlighting}[]
\NormalTok{src\_geo.set\_index(}\StringTok{\textquotesingle{}page\_id\textquotesingle{}}\NormalTok{, inplace}\OperatorTok{=}\VariableTok{True}\NormalTok{)}
\end{Highlighting}
\end{Shaded}

\leavevmode\vadjust pre{\hypertarget{fdcaba35-14b0-4462-abc2-0bf5afa44edd}{}}%
Expand, then put 1 in UNKNOWN for everything that's missing:

\hypertarget{c612164b-9ba6-4abf-9fce-aa8a13fa3b8c}{}
\begin{Shaded}
\begin{Highlighting}[]
\NormalTok{src\_geo\_align }\OperatorTok{=}\NormalTok{ src\_geo.reindex(all\_pages, fill\_value}\OperatorTok{=}\DecValTok{0}\NormalTok{)}
\NormalTok{src\_geo\_align.loc[src\_geo\_align.}\BuiltInTok{sum}\NormalTok{(}\StringTok{\textquotesingle{}columns\textquotesingle{}}\NormalTok{) }\OperatorTok{==} \DecValTok{0}\NormalTok{, UNKNOWN] }\OperatorTok{=} \DecValTok{1}
\NormalTok{src\_geo\_align}
\end{Highlighting}
\end{Shaded}

\begin{verbatim}
          Northern America  @UNKNOWN  Northern Europe  Western Asia  \
12                    50.0      42.0             40.0           2.0   
25                    42.0     152.0             16.0           NaN   
39                    24.0      25.0              6.0           NaN   
290                   15.0      13.0              3.0           NaN   
303                  202.0      23.0              9.0           NaN   
...                    ...       ...              ...           ...   
70194419               NaN       1.0              NaN           NaN   
70194480               NaN       1.0              NaN           NaN   
70194481               7.0       1.0              NaN           NaN   
70194489               NaN       2.0              NaN           NaN   
70194530               8.0       NaN              NaN           NaN   

          Western Europe  Western Africa  Southern Europe  \
12                   NaN             NaN              NaN   
25                   3.0             2.0              NaN   
39                   5.0             NaN              NaN   
290                  1.0             NaN              NaN   
303                  4.0             NaN              NaN   
...                  ...             ...              ...   
70194419             NaN             NaN              NaN   
70194480             NaN             NaN              NaN   
70194481             NaN             NaN              NaN   
70194489             NaN             NaN              NaN   
70194530             NaN             NaN              NaN   

          Australia and New Zealand  Central America  Eastern Asia  ...  \
12                              NaN              NaN           NaN  ...   
25                              NaN              NaN           NaN  ...   
39                              NaN              NaN           NaN  ...   
290                             NaN              NaN           NaN  ...   
303                             NaN              NaN           NaN  ...   
...                             ...              ...           ...  ...   
70194419                        NaN              NaN           NaN  ...   
70194480                        NaN              NaN           NaN  ...   
70194481                        NaN              NaN           NaN  ...   
70194489                        1.0              NaN           NaN  ...   
70194530                        NaN              NaN           NaN  ...   

          Southern Asia  Polynesia  South-eastern Asia  Central Asia  \
12                  NaN        NaN                 NaN           NaN   
25                  NaN        NaN                 NaN           NaN   
39                  NaN        NaN                 NaN           NaN   
290                 NaN        NaN                 NaN           NaN   
303                 NaN        NaN                 NaN           NaN   
...                 ...        ...                 ...           ...   
70194419            NaN        NaN                 NaN           NaN   
70194480            NaN        NaN                 NaN           NaN   
70194481            NaN        NaN                 NaN           NaN   
70194489            NaN        NaN                 NaN           NaN   
70194530            NaN        NaN                 NaN           NaN   

          Caribbean  Southern Africa  Middle Africa  Antarctica  Melanesia  \
12              NaN              NaN            NaN         NaN        NaN   
25              NaN              NaN            NaN         NaN        NaN   
39              NaN              NaN            NaN         NaN        NaN   
290             NaN              NaN            NaN         NaN        NaN   
303             NaN              NaN            NaN         NaN        NaN   
...             ...              ...            ...         ...        ...   
70194419        NaN              NaN            NaN         NaN        NaN   
70194480        NaN              NaN            NaN         NaN        NaN   
70194481        NaN              NaN            NaN         NaN        NaN   
70194489        NaN              NaN            NaN         NaN        NaN   
70194530        NaN              NaN            NaN         NaN        NaN   

          Micronesia  
12               NaN  
25               NaN  
39               NaN  
290              NaN  
303              NaN  
...              ...  
70194419         NaN  
70194480         NaN  
70194481         NaN  
70194489         NaN  
70194530         NaN  

[6460210 rows x 24 columns]
\end{verbatim}

\leavevmode\vadjust pre{\hypertarget{873f7e6f-157d-4fdd-b936-eb23e3a0198b}{}}%
Collapse Oceania:

\hypertarget{521f5236-feb7-457b-9cba-38eda3a94043}{}
\begin{Shaded}
\begin{Highlighting}[]
\NormalTok{ocean }\OperatorTok{=}\NormalTok{ src\_geo\_align.loc[:, oc\_regions].}\BuiltInTok{sum}\NormalTok{(axis}\OperatorTok{=}\StringTok{\textquotesingle{}columns\textquotesingle{}}\NormalTok{)}
\NormalTok{src\_geo\_align }\OperatorTok{=}\NormalTok{ src\_geo\_align.drop(columns}\OperatorTok{=}\NormalTok{oc\_regions)}
\NormalTok{src\_geo\_align[}\StringTok{\textquotesingle{}Oceania\textquotesingle{}}\NormalTok{] }\OperatorTok{=}\NormalTok{ ocean}
\end{Highlighting}
\end{Shaded}

\leavevmode\vadjust pre{\hypertarget{ec8925c0-cd9e-4692-a86c-7f597e21d0a8}{}}%
And normalize.

\hypertarget{8ff412c9-0b13-4599-a4df-466ef01aa49e}{}
\begin{Shaded}
\begin{Highlighting}[]
\NormalTok{src\_geo\_align }\OperatorTok{=}\NormalTok{ norm\_align\_matrix(src\_geo\_align)}
\end{Highlighting}
\end{Shaded}

\hypertarget{0b073c76-e0b9-4f4e-88d6-e86f2905c337}{}
\begin{Shaded}
\begin{Highlighting}[]
\NormalTok{src\_geo\_align.sort\_index(axis}\OperatorTok{=}\StringTok{\textquotesingle{}columns\textquotesingle{}}\NormalTok{, inplace}\OperatorTok{=}\VariableTok{True}\NormalTok{)}
\NormalTok{src\_geo\_align.info()}
\end{Highlighting}
\end{Shaded}

\begin{verbatim}
<class 'pandas.core.frame.DataFrame'>
Int64Index: 6460210 entries, 12 to 70194530
Data columns (total 21 columns):
 #   Column              Dtype  
---  ------              -----  
 0   @UNKNOWN            float64
 1   Antarctica          float64
 2   Caribbean           float64
 3   Central America     float64
 4   Central Asia        float64
 5   Eastern Africa      float64
 6   Eastern Asia        float64
 7   Eastern Europe      float64
 8   Middle Africa       float64
 9   Northern Africa     float64
 10  Northern America    float64
 11  Northern Europe     float64
 12  Oceania             float64
 13  South America       float64
 14  South-eastern Asia  float64
 15  Southern Africa     float64
 16  Southern Asia       float64
 17  Southern Europe     float64
 18  Western Africa      float64
 19  Western Asia        float64
 20  Western Europe      float64
dtypes: float64(21)
memory usage: 1.1 GB
\end{verbatim}

\leavevmode\vadjust pre{\hypertarget{d6ecec2b-5581-4181-927a-069ffb9e4ae2}{}}%
Xarray:

\hypertarget{8ff69491-5d0b-48f5-916b-cfe25aea838d}{}
\begin{Shaded}
\begin{Highlighting}[]
\NormalTok{src\_geo\_xr }\OperatorTok{=}\NormalTok{ xr.DataArray(src\_geo\_align, dims}\OperatorTok{=}\NormalTok{[}\StringTok{\textquotesingle{}page\textquotesingle{}}\NormalTok{, }\StringTok{\textquotesingle{}src\_geo\textquotesingle{}}\NormalTok{])}
\NormalTok{src\_geo\_xr}
\end{Highlighting}
\end{Shaded}

\begin{verbatim}
<xarray.DataArray (page: 6460210, src_geo: 21)>
array([[0.31343284, 0.        , 0.        , ..., 0.        , 0.01492537,
        0.        ],
       [0.70697674, 0.        , 0.        , ..., 0.00930233, 0.        ,
        0.01395349],
       [0.41666667, 0.        , 0.        , ..., 0.        , 0.        ,
        0.08333333],
       ...,
       [0.125     , 0.        , 0.        , ..., 0.        , 0.        ,
        0.        ],
       [0.66666667, 0.        , 0.        , ..., 0.        , 0.        ,
        0.        ],
       [0.        , 0.        , 0.        , ..., 0.        , 0.        ,
        0.        ]])
Coordinates:
  * page     (page) int64 12 25 39 290 ... 70194480 70194481 70194489 70194530
  * src_geo  (src_geo) object '@UNKNOWN' 'Antarctica' ... 'Western Europe'
\end{verbatim}

\leavevmode\vadjust pre{\hypertarget{7abf4c04-d720-4185-a886-fa699f3c00a9}{}}%
And save:

\hypertarget{e2483bf5-d332-4a06-83ba-ba0a098a8cbb}{}
\begin{Shaded}
\begin{Highlighting}[]
\NormalTok{output.save\_table(src\_geo\_align, }\StringTok{\textquotesingle{}page{-}src{-}geo{-}align\textquotesingle{}}\NormalTok{, parquet}\OperatorTok{=}\VariableTok{True}\NormalTok{)}
\end{Highlighting}
\end{Shaded}

\begin{verbatim}
INFO:wptrec.save:saving CSV to data\metric-tables\page-src-geo-align.csv.gz
INFO:wptrec.save:data\metric-tables\page-src-geo-align.csv.gz: 43.69 MiB
INFO:wptrec.save:saving Parquet to data\metric-tables\page-src-geo-align.parquet
INFO:wptrec.save:data\metric-tables\page-src-geo-align.parquet: 28.94 MiB
\end{verbatim}

\hypertarget{7a2f0dd8}{}
\hypertarget{gender}{%
\subsubsection{Gender}\label{gender}}

Now let's work on extracting gender - this is going work a lot like page
geography.

\hypertarget{db3167ce}{}
\begin{Shaded}
\begin{Highlighting}[]
\NormalTok{gender.head()}
\end{Highlighting}
\end{Shaded}

\begin{verbatim}
   page_id  gender
0      307    male
1      308    male
2      339  female
3      340    male
4      344    male
\end{verbatim}

\leavevmode\vadjust pre{\hypertarget{2bcf989e-123b-4d4c-8269-8bcd73f590c7}{}}%
And summarize:

\hypertarget{f1a2c728-8f08-4730-b601-9f281054bf00}{}
\begin{Shaded}
\begin{Highlighting}[]
\NormalTok{gender[}\StringTok{\textquotesingle{}gender\textquotesingle{}}\NormalTok{].value\_counts()}
\end{Highlighting}
\end{Shaded}

\begin{verbatim}
male                        1495445
female                       353301
transgender female              636
non-binary                      329
transgender male                197
intersex                         94
eunuch                           70
genderfluid                      29
genderqueer                      27
cisgender female                 18
two-spiriit                      11
travesti                         10
transgender person               10
cisgender male                    7
agender                           6
transmasculine                    6
neutral sex                       5
transfeminine                     4
bigender                          4
third gender                      2
demiboy                           2
fa'afafine                        2
neutrois                          1
assigned female at birth          1
māhū                              1
hijra                             1
Name: gender, dtype: int64
\end{verbatim}

\leavevmode\vadjust pre{\hypertarget{7c4803c4}{}}%
Now, we're going to do a little more work to reduce the dimensionality
of the space. Points:

\begin{enumerate}
\tightlist
\item
  Trans men are men
\item
  Trans women are women
\item
  Cis/trans status is an adjective that can be dropped for the present
  purposes
\end{enumerate}

The result is that we will collapse "transgender female" and "cisgender
female" into "female".

The \textbf{downside} to this is that trans men are probabily
significantly under-represented, but are now being collapsed into the
dominant group.

\hypertarget{72220c59}{}
\begin{Shaded}
\begin{Highlighting}[]
\NormalTok{pgcol }\OperatorTok{=}\NormalTok{ gender[}\StringTok{\textquotesingle{}gender\textquotesingle{}}\NormalTok{]}
\NormalTok{pgcol }\OperatorTok{=}\NormalTok{ pgcol.}\BuiltInTok{str}\NormalTok{.replace(}\VerbatimStringTok{r\textquotesingle{}(?:tran|ci)sgender\textbackslash{}s+((?:fe)?male)\textquotesingle{}}\NormalTok{, }\VerbatimStringTok{r\textquotesingle{}\textbackslash{}1\textquotesingle{}}\NormalTok{, regex}\OperatorTok{=}\VariableTok{True}\NormalTok{)}
\NormalTok{pgcol.value\_counts()}
\end{Highlighting}
\end{Shaded}

\begin{verbatim}
male                        1495649
female                       353955
non-binary                      329
intersex                         94
eunuch                           70
genderfluid                      29
genderqueer                      27
two-spiriit                      11
transgender person               10
travesti                         10
agender                           6
transmasculine                    6
neutral sex                       5
transfeminine                     4
bigender                          4
third gender                      2
demiboy                           2
fa'afafine                        2
māhū                              1
hijra                             1
neutrois                          1
assigned female at birth          1
Name: gender, dtype: int64
\end{verbatim}

\leavevmode\vadjust pre{\hypertarget{7c67aa25}{}}%
Now, we're going to group the remaining gender identities together under
the label 'NB'. As noted above, this is a debatable exercise that
collapses a lot of identity.

\hypertarget{96069118}{}
\begin{Shaded}
\begin{Highlighting}[]
\NormalTok{gender\_labels }\OperatorTok{=}\NormalTok{ [UNKNOWN, }\StringTok{\textquotesingle{}female\textquotesingle{}}\NormalTok{, }\StringTok{\textquotesingle{}male\textquotesingle{}}\NormalTok{, }\StringTok{\textquotesingle{}NB\textquotesingle{}}\NormalTok{]}
\NormalTok{pgcol[}\OperatorTok{\textasciitilde{}}\NormalTok{pgcol.isin(gender\_labels)] }\OperatorTok{=} \StringTok{\textquotesingle{}NB\textquotesingle{}}
\NormalTok{pgcol.value\_counts()}
\end{Highlighting}
\end{Shaded}

\begin{verbatim}
male      1495649
female     353955
NB            615
Name: gender, dtype: int64
\end{verbatim}

\leavevmode\vadjust pre{\hypertarget{796a4fbf}{}}%
Now put this column back in the frame and deduplicate.

\hypertarget{b3b6b661}{}
\begin{Shaded}
\begin{Highlighting}[]
\NormalTok{page\_gender }\OperatorTok{=}\NormalTok{ gender.assign(gender}\OperatorTok{=}\NormalTok{pgcol)}
\NormalTok{page\_gender }\OperatorTok{=}\NormalTok{ page\_gender.drop\_duplicates()}
\end{Highlighting}
\end{Shaded}

\hypertarget{f05206d1-9b41-4dff-b800-683213f7d709}{}
\begin{Shaded}
\begin{Highlighting}[]
\KeywordTok{del}\NormalTok{ pgcol}
\end{Highlighting}
\end{Shaded}

\leavevmode\vadjust pre{\hypertarget{179d9a6a-9981-4cdd-bc44-9e67581c9239}{}}%
Now we need to add unknown genders.

\hypertarget{3794557e-1edb-4a15-b988-7ffa9212fc42}{}
\begin{Shaded}
\begin{Highlighting}[]
\NormalTok{kg\_mask }\OperatorTok{=}\NormalTok{ all\_pages.isin(page\_gender[}\StringTok{\textquotesingle{}page\_id\textquotesingle{}}\NormalTok{])}
\NormalTok{unknown }\OperatorTok{=}\NormalTok{ all\_pages[}\OperatorTok{\textasciitilde{}}\NormalTok{kg\_mask]}
\NormalTok{page\_gender }\OperatorTok{=}\NormalTok{ pd.concat([}
\NormalTok{    page\_gender,}
\NormalTok{    pd.DataFrame(\{}\StringTok{\textquotesingle{}page\_id\textquotesingle{}}\NormalTok{: unknown, }\StringTok{\textquotesingle{}gender\textquotesingle{}}\NormalTok{: UNKNOWN\})}
\NormalTok{], ignore\_index}\OperatorTok{=}\VariableTok{True}\NormalTok{)}
\NormalTok{page\_gender}
\end{Highlighting}
\end{Shaded}

\begin{verbatim}
          page_id    gender
0             307      male
1             308      male
2             339    female
3             340      male
4             344      male
...           ...       ...
6460607  70194419  @UNKNOWN
6460608  70194480  @UNKNOWN
6460609  70194481  @UNKNOWN
6460610  70194489  @UNKNOWN
6460611  70194530  @UNKNOWN

[6460612 rows x 2 columns]
\end{verbatim}

\leavevmode\vadjust pre{\hypertarget{d7141688}{}}%
And make an alignment matrix:

\hypertarget{4d25abe0}{}
\begin{Shaded}
\begin{Highlighting}[]
\NormalTok{gender\_align }\OperatorTok{=}\NormalTok{ page\_gender.reset\_index().assign(x}\OperatorTok{=}\DecValTok{1}\NormalTok{).pivot(index}\OperatorTok{=}\StringTok{\textquotesingle{}page\_id\textquotesingle{}}\NormalTok{, columns}\OperatorTok{=}\StringTok{\textquotesingle{}gender\textquotesingle{}}\NormalTok{, values}\OperatorTok{=}\StringTok{\textquotesingle{}x\textquotesingle{}}\NormalTok{)}
\NormalTok{gender\_align.fillna(}\DecValTok{0}\NormalTok{, inplace}\OperatorTok{=}\VariableTok{True}\NormalTok{)}
\NormalTok{gender\_align }\OperatorTok{=}\NormalTok{ gender\_align.reindex(columns}\OperatorTok{=}\NormalTok{gender\_labels)}
\NormalTok{gender\_align.head()}
\end{Highlighting}
\end{Shaded}

\begin{verbatim}
gender   @UNKNOWN  female  male   NB
page_id                             
12            1.0     0.0   0.0  0.0
25            1.0     0.0   0.0  0.0
39            1.0     0.0   0.0  0.0
290           1.0     0.0   0.0  0.0
303           1.0     0.0   0.0  0.0
\end{verbatim}

\leavevmode\vadjust pre{\hypertarget{eb409647}{}}%
Let's see how frequent each of the genders is:

\hypertarget{ba2c1c76}{}
\begin{Shaded}
\begin{Highlighting}[]
\NormalTok{gender\_align.}\BuiltInTok{sum}\NormalTok{(axis}\OperatorTok{=}\DecValTok{0}\NormalTok{).sort\_values(ascending}\OperatorTok{=}\VariableTok{False}\NormalTok{)}
\end{Highlighting}
\end{Shaded}

\begin{verbatim}
gender
@UNKNOWN    4610461.0
male        1495647.0
female       353933.0
NB              571.0
dtype: float64
\end{verbatim}

\leavevmode\vadjust pre{\hypertarget{5f8862b2}{}}%
And convert to an xarray:

\hypertarget{4cb8ad72}{}
\begin{Shaded}
\begin{Highlighting}[]
\NormalTok{gender\_xr }\OperatorTok{=}\NormalTok{ xr.DataArray(gender\_align, dims}\OperatorTok{=}\NormalTok{[}\StringTok{\textquotesingle{}page\textquotesingle{}}\NormalTok{, }\StringTok{\textquotesingle{}gender\textquotesingle{}}\NormalTok{])}
\NormalTok{gender\_xr}
\end{Highlighting}
\end{Shaded}

\begin{verbatim}
<xarray.DataArray (page: 6460210, gender: 4)>
array([[1., 0., 0., 0.],
       [1., 0., 0., 0.],
       [1., 0., 0., 0.],
       ...,
       [1., 0., 0., 0.],
       [1., 0., 0., 0.],
       [1., 0., 0., 0.]])
Coordinates:
  * page     (page) int64 12 25 39 290 ... 70194480 70194481 70194489 70194530
  * gender   (gender) object '@UNKNOWN' 'female' 'male' 'NB'
\end{verbatim}

\hypertarget{807766f9}{}
\begin{Shaded}
\begin{Highlighting}[]
\NormalTok{binarysize(gender\_xr.nbytes)}
\end{Highlighting}
\end{Shaded}

\begin{verbatim}
'206.73 MiB'
\end{verbatim}

\hypertarget{db435c52-9085-45ef-a15d-c917d100c036}{}
\begin{Shaded}
\begin{Highlighting}[]
\NormalTok{output.save\_table(gender\_align, }\StringTok{\textquotesingle{}page{-}gender{-}align\textquotesingle{}}\NormalTok{, parquet}\OperatorTok{=}\VariableTok{True}\NormalTok{)}
\end{Highlighting}
\end{Shaded}

\begin{verbatim}
INFO:wptrec.save:saving CSV to data\metric-tables\page-gender-align.csv.gz
INFO:wptrec.save:data\metric-tables\page-gender-align.csv.gz: 18.80 MiB
INFO:wptrec.save:saving Parquet to data\metric-tables\page-gender-align.parquet
INFO:wptrec.save:data\metric-tables\page-gender-align.parquet: 9.33 MiB
\end{verbatim}

\hypertarget{6f5acd14-f577-43d7-a286-b9dfa1a6e6f4}{}
\hypertarget{occupation}{%
\subsubsection{Occupation}\label{occupation}}

Occupation works like gender, but without the need for processing.

Convert to a matrix:

\hypertarget{5a7d00cf-8f6e-4ad4-95e2-e5eb5ebd4906}{}
\begin{Shaded}
\begin{Highlighting}[]
\NormalTok{occ\_align }\OperatorTok{=}\NormalTok{ occupations.assign(x}\OperatorTok{=}\DecValTok{1}\NormalTok{).pivot(index}\OperatorTok{=}\StringTok{\textquotesingle{}page\_id\textquotesingle{}}\NormalTok{, columns}\OperatorTok{=}\StringTok{\textquotesingle{}occ\textquotesingle{}}\NormalTok{, values}\OperatorTok{=}\StringTok{\textquotesingle{}x\textquotesingle{}}\NormalTok{)}
\NormalTok{occ\_align.head()}
\end{Highlighting}
\end{Shaded}

\begin{verbatim}
occ      activist  agricultural worker  artist  athlete  biologist  \
page_id                                                              
307           NaN                  1.0     NaN      NaN        NaN   
308           NaN                  NaN     NaN      NaN        1.0   
339           NaN                  NaN     NaN      NaN        NaN   
340           NaN                  NaN     NaN      NaN        NaN   
344           NaN                  NaN     1.0      NaN        NaN   

occ      businessperson  chemist  civil servant  clergyperson  \
page_id                                                         
307                 NaN      NaN            NaN           NaN   
308                 NaN      NaN            NaN           NaN   
339                 NaN      NaN            NaN           NaN   
340                 1.0      NaN            NaN           NaN   
344                 1.0      NaN            NaN           NaN   

occ      computer scientist  ...  military personnel  musician  \
page_id                      ...                                 
307                     NaN  ...                 1.0       NaN   
308                     NaN  ...                 NaN       NaN   
339                     NaN  ...                 NaN       NaN   
340                     NaN  ...                 NaN       NaN   
344                     NaN  ...                 NaN       NaN   

occ      performing artist  physicist  politician  scientist  \
page_id                                                        
307                    NaN        NaN         1.0        NaN   
308                    NaN        1.0         NaN        1.0   
339                    NaN        NaN         NaN        NaN   
340                    NaN        NaN         NaN        NaN   
344                    NaN        NaN         NaN        NaN   

occ      social scientist  sportsperson (non-athlete)  \
page_id                                                 
307                   NaN                         NaN   
308                   NaN                         NaN   
339                   NaN                         NaN   
340                   NaN                         NaN   
344                   NaN                         NaN   

occ      transportation occupation  writer  
page_id                                     
307                            NaN     1.0  
308                            NaN     1.0  
339                            NaN     1.0  
340                            NaN     NaN  
344                            NaN     1.0  

[5 rows x 32 columns]
\end{verbatim}

\leavevmode\vadjust pre{\hypertarget{3631ef5e-3cfa-4b00-a3a4-74297a099798}{}}%
Set up unknown and merge:

\hypertarget{d9df4787-dc7b-4605-83ef-b597eb2a3c38}{}
\begin{Shaded}
\begin{Highlighting}[]
\NormalTok{occ\_unk }\OperatorTok{=}\NormalTok{ pd.Series(}\FloatTok{1.0}\NormalTok{, index}\OperatorTok{=}\NormalTok{all\_pages)}
\NormalTok{occ\_unk.index.name }\OperatorTok{=} \StringTok{\textquotesingle{}page\_id\textquotesingle{}}
\NormalTok{occ\_kmask }\OperatorTok{=}\NormalTok{ all\_pages.isin(occ\_align.index)}
\NormalTok{occ\_kmask.index }\OperatorTok{=}\NormalTok{ all\_pages}
\NormalTok{occ\_unk[occ\_kmask] }\OperatorTok{=} \DecValTok{0}
\NormalTok{occ\_align }\OperatorTok{=}\NormalTok{ occ\_unk.to\_frame(UNKNOWN).join(occ\_align, how}\OperatorTok{=}\StringTok{\textquotesingle{}left\textquotesingle{}}\NormalTok{)}
\NormalTok{occ\_align }\OperatorTok{=}\NormalTok{ norm\_align\_matrix(occ\_align)}
\NormalTok{occ\_align.head()}
\end{Highlighting}
\end{Shaded}

\begin{verbatim}
         @UNKNOWN  activist  agricultural worker  artist  athlete  biologist  \
page_id                                                                        
12            1.0       0.0                  0.0     0.0      0.0        0.0   
25            1.0       0.0                  0.0     0.0      0.0        0.0   
39            1.0       0.0                  0.0     0.0      0.0        0.0   
290           1.0       0.0                  0.0     0.0      0.0        0.0   
303           1.0       0.0                  0.0     0.0      0.0        0.0   

         businessperson  chemist  civil servant  clergyperson  ...  \
page_id                                                        ...   
12                  0.0      0.0            0.0           0.0  ...   
25                  0.0      0.0            0.0           0.0  ...   
39                  0.0      0.0            0.0           0.0  ...   
290                 0.0      0.0            0.0           0.0  ...   
303                 0.0      0.0            0.0           0.0  ...   

         military personnel  musician  performing artist  physicist  \
page_id                                                               
12                      0.0       0.0                0.0        0.0   
25                      0.0       0.0                0.0        0.0   
39                      0.0       0.0                0.0        0.0   
290                     0.0       0.0                0.0        0.0   
303                     0.0       0.0                0.0        0.0   

         politician  scientist  social scientist  sportsperson (non-athlete)  \
page_id                                                                        
12              0.0        0.0               0.0                         0.0   
25              0.0        0.0               0.0                         0.0   
39              0.0        0.0               0.0                         0.0   
290             0.0        0.0               0.0                         0.0   
303             0.0        0.0               0.0                         0.0   

         transportation occupation  writer  
page_id                                     
12                             0.0     0.0  
25                             0.0     0.0  
39                             0.0     0.0  
290                            0.0     0.0  
303                            0.0     0.0  

[5 rows x 33 columns]
\end{verbatim}

\hypertarget{6daf02d9-fff1-4064-ad34-2ec7bf2fb48d}{}
\begin{Shaded}
\begin{Highlighting}[]
\NormalTok{occ\_xr }\OperatorTok{=}\NormalTok{ xr.DataArray(occ\_align, dims}\OperatorTok{=}\NormalTok{[}\StringTok{\textquotesingle{}page\textquotesingle{}}\NormalTok{, }\StringTok{\textquotesingle{}occ\textquotesingle{}}\NormalTok{])}
\NormalTok{occ\_xr}
\end{Highlighting}
\end{Shaded}

\begin{verbatim}
<xarray.DataArray (page: 6460210, occ: 33)>
array([[1., 0., 0., ..., 0., 0., 0.],
       [1., 0., 0., ..., 0., 0., 0.],
       [1., 0., 0., ..., 0., 0., 0.],
       ...,
       [1., 0., 0., ..., 0., 0., 0.],
       [1., 0., 0., ..., 0., 0., 0.],
       [1., 0., 0., ..., 0., 0., 0.]])
Coordinates:
  * page     (page) int64 12 25 39 290 ... 70194480 70194481 70194489 70194530
  * occ      (occ) object '@UNKNOWN' 'activist' ... 'writer'
\end{verbatim}

\leavevmode\vadjust pre{\hypertarget{030d7154-8052-43d1-b168-a5939e1b1445}{}}%
And save:

\hypertarget{85d85173-209c-4077-8460-3a3a3bdbaced}{}
\begin{Shaded}
\begin{Highlighting}[]
\NormalTok{output.save\_table(occ\_align, }\StringTok{\textquotesingle{}page{-}occ{-}align\textquotesingle{}}\NormalTok{, parquet}\OperatorTok{=}\VariableTok{True}\NormalTok{)}
\end{Highlighting}
\end{Shaded}

\begin{verbatim}
INFO:wptrec.save:saving CSV to data\metric-tables\page-occ-align.csv.gz
INFO:wptrec.save:data\metric-tables\page-occ-align.csv.gz: 26.18 MiB
INFO:wptrec.save:saving Parquet to data\metric-tables\page-occ-align.parquet
INFO:wptrec.save:data\metric-tables\page-occ-align.parquet: 12.67 MiB
\end{verbatim}

\hypertarget{65e2d00e-228e-48f0-802d-fcde2f18b3d5}{}
\hypertarget{other-attributes}{%
\subsubsection{Other Attributes}\label{other-attributes}}

The other attributes don't require as much re-processing - they can be
used as-is as categorical variables. Let's save!

\hypertarget{a09da45f-473c-422e-b903-035f36893207}{}
\begin{Shaded}
\begin{Highlighting}[]
\NormalTok{pages }\OperatorTok{=}\NormalTok{ cat\_attrs.set\_index(}\StringTok{\textquotesingle{}page\_id\textquotesingle{}}\NormalTok{)}
\NormalTok{pages}
\end{Highlighting}
\end{Shaded}

\begin{verbatim}
         first_letter_category creation_date_category  \
page_id                                                 
12                         a-d              2001-2006   
25                         a-d              2001-2006   
39                         a-d              2001-2006   
290                        a-d              2001-2006   
303                        a-d              2001-2006   
...                        ...                    ...   
70194419                   l-r              2017-2022   
70194480                   a-d              2017-2022   
70194481                   a-d              2017-2022   
70194489                   l-r              2017-2022   
70194530                   a-d              2017-2022   

         relative_pageviews_category num_sitelinks_category  
page_id                                                      
12                              High           5+ languages  
25                              High           5+ languages  
39                              High           5+ languages  
290                             High           5+ languages  
303                             High           5+ languages  
...                              ...                    ...  
70194419                         Low          2-4 languages  
70194480                         Low           English only  
70194481                         Low           English only  
70194489                         Low          2-4 languages  
70194530                         Low           English only  

[6460210 rows x 4 columns]
\end{verbatim}

\leavevmode\vadjust pre{\hypertarget{92943334-0998-45e3-a98d-5728ef413c95}{}}%
Now each of these needs to become another table. The
\texttt{get\_dummies} function is our friend.

\hypertarget{4580489b-3b19-42ea-83b3-e5f588164a87}{}
\begin{Shaded}
\begin{Highlighting}[]
\NormalTok{alpha\_align }\OperatorTok{=}\NormalTok{ pd.get\_dummies(pages[}\StringTok{\textquotesingle{}first\_letter\_category\textquotesingle{}}\NormalTok{])}
\end{Highlighting}
\end{Shaded}

\hypertarget{02f8c370-06e2-49f9-b312-239c5d3de943}{}
\begin{Shaded}
\begin{Highlighting}[]
\NormalTok{output.save\_table(alpha\_align, }\StringTok{\textquotesingle{}page{-}alpha{-}align\textquotesingle{}}\NormalTok{, parquet}\OperatorTok{=}\VariableTok{True}\NormalTok{)}
\end{Highlighting}
\end{Shaded}

\begin{verbatim}
INFO:wptrec.save:saving CSV to data\metric-tables\page-alpha-align.csv.gz
INFO:wptrec.save:data\metric-tables\page-alpha-align.csv.gz: 19.47 MiB
INFO:wptrec.save:saving Parquet to data\metric-tables\page-alpha-align.parquet
INFO:wptrec.save:data\metric-tables\page-alpha-align.parquet: 10.52 MiB
\end{verbatim}

\hypertarget{f98da113-53e0-472e-aae5-52776526b4a8}{}
\begin{Shaded}
\begin{Highlighting}[]
\NormalTok{alpha\_xr }\OperatorTok{=}\NormalTok{ xr.DataArray(alpha\_align, dims}\OperatorTok{=}\NormalTok{[}\StringTok{\textquotesingle{}page\textquotesingle{}}\NormalTok{, }\StringTok{\textquotesingle{}alpha\textquotesingle{}}\NormalTok{])}
\end{Highlighting}
\end{Shaded}

\hypertarget{421703d8-6294-46d8-8cbe-2f214958a2db}{}
\begin{Shaded}
\begin{Highlighting}[]
\NormalTok{age\_align }\OperatorTok{=}\NormalTok{ pd.get\_dummies(pages[}\StringTok{\textquotesingle{}creation\_date\_category\textquotesingle{}}\NormalTok{])}
\NormalTok{output.save\_table(age\_align, }\StringTok{\textquotesingle{}page{-}age{-}align\textquotesingle{}}\NormalTok{, parquet}\OperatorTok{=}\VariableTok{True}\NormalTok{)}
\end{Highlighting}
\end{Shaded}

\begin{verbatim}
INFO:wptrec.save:saving CSV to data\metric-tables\page-age-align.csv.gz
INFO:wptrec.save:data\metric-tables\page-age-align.csv.gz: 17.29 MiB
INFO:wptrec.save:saving Parquet to data\metric-tables\page-age-align.parquet
INFO:wptrec.save:data\metric-tables\page-age-align.parquet: 7.53 MiB
\end{verbatim}

\hypertarget{b6d9bb89-9c4a-4a4c-81c7-21993b01d557}{}
\begin{Shaded}
\begin{Highlighting}[]
\NormalTok{age\_xr }\OperatorTok{=}\NormalTok{ xr.DataArray(age\_align, dims}\OperatorTok{=}\NormalTok{[}\StringTok{\textquotesingle{}page\textquotesingle{}}\NormalTok{, }\StringTok{\textquotesingle{}age\textquotesingle{}}\NormalTok{])}
\end{Highlighting}
\end{Shaded}

\hypertarget{e5120b4f-5941-4727-b3c4-a2550616fb2e}{}
\begin{Shaded}
\begin{Highlighting}[]
\NormalTok{pop\_align }\OperatorTok{=}\NormalTok{ pd.get\_dummies(pages[}\StringTok{\textquotesingle{}relative\_pageviews\_category\textquotesingle{}}\NormalTok{])}
\NormalTok{output.save\_table(pop\_align, }\StringTok{\textquotesingle{}page{-}pop{-}align\textquotesingle{}}\NormalTok{, parquet}\OperatorTok{=}\VariableTok{True}\NormalTok{)}
\end{Highlighting}
\end{Shaded}

\begin{verbatim}
INFO:wptrec.save:saving CSV to data\metric-tables\page-pop-align.csv.gz
INFO:wptrec.save:data\metric-tables\page-pop-align.csv.gz: 18.69 MiB
INFO:wptrec.save:saving Parquet to data\metric-tables\page-pop-align.parquet
INFO:wptrec.save:data\metric-tables\page-pop-align.parquet: 9.52 MiB
\end{verbatim}

\hypertarget{45633434-0a82-4636-883b-d97076144695}{}
\begin{Shaded}
\begin{Highlighting}[]
\NormalTok{pop\_xr }\OperatorTok{=}\NormalTok{ xr.DataArray(pop\_align, dims}\OperatorTok{=}\NormalTok{[}\StringTok{\textquotesingle{}page\textquotesingle{}}\NormalTok{, }\StringTok{\textquotesingle{}pop\textquotesingle{}}\NormalTok{])}
\end{Highlighting}
\end{Shaded}

\hypertarget{d3ade1ad-60f1-4a01-961e-ae79a6fe2e7b}{}
\begin{Shaded}
\begin{Highlighting}[]
\NormalTok{langs\_align }\OperatorTok{=}\NormalTok{ pd.get\_dummies(pages[}\StringTok{\textquotesingle{}num\_sitelinks\_category\textquotesingle{}}\NormalTok{])}
\NormalTok{output.save\_table(langs\_align, }\StringTok{\textquotesingle{}page{-}langs{-}align\textquotesingle{}}\NormalTok{, parquet}\OperatorTok{=}\VariableTok{True}\NormalTok{)}
\end{Highlighting}
\end{Shaded}

\begin{verbatim}
INFO:wptrec.save:saving CSV to data\metric-tables\page-langs-align.csv.gz
INFO:wptrec.save:data\metric-tables\page-langs-align.csv.gz: 18.64 MiB
INFO:wptrec.save:saving Parquet to data\metric-tables\page-langs-align.parquet
INFO:wptrec.save:data\metric-tables\page-langs-align.parquet: 9.80 MiB
\end{verbatim}

\hypertarget{90bdfa6c-b880-4656-b71c-8fcc97aaaad6}{}
\begin{Shaded}
\begin{Highlighting}[]
\NormalTok{langs\_xr }\OperatorTok{=}\NormalTok{ xr.DataArray(langs\_align, dims}\OperatorTok{=}\NormalTok{[}\StringTok{\textquotesingle{}page\textquotesingle{}}\NormalTok{, }\StringTok{\textquotesingle{}langs\textquotesingle{}}\NormalTok{])}
\end{Highlighting}
\end{Shaded}

\hypertarget{c1e570c8-c90a-4157-adcc-8f125714f266}{}
\hypertarget{working-with-alignments}{%
\subsection{Working with Alignments}\label{working-with-alignments}}

At this point, we have computed an alignment matrix for each of our
attributes, and extracted the qrels.

We will use the data saved from this in separate notebooks to compute
targets and alignments for tasks.

\hypertarget{15c0801d}{}
\hypertarget{task-1-alignment}{%
\section{Task 1 Alignment}\label{task-1-alignment}}

This notebook computes the target distributions and retrieved page
alignments for \textbf{Task 1}. It depends on the output of the
PageAlignments notebook.

\leavevmode\vadjust pre{\hypertarget{8966b1ad}{}}%
This notebook can be run in two modes: 'train', to process the training
topics, and 'eval' for the eval topics.

\hypertarget{c9334929}{}
\begin{Shaded}
\begin{Highlighting}[]
\NormalTok{DATA\_MODE }\OperatorTok{=} \StringTok{\textquotesingle{}eval\textquotesingle{}}
\end{Highlighting}
\end{Shaded}

\hypertarget{f0aacdf1}{}
\hypertarget{setup}{%
\subsection{Setup}\label{setup}}

We begin by loading necessary libraries:

\hypertarget{40262d18-b3bc-4b99-be93-467e750a3f1e}{}
\begin{Shaded}
\begin{Highlighting}[]
\ImportTok{import}\NormalTok{ sys}
\ImportTok{import}\NormalTok{ warnings}
\ImportTok{from}\NormalTok{ collections }\ImportTok{import}\NormalTok{ namedtuple}
\ImportTok{from}\NormalTok{ functools }\ImportTok{import} \BuiltInTok{reduce}
\ImportTok{from}\NormalTok{ itertools }\ImportTok{import}\NormalTok{ product}
\ImportTok{import}\NormalTok{ operator}
\ImportTok{from}\NormalTok{ pathlib }\ImportTok{import}\NormalTok{ Path}
\end{Highlighting}
\end{Shaded}

\hypertarget{f974e9ef}{}
\begin{Shaded}
\begin{Highlighting}[]
\ImportTok{import}\NormalTok{ pandas }\ImportTok{as}\NormalTok{ pd}
\ImportTok{import}\NormalTok{ xarray }\ImportTok{as}\NormalTok{ xr}
\ImportTok{import}\NormalTok{ numpy }\ImportTok{as}\NormalTok{ np}
\ImportTok{import}\NormalTok{ matplotlib.pyplot }\ImportTok{as}\NormalTok{ plt}
\ImportTok{import}\NormalTok{ seaborn }\ImportTok{as}\NormalTok{ sns}
\ImportTok{import}\NormalTok{ gzip}
\ImportTok{import}\NormalTok{ json}
\ImportTok{from}\NormalTok{ natural.size }\ImportTok{import}\NormalTok{ binarysize}
\ImportTok{from}\NormalTok{ natural.number }\ImportTok{import}\NormalTok{ number}
\end{Highlighting}
\end{Shaded}

\leavevmode\vadjust pre{\hypertarget{ec6577b9}{}}%
Set up progress bar and logging support:

\hypertarget{4bf51f31}{}
\begin{Shaded}
\begin{Highlighting}[]
\ImportTok{from}\NormalTok{ tqdm.auto }\ImportTok{import}\NormalTok{ tqdm}
\NormalTok{tqdm.pandas(leave}\OperatorTok{=}\VariableTok{False}\NormalTok{)}
\end{Highlighting}
\end{Shaded}

\hypertarget{1c8c4424}{}
\begin{Shaded}
\begin{Highlighting}[]
\ImportTok{import}\NormalTok{ sys, logging}
\NormalTok{logging.basicConfig(level}\OperatorTok{=}\NormalTok{logging.INFO, stream}\OperatorTok{=}\NormalTok{sys.stderr)}
\NormalTok{log }\OperatorTok{=}\NormalTok{ logging.getLogger(}\StringTok{\textquotesingle{}Task1Alignment\textquotesingle{}}\NormalTok{)}
\end{Highlighting}
\end{Shaded}

\leavevmode\vadjust pre{\hypertarget{ab2e4d09}{}}%
And set up an output directory:

\hypertarget{0a62d312}{}
\begin{Shaded}
\begin{Highlighting}[]
\ImportTok{from}\NormalTok{ wptrec.save }\ImportTok{import}\NormalTok{ OutRepo}
\NormalTok{output }\OperatorTok{=}\NormalTok{ OutRepo(}\StringTok{\textquotesingle{}data/metric{-}tables\textquotesingle{}}\NormalTok{)}
\end{Highlighting}
\end{Shaded}

\hypertarget{43db5240}{}
\hypertarget{data-and-helpers}{%
\subsection{Data and Helpers}\label{data-and-helpers}}

Most data loading is outsourced to \texttt{MetricInputs}. First we save
the data mode where metric inputs can find it:

\hypertarget{0d52904c-1821-454f-940d-fd62266bc18f}{}
\begin{Shaded}
\begin{Highlighting}[]
\ImportTok{import}\NormalTok{ wptrec}
\NormalTok{wptrec.DATA\_MODE }\OperatorTok{=}\NormalTok{ DATA\_MODE}
\end{Highlighting}
\end{Shaded}

\hypertarget{d2e4ddff-3feb-4883-8ad2-a3788e288890}{}
\begin{Shaded}
\begin{Highlighting}[]
\ImportTok{from}\NormalTok{ MetricInputs }\ImportTok{import} \OperatorTok{*}
\end{Highlighting}
\end{Shaded}

\begin{verbatim}
INFO:MetricInputs:reading data\metric-tables\page-sub-geo-align.parquet
INFO:MetricInputs:reading data\metric-tables\page-src-geo-align.parquet
INFO:MetricInputs:reading data\metric-tables\page-gender-align.parquet
INFO:MetricInputs:reading data\metric-tables\page-occ-align.parquet
INFO:MetricInputs:reading data\metric-tables\page-alpha-align.parquet
INFO:MetricInputs:reading data\metric-tables\page-age-align.parquet
INFO:MetricInputs:reading data\metric-tables\page-pop-align.parquet
INFO:MetricInputs:reading data\metric-tables\page-langs-align.parquet
\end{verbatim}

\hypertarget{2e3514ff-ca3c-4022-9e88-e217f6d82255}{}
\begin{Shaded}
\begin{Highlighting}[]
\NormalTok{dimensions}
\end{Highlighting}
\end{Shaded}

\begin{verbatim}
[<dimension "sub-geo": 21 levels>,
 <dimension "src-geo": 21 levels>,
 <dimension "gender": 4 levels>,
 <dimension "occ": 33 levels>,
 <dimension "alpha": 4 levels>,
 <dimension "age": 4 levels>,
 <dimension "pop": 4 levels>,
 <dimension "langs": 3 levels>]
\end{verbatim}

\hypertarget{7f0a206b-ae88-4166-afee-c4a1ebab01e0}{}
\hypertarget{qrel-join}{%
\subsubsection{qrel join}\label{qrel-join}}

We want a function to join alignments with qrels:

\hypertarget{b1366ace-1242-419f-9074-8cf8ec172603}{}
\begin{Shaded}
\begin{Highlighting}[]
\KeywordTok{def}\NormalTok{ qr\_join(align):}
    \ControlFlowTok{return}\NormalTok{ qrels.join(align, on}\OperatorTok{=}\StringTok{\textquotesingle{}page\_id\textquotesingle{}}\NormalTok{).set\_index([}\StringTok{\textquotesingle{}topic\_id\textquotesingle{}}\NormalTok{, }\StringTok{\textquotesingle{}page\_id\textquotesingle{}}\NormalTok{])}
\end{Highlighting}
\end{Shaded}

\hypertarget{947d54d8-7f93-4464-9abf-442454eeb7ce}{}
\hypertarget{norm_dist}{%
\subsubsection{norm\_dist}\label{norm_dist}}

And a function to normalize to a distribution:

\hypertarget{f86b07d6-0593-4743-83aa-20006c97360d}{}
\begin{Shaded}
\begin{Highlighting}[]
\KeywordTok{def}\NormalTok{ norm\_dist\_df(mat):}
\NormalTok{    sums }\OperatorTok{=}\NormalTok{ mat.}\BuiltInTok{sum}\NormalTok{(}\StringTok{\textquotesingle{}columns\textquotesingle{}}\NormalTok{)}
    \ControlFlowTok{return}\NormalTok{ mat.divide(sums, }\StringTok{\textquotesingle{}rows\textquotesingle{}}\NormalTok{)}
\end{Highlighting}
\end{Shaded}

\hypertarget{0e1b1422}{}
\hypertarget{prep-overview}{%
\subsection{Prep Overview}\label{prep-overview}}

Now that we have our alignments and qrels, we are ready to prepare the
Task 1 metrics.

We're first going to prepare the target distributions; then we will
compute the alignments for the retrieved pages.

\hypertarget{510de4b9-4ea6-4824-9b1d-72cf7b78a24b}{}
\hypertarget{subject-geography}{%
\subsection{Subject Geography}\label{subject-geography}}

Subject geography targets the average of the relevant set alignments and
the world population.

\hypertarget{d08efb3c-50ba-4027-8b62-bf8d88268476}{}
\begin{Shaded}
\begin{Highlighting}[]
\NormalTok{qr\_sub\_geo\_align }\OperatorTok{=}\NormalTok{ qr\_join(sub\_geo\_align)}
\NormalTok{qr\_sub\_geo\_align}
\end{Highlighting}
\end{Shaded}

\begin{verbatim}
                   @UNKNOWN  Antarctica  Caribbean  Central America  \
topic_id page_id                                                      
187      682            1.0         0.0        0.0              0.0   
         954            0.0         0.0        0.0              0.0   
         1170           1.0         0.0        0.0              0.0   
         1315           1.0         0.0        0.0              0.0   
         1322           0.0         0.0        0.0              0.0   
...                     ...         ...        ...              ...   
2872     69877511       1.0         0.0        0.0              0.0   
         69878912       1.0         0.0        0.0              0.0   
         69879322       1.0         0.0        0.0              0.0   
         69881345       0.0         0.0        0.0              0.0   
         69883661       1.0         0.0        0.0              0.0   

                   Central Asia  Eastern Africa  Eastern Asia  Eastern Europe  \
topic_id page_id                                                                
187      682                0.0             0.0           0.0             0.0   
         954                0.0             0.0           0.0             0.0   
         1170               0.0             0.0           0.0             0.0   
         1315               0.0             0.0           0.0             0.0   
         1322               0.0             0.0           0.0             0.0   
...                         ...             ...           ...             ...   
2872     69877511           0.0             0.0           0.0             0.0   
         69878912           0.0             0.0           0.0             0.0   
         69879322           0.0             0.0           0.0             0.0   
         69881345           0.0             0.0           0.0             0.0   
         69883661           0.0             0.0           0.0             0.0   

                   Middle Africa  Northern Africa  ...  Northern Europe  \
topic_id page_id                                   ...                    
187      682                 0.0              0.0  ...              0.0   
         954                 0.0              0.0  ...              0.0   
         1170                0.0              0.0  ...              0.0   
         1315                0.0              0.0  ...              0.0   
         1322                0.0              0.0  ...              0.0   
...                          ...              ...  ...              ...   
2872     69877511            0.0              0.0  ...              0.0   
         69878912            0.0              0.0  ...              0.0   
         69879322            0.0              0.0  ...              0.0   
         69881345            0.0              0.0  ...              0.0   
         69883661            0.0              0.0  ...              0.0   

                   Oceania  South America  South-eastern Asia  \
topic_id page_id                                                
187      682           0.0            0.0                 0.0   
         954           0.0            0.0                 0.0   
         1170          0.0            0.0                 0.0   
         1315          0.0            0.0                 0.0   
         1322          0.0            0.0                 0.0   
...                    ...            ...                 ...   
2872     69877511      0.0            0.0                 0.0   
         69878912      0.0            0.0                 0.0   
         69879322      0.0            0.0                 0.0   
         69881345      0.0            0.0                 1.0   
         69883661      0.0            0.0                 0.0   

                   Southern Africa  Southern Asia  Southern Europe  \
topic_id page_id                                                     
187      682                   0.0            0.0              0.0   
         954                   0.0            0.0              0.0   
         1170                  0.0            0.0              0.0   
         1315                  0.0            0.0              0.0   
         1322                  0.0            0.0              1.0   
...                            ...            ...              ...   
2872     69877511              0.0            0.0              0.0   
         69878912              0.0            0.0              0.0   
         69879322              0.0            0.0              0.0   
         69881345              0.0            0.0              0.0   
         69883661              0.0            0.0              0.0   

                   Western Africa  Western Asia  Western Europe  
topic_id page_id                                                 
187      682                  0.0           0.0             0.0  
         954                  0.0           0.0             1.0  
         1170                 0.0           0.0             0.0  
         1315                 0.0           0.0             0.0  
         1322                 0.0           0.0             0.0  
...                           ...           ...             ...  
2872     69877511             0.0           0.0             0.0  
         69878912             0.0           0.0             0.0  
         69879322             0.0           0.0             0.0  
         69881345             0.0           0.0             0.0  
         69883661             0.0           0.0             0.0  

[2737612 rows x 21 columns]
\end{verbatim}

\leavevmode\vadjust pre{\hypertarget{4952038e}{}}%
For purely geographic fairness, we just need to average the unknowns
with the world pop:

\hypertarget{5f07bc9b}{}
\begin{Shaded}
\begin{Highlighting}[]
\NormalTok{qr\_sub\_geo\_tgt }\OperatorTok{=}\NormalTok{ qr\_sub\_geo\_align.groupby(}\StringTok{\textquotesingle{}topic\_id\textquotesingle{}}\NormalTok{).mean()}
\NormalTok{qr\_sub\_geo\_fk }\OperatorTok{=}\NormalTok{ qr\_sub\_geo\_tgt.iloc[:, }\DecValTok{1}\NormalTok{:].}\BuiltInTok{sum}\NormalTok{(}\StringTok{\textquotesingle{}columns\textquotesingle{}}\NormalTok{)}
\NormalTok{qr\_sub\_geo\_tgt.iloc[:, }\DecValTok{1}\NormalTok{:] }\OperatorTok{*=} \FloatTok{0.5}
\NormalTok{qr\_sub\_geo\_tgt.iloc[:, }\DecValTok{1}\NormalTok{:] }\OperatorTok{+=}\NormalTok{ qr\_sub\_geo\_fk.}\BuiltInTok{apply}\NormalTok{(}\KeywordTok{lambda}\NormalTok{ k: world\_pop }\OperatorTok{*}\NormalTok{ k }\OperatorTok{*} \FloatTok{0.5}\NormalTok{)}
\NormalTok{qr\_sub\_geo\_tgt.head()}
\end{Highlighting}
\end{Shaded}

\begin{verbatim}
          @UNKNOWN    Antarctica  Caribbean  Central America  Central Asia  \
topic_id                                                                     
187       0.161757  6.472220e-08   0.004007         0.012384      0.004401   
270       0.242805  5.846440e-08   0.017378         0.014851      0.005852   
359       0.183666  6.303060e-08   0.017007         0.014391      0.003689   
365       0.201370  6.166361e-08   0.007572         0.012774      0.004079   
400       0.258172  5.727783e-08   0.004827         0.013104      0.003552   

          Eastern Africa  Eastern Asia  Eastern Europe  Middle Africa  \
topic_id                                                                
187             0.022830      0.112412        0.033440       0.008264   
270             0.037144      0.106411        0.053948       0.009914   
359             0.021289      0.118833        0.017016       0.007747   
365             0.022296      0.104172        0.035950       0.011613   
400             0.020758      0.101462        0.027533       0.007496   

          Northern Africa  ...  Northern Europe   Oceania  South America  \
topic_id                   ...                                             
187              0.014711  ...         0.133172  0.020594       0.030093   
270              0.017165  ...         0.058914  0.020977       0.038029   
359              0.011968  ...         0.006663  0.005588       0.029521   
365              0.015012  ...         0.029218  0.016421       0.036189   
400              0.012440  ...         0.076621  0.023341       0.030668   

          South-eastern Asia  Southern Africa  Southern Asia  Southern Europe  \
topic_id                                                                        
187                 0.043274         0.004694       0.116350         0.059294   
270                 0.038750         0.008852       0.101007         0.044103   
359                 0.035681         0.003675       0.099904         0.010362   
365                 0.053554         0.003956       0.100548         0.065794   
400                 0.036634         0.005453       0.101073         0.027173   

          Western Africa  Western Asia  Western Europe  
topic_id                                                
187             0.020306      0.023583        0.058312  
270             0.026599      0.022927        0.055952  
359             0.018935      0.014239        0.012154  
365             0.024046      0.029213        0.031859  
400             0.018965      0.018795        0.056502  

[5 rows x 21 columns]
\end{verbatim}

\leavevmode\vadjust pre{\hypertarget{0281b0b7-6115-49ef-9338-c5bfc4b416cd}{}}%
Make sure the rows are distributions:

\hypertarget{52caf011-7325-49db-afad-c259eac8635f}{}
\begin{Shaded}
\begin{Highlighting}[]
\NormalTok{qr\_sub\_geo\_tgt.}\BuiltInTok{sum}\NormalTok{(}\StringTok{\textquotesingle{}columns\textquotesingle{}}\NormalTok{).describe()}
\end{Highlighting}
\end{Shaded}

\begin{verbatim}
count    5.000000e+01
mean     1.000000e+00
std      1.409697e-16
min      1.000000e+00
25%      1.000000e+00
50%      1.000000e+00
75%      1.000000e+00
max      1.000000e+00
dtype: float64
\end{verbatim}

\leavevmode\vadjust pre{\hypertarget{514e7c86-079f-414d-8576-ea0283aea166}{}}%
Everything is 1, we're good to go!

\hypertarget{3b0cc18d-f55d-4eb3-b775-b205c10efb4f}{}
\begin{Shaded}
\begin{Highlighting}[]
\NormalTok{output.save\_table(qr\_sub\_geo\_tgt, }\SpecialStringTok{f\textquotesingle{}task1{-}}\SpecialCharTok{\{}\NormalTok{DATA\_MODE}\SpecialCharTok{\}}\SpecialStringTok{{-}sub{-}geo{-}target\textquotesingle{}}\NormalTok{, parquet}\OperatorTok{=}\VariableTok{True}\NormalTok{)}
\end{Highlighting}
\end{Shaded}

\begin{verbatim}
INFO:wptrec.save:saving CSV to data\metric-tables\task1-eval-sub-geo-target.csv.gz
INFO:wptrec.save:data\metric-tables\task1-eval-sub-geo-target.csv.gz: 10.66 KiB
INFO:wptrec.save:saving Parquet to data\metric-tables\task1-eval-sub-geo-target.parquet
INFO:wptrec.save:data\metric-tables\task1-eval-sub-geo-target.parquet: 25.97 KiB
\end{verbatim}

\hypertarget{ee72b69f-949f-46ed-9ece-2e795963d4c3}{}
\hypertarget{source-geography}{%
\subsection{Source Geography}\label{source-geography}}

Source geography works the same way.

\hypertarget{6d102833}{}
\begin{Shaded}
\begin{Highlighting}[]
\NormalTok{qr\_src\_geo\_align }\OperatorTok{=}\NormalTok{ qr\_join(src\_geo\_align)}
\NormalTok{qr\_src\_geo\_align}
\end{Highlighting}
\end{Shaded}

\begin{verbatim}
                   @UNKNOWN  Antarctica  Caribbean  Central America  \
topic_id page_id                                                      
187      682       0.400000         0.0        0.0              0.0   
         954       0.257143         0.0        0.0              0.0   
         1170      0.368421         0.0        0.0              0.0   
         1315      0.375000         0.0        0.0              0.0   
         1322      0.428571         0.0        0.0              0.0   
...                     ...         ...        ...              ...   
2872     69877511  1.000000         0.0        0.0              0.0   
         69878912  0.366667         0.0        0.0              0.0   
         69879322  0.200000         0.0        0.0              0.0   
         69881345  0.500000         0.0        0.0              0.0   
         69883661  0.000000         0.0        0.0              0.0   

                   Central Asia  Eastern Africa  Eastern Asia  Eastern Europe  \
topic_id page_id                                                                
187      682                0.0             0.0           0.0             0.0   
         954                0.0             0.0           0.0             0.0   
         1170               0.0             0.0           0.0             0.0   
         1315               0.0             0.0           0.0             0.0   
         1322               0.0             0.0           0.0             0.0   
...                         ...             ...           ...             ...   
2872     69877511           0.0             0.0           0.0             0.0   
         69878912           0.0             0.0           0.0             0.0   
         69879322           0.0             0.0           0.0             0.0   
         69881345           0.0             0.0           0.0             0.0   
         69883661           0.0             0.0           0.0             0.0   

                   Middle Africa  Northern Africa  ...  Northern Europe  \
topic_id page_id                                   ...                    
187      682                 0.0              0.0  ...         0.150000   
         954                 0.0              0.0  ...         0.285714   
         1170                0.0              0.0  ...         0.052632   
         1315                0.0              0.0  ...         0.000000   
         1322                0.0              0.0  ...         0.000000   
...                          ...              ...  ...              ...   
2872     69877511            0.0              0.0  ...         0.000000   
         69878912            0.0              0.0  ...         0.000000   
         69879322            0.0              0.0  ...         0.000000   
         69881345            0.0              0.0  ...         0.000000   
         69883661            0.0              0.0  ...         0.000000   

                    Oceania  South America  South-eastern Asia  \
topic_id page_id                                                 
187      682       0.000000            0.0                 0.0   
         954       0.000000            0.0                 0.0   
         1170      0.052632            0.0                 0.0   
         1315      0.000000            0.0                 0.0   
         1322      0.000000            0.0                 0.0   
...                     ...            ...                 ...   
2872     69877511  0.000000            0.0                 0.0   
         69878912  0.000000            0.0                 0.1   
         69879322  0.000000            0.0                 0.0   
         69881345  0.000000            0.0                 0.5   
         69883661  0.000000            0.0                 0.0   

                   Southern Africa  Southern Asia  Southern Europe  \
topic_id page_id                                                     
187      682                   0.0            0.0         0.000000   
         954                   0.0            0.0         0.000000   
         1170                  0.0            0.0         0.000000   
         1315                  0.0            0.0         0.000000   
         1322                  0.0            0.0         0.571429   
...                            ...            ...              ...   
2872     69877511              0.0            0.0         0.000000   
         69878912              0.0            0.0         0.000000   
         69879322              0.0            0.0         0.000000   
         69881345              0.0            0.0         0.000000   
         69883661              0.0            0.0         0.000000   

                   Western Africa  Western Asia  Western Europe  
topic_id page_id                                                 
187      682                  0.0         0.000        0.050000  
         954                  0.0         0.000        0.171429  
         1170                 0.0         0.000        0.000000  
         1315                 0.0         0.125        0.000000  
         1322                 0.0         0.000        0.000000  
...                           ...           ...             ...  
2872     69877511             0.0         0.000        0.000000  
         69878912             0.0         0.000        0.000000  
         69879322             0.0         0.600        0.000000  
         69881345             0.0         0.000        0.000000  
         69883661             0.0         0.000        0.000000  

[2737612 rows x 21 columns]
\end{verbatim}

\leavevmode\vadjust pre{\hypertarget{792e1a11-a20d-46c2-b145-6bdee219db9a}{}}%
And repeat:

\hypertarget{cdbc0bf3-cfc4-4be2-aef7-b17f2ff33a38}{}
\begin{Shaded}
\begin{Highlighting}[]
\NormalTok{qr\_src\_geo\_tgt }\OperatorTok{=}\NormalTok{ qr\_src\_geo\_align.groupby(}\StringTok{\textquotesingle{}topic\_id\textquotesingle{}}\NormalTok{).mean()}
\NormalTok{qr\_src\_geo\_fk }\OperatorTok{=}\NormalTok{ qr\_src\_geo\_tgt.iloc[:, }\DecValTok{1}\NormalTok{:].}\BuiltInTok{sum}\NormalTok{(}\StringTok{\textquotesingle{}columns\textquotesingle{}}\NormalTok{)}
\NormalTok{qr\_src\_geo\_tgt.iloc[:, }\DecValTok{1}\NormalTok{:] }\OperatorTok{*=} \FloatTok{0.5}
\NormalTok{qr\_src\_geo\_tgt.iloc[:, }\DecValTok{1}\NormalTok{:] }\OperatorTok{+=}\NormalTok{ qr\_src\_geo\_fk.}\BuiltInTok{apply}\NormalTok{(}\KeywordTok{lambda}\NormalTok{ k: world\_pop }\OperatorTok{*}\NormalTok{ k }\OperatorTok{*} \FloatTok{0.5}\NormalTok{)}
\NormalTok{qr\_src\_geo\_tgt.head()}
\end{Highlighting}
\end{Shaded}

\begin{verbatim}
          @UNKNOWN    Antarctica  Caribbean  Central America  Central Asia  \
topic_id                                                                     
187       0.391787  4.696121e-08   0.002250         0.008070      0.002876   
270       0.420047  4.477917e-08   0.003611         0.008171      0.002702   
359       0.372489  4.845126e-08   0.003072         0.008260      0.002821   
365       0.364985  4.903066e-08   0.010223         0.008492      0.002984   
400       0.422769  2.798744e-07   0.002478         0.008311      0.002702   

          Eastern Africa  Eastern Asia  Eastern Europe  Middle Africa  \
topic_id                                                                
187             0.016153      0.077938        0.019365       0.005790   
270             0.015759      0.073673        0.019488       0.005524   
359             0.016384      0.084042        0.013101       0.005947   
365             0.017147      0.082518        0.018251       0.007674   
400             0.015381      0.074893        0.018031       0.005497   

          Northern Africa  ...  Northern Europe   Oceania  South America  \
topic_id                   ...                                             
187              0.009195  ...         0.110871  0.011692       0.019483   
270              0.008721  ...         0.044787  0.010542       0.020577   
359              0.009209  ...         0.007908  0.003628       0.018333   
365              0.009672  ...         0.021657  0.012542       0.020322   
400              0.008827  ...         0.069702  0.019709       0.019562   

          South-eastern Asia  Southern Africa  Southern Asia  Southern Europe  \
topic_id                                                                        
187                 0.029280         0.003079       0.081422         0.019888   
270                 0.026281         0.003505       0.073534         0.018938   
359                 0.027301         0.002669       0.076759         0.007120   
365                 0.038885         0.002730       0.078353         0.039960   
400                 0.027291         0.003381       0.078346         0.015888   

          Western Africa  Western Asia  Western Europe  
topic_id                                                
187             0.014278      0.013633        0.033541  
270             0.013802      0.011568        0.061875  
359             0.014524      0.010901        0.010185  
365             0.015051      0.020196        0.029345  
400             0.013821      0.012813        0.025499  

[5 rows x 21 columns]
\end{verbatim}

\leavevmode\vadjust pre{\hypertarget{56882b42}{}}%
Make sure the rows are distributions:

\hypertarget{73010603}{}
\begin{Shaded}
\begin{Highlighting}[]
\NormalTok{qr\_src\_geo\_tgt.}\BuiltInTok{sum}\NormalTok{(}\StringTok{\textquotesingle{}columns\textquotesingle{}}\NormalTok{).describe()}
\end{Highlighting}
\end{Shaded}

\begin{verbatim}
count    5.000000e+01
mean     1.000000e+00
std      1.218255e-16
min      1.000000e+00
25%      1.000000e+00
50%      1.000000e+00
75%      1.000000e+00
max      1.000000e+00
dtype: float64
\end{verbatim}

\leavevmode\vadjust pre{\hypertarget{a2a02e90}{}}%
Everything is 1, we're good to go!

\hypertarget{2f6de863}{}
\begin{Shaded}
\begin{Highlighting}[]
\NormalTok{output.save\_table(qr\_src\_geo\_tgt, }\SpecialStringTok{f\textquotesingle{}task1{-}}\SpecialCharTok{\{}\NormalTok{DATA\_MODE}\SpecialCharTok{\}}\SpecialStringTok{{-}src{-}geo{-}target\textquotesingle{}}\NormalTok{, parquet}\OperatorTok{=}\VariableTok{True}\NormalTok{)}
\end{Highlighting}
\end{Shaded}

\begin{verbatim}
INFO:wptrec.save:saving CSV to data\metric-tables\task1-eval-src-geo-target.csv.gz
INFO:wptrec.save:data\metric-tables\task1-eval-src-geo-target.csv.gz: 10.64 KiB
INFO:wptrec.save:saving Parquet to data\metric-tables\task1-eval-src-geo-target.parquet
INFO:wptrec.save:data\metric-tables\task1-eval-src-geo-target.parquet: 25.97 KiB
\end{verbatim}

\hypertarget{6cd73da4-ed2b-456c-8cbd-aacbef8d8be0}{}
\hypertarget{gender}{%
\subsection{Gender}\label{gender}}

Now we're going to grab the gender alignments. Again, we ignore UNKNOWN.

\hypertarget{326b87c0-91e6-4b65-8020-cadc23d21a9c}{}
\begin{Shaded}
\begin{Highlighting}[]
\NormalTok{qr\_gender\_align }\OperatorTok{=}\NormalTok{ qr\_join(gender\_align)}
\NormalTok{qr\_gender\_align.head()}
\end{Highlighting}
\end{Shaded}

\begin{verbatim}
                  @UNKNOWN  female  male   NB
topic_id page_id                             
187      682           1.0     0.0   0.0  0.0
         954           0.0     0.0   1.0  0.0
         1170          1.0     0.0   0.0  0.0
         1315          1.0     0.0   0.0  0.0
         1322          1.0     0.0   0.0  0.0
\end{verbatim}

\hypertarget{6dc2fdc7-04b4-4161-92c6-ca996923b820}{}
\begin{Shaded}
\begin{Highlighting}[]
\NormalTok{qr\_gender\_tgt }\OperatorTok{=}\NormalTok{ qr\_gender\_align.groupby(}\StringTok{\textquotesingle{}topic\_id\textquotesingle{}}\NormalTok{).mean()}
\NormalTok{qr\_gender\_fk }\OperatorTok{=}\NormalTok{ qr\_gender\_tgt.iloc[:, }\DecValTok{1}\NormalTok{:].}\BuiltInTok{sum}\NormalTok{(}\StringTok{\textquotesingle{}columns\textquotesingle{}}\NormalTok{)}
\NormalTok{qr\_gender\_tgt.iloc[:, }\DecValTok{1}\NormalTok{:] }\OperatorTok{*=} \FloatTok{0.5}
\NormalTok{qr\_gender\_tgt.iloc[:, }\DecValTok{1}\NormalTok{:] }\OperatorTok{+=}\NormalTok{ qr\_gender\_fk.}\BuiltInTok{apply}\NormalTok{(}\KeywordTok{lambda}\NormalTok{ k: gender\_tgt }\OperatorTok{*}\NormalTok{ k }\OperatorTok{*} \FloatTok{0.5}\NormalTok{)}
\NormalTok{qr\_gender\_tgt.head()}
\end{Highlighting}
\end{Shaded}

\begin{verbatim}
          @UNKNOWN    female      male        NB
topic_id                                        
187       0.888195  0.033910  0.077336  0.000574
270       0.371833  0.257322  0.367774  0.003231
359       0.340156  0.170558  0.486007  0.003299
365       0.424643  0.183396  0.389116  0.002877
400       0.011697  0.408054  0.575302  0.005275
\end{verbatim}

\hypertarget{b74608a4-ec64-48c9-8702-060723194a62}{}
\begin{Shaded}
\begin{Highlighting}[]
\NormalTok{output.save\_table(qr\_gender\_tgt, }\SpecialStringTok{f\textquotesingle{}task1{-}}\SpecialCharTok{\{}\NormalTok{DATA\_MODE}\SpecialCharTok{\}}\SpecialStringTok{{-}gender{-}target\textquotesingle{}}\NormalTok{, parquet}\OperatorTok{=}\VariableTok{True}\NormalTok{)}
\end{Highlighting}
\end{Shaded}

\begin{verbatim}
INFO:wptrec.save:saving CSV to data\metric-tables\task1-eval-gender-target.csv.gz
INFO:wptrec.save:data\metric-tables\task1-eval-gender-target.csv.gz: 2.22 KiB
INFO:wptrec.save:saving Parquet to data\metric-tables\task1-eval-gender-target.parquet
INFO:wptrec.save:data\metric-tables\task1-eval-gender-target.parquet: 6.90 KiB
\end{verbatim}

\hypertarget{22428b3f-934d-4edc-b8d3-097474240802}{}
\hypertarget{remaining-attributes}{%
\subsection{Remaining Attributes}\label{remaining-attributes}}

The remaining attributes don't need any further processing, as they
aren't averaged.

\hypertarget{03193ea0-0d82-47f2-af6c-faf524c9c632}{}
\begin{Shaded}
\begin{Highlighting}[]
\NormalTok{qr\_occ\_align }\OperatorTok{=}\NormalTok{ qr\_join(occ\_align)}
\NormalTok{qr\_occ\_tgt }\OperatorTok{=}\NormalTok{ qr\_occ\_align.groupby(}\StringTok{\textquotesingle{}topic\_id\textquotesingle{}}\NormalTok{).}\BuiltInTok{sum}\NormalTok{()}
\NormalTok{qr\_occ\_tgt }\OperatorTok{=}\NormalTok{ norm\_dist\_df(qr\_occ\_tgt)}
\NormalTok{qr\_occ\_tgt.head()}
\end{Highlighting}
\end{Shaded}

\begin{verbatim}
          @UNKNOWN  activist  agricultural worker    artist   athlete  \
topic_id                                                                
187       0.891108  0.000192             0.000049  0.005105  0.000383   
270       0.379033  0.000143             0.000153  0.000569  0.597543   
359       0.355009  0.000216             0.000048  0.000564  0.587417   
365       0.427646  0.000081             0.000016  0.000186  0.499385   
400       0.044346  0.004397             0.000387  0.316302  0.003669   

          biologist  businessperson   chemist  civil servant  clergyperson  \
topic_id                                                                     
187        0.000193        0.002763  0.000005       0.000194      0.000081   
270        0.000145        0.001116  0.000123       0.000671      0.000110   
359        0.000045        0.004931  0.000062       0.000336      0.000046   
365        0.000023        0.001868  0.000047       0.000207      0.000094   
400        0.001530        0.019926  0.000269       0.002284      0.001724   

          ...  military personnel  musician  performing artist  physicist  \
topic_id  ...                                                               
187       ...            0.000335  0.000128           0.000110   0.000052   
270       ...            0.000867  0.000404           0.001072   0.000024   
359       ...            0.001501  0.000922           0.002827   0.000010   
365       ...            0.000696  0.000274           0.001756   0.000000   
400       ...            0.002074  0.010823           0.128105   0.000393   

          politician  scientist  social scientist  sportsperson (non-athlete)  \
topic_id                                                                        
187         0.001044   0.001168          0.000461                    0.000040   
270         0.002388   0.000277          0.000275                    0.008550   
359         0.001808   0.000037          0.000045                    0.031237   
365         0.001031   0.000063          0.000070                    0.061864   
400         0.007384   0.003000          0.003345                    0.001635   

          transportation occupation    writer  
topic_id                                       
187                        0.000031  0.001421  
270                        0.000281  0.000811  
359                        0.000059  0.001414  
365                        0.000094  0.000777  
400                        0.000520  0.249432  

[5 rows x 33 columns]
\end{verbatim}

\hypertarget{bdf9e942-4da6-4a15-9e9b-3401be54d10e}{}
\begin{Shaded}
\begin{Highlighting}[]
\NormalTok{output.save\_table(qr\_occ\_tgt, }\SpecialStringTok{f\textquotesingle{}task1{-}}\SpecialCharTok{\{}\NormalTok{DATA\_MODE}\SpecialCharTok{\}}\SpecialStringTok{{-}occ{-}target\textquotesingle{}}\NormalTok{, parquet}\OperatorTok{=}\VariableTok{True}\NormalTok{)}
\end{Highlighting}
\end{Shaded}

\begin{verbatim}
INFO:wptrec.save:saving CSV to data\metric-tables\task1-eval-occ-target.csv.gz
INFO:wptrec.save:data\metric-tables\task1-eval-occ-target.csv.gz: 14.99 KiB
INFO:wptrec.save:saving Parquet to data\metric-tables\task1-eval-occ-target.parquet
INFO:wptrec.save:data\metric-tables\task1-eval-occ-target.parquet: 38.59 KiB
\end{verbatim}

\hypertarget{733266c6-f7b7-48c2-988a-4edca70f62d8}{}
\begin{Shaded}
\begin{Highlighting}[]
\NormalTok{qr\_age\_align }\OperatorTok{=}\NormalTok{ qr\_join(age\_align)}
\NormalTok{qr\_age\_tgt }\OperatorTok{=}\NormalTok{ norm\_dist\_df(qr\_age\_align.groupby(}\StringTok{\textquotesingle{}topic\_id\textquotesingle{}}\NormalTok{).}\BuiltInTok{sum}\NormalTok{())}
\NormalTok{output.save\_table(qr\_age\_tgt, }\SpecialStringTok{f\textquotesingle{}task1{-}}\SpecialCharTok{\{}\NormalTok{DATA\_MODE}\SpecialCharTok{\}}\SpecialStringTok{{-}age{-}target\textquotesingle{}}\NormalTok{, parquet}\OperatorTok{=}\VariableTok{True}\NormalTok{)}
\end{Highlighting}
\end{Shaded}

\begin{verbatim}
INFO:wptrec.save:saving CSV to data\metric-tables\task1-eval-age-target.csv.gz
INFO:wptrec.save:data\metric-tables\task1-eval-age-target.csv.gz: 2.13 KiB
INFO:wptrec.save:saving Parquet to data\metric-tables\task1-eval-age-target.parquet
INFO:wptrec.save:data\metric-tables\task1-eval-age-target.parquet: 6.23 KiB
\end{verbatim}

\hypertarget{b897d166-b0c0-4f7a-9e08-17c2c8208412}{}
\begin{Shaded}
\begin{Highlighting}[]
\NormalTok{qr\_alpha\_align }\OperatorTok{=}\NormalTok{ qr\_join(alpha\_align)}
\NormalTok{qr\_alpha\_tgt }\OperatorTok{=}\NormalTok{ norm\_dist\_df(qr\_alpha\_align.groupby(}\StringTok{\textquotesingle{}topic\_id\textquotesingle{}}\NormalTok{).}\BuiltInTok{sum}\NormalTok{())}
\NormalTok{output.save\_table(qr\_alpha\_tgt, }\SpecialStringTok{f\textquotesingle{}task1{-}}\SpecialCharTok{\{}\NormalTok{DATA\_MODE}\SpecialCharTok{\}}\SpecialStringTok{{-}alpha{-}target\textquotesingle{}}\NormalTok{, parquet}\OperatorTok{=}\VariableTok{True}\NormalTok{)}
\end{Highlighting}
\end{Shaded}

\begin{verbatim}
INFO:wptrec.save:saving CSV to data\metric-tables\task1-eval-alpha-target.csv.gz
INFO:wptrec.save:data\metric-tables\task1-eval-alpha-target.csv.gz: 2.11 KiB
INFO:wptrec.save:saving Parquet to data\metric-tables\task1-eval-alpha-target.parquet
INFO:wptrec.save:data\metric-tables\task1-eval-alpha-target.parquet: 5.10 KiB
\end{verbatim}

\hypertarget{a4705eb1-02b6-479d-b833-c9de2decd730}{}
\begin{Shaded}
\begin{Highlighting}[]
\NormalTok{qr\_langs\_align }\OperatorTok{=}\NormalTok{ qr\_join(langs\_align)}
\NormalTok{qr\_langs\_tgt }\OperatorTok{=}\NormalTok{ norm\_dist\_df(qr\_langs\_align.groupby(}\StringTok{\textquotesingle{}topic\_id\textquotesingle{}}\NormalTok{).}\BuiltInTok{sum}\NormalTok{())}
\NormalTok{output.save\_table(qr\_langs\_tgt, }\SpecialStringTok{f\textquotesingle{}task1{-}}\SpecialCharTok{\{}\NormalTok{DATA\_MODE}\SpecialCharTok{\}}\SpecialStringTok{{-}langs{-}target\textquotesingle{}}\NormalTok{, parquet}\OperatorTok{=}\VariableTok{True}\NormalTok{)}
\end{Highlighting}
\end{Shaded}

\begin{verbatim}
INFO:wptrec.save:saving CSV to data\metric-tables\task1-eval-langs-target.csv.gz
INFO:wptrec.save:data\metric-tables\task1-eval-langs-target.csv.gz: 1.67 KiB
INFO:wptrec.save:saving Parquet to data\metric-tables\task1-eval-langs-target.parquet
INFO:wptrec.save:data\metric-tables\task1-eval-langs-target.parquet: 5.20 KiB
\end{verbatim}

\hypertarget{f5f336dc-cc67-44dd-98ae-3673c6de4ef7}{}
\begin{Shaded}
\begin{Highlighting}[]
\NormalTok{qr\_pop\_align }\OperatorTok{=}\NormalTok{ qr\_join(pop\_align)}
\NormalTok{qr\_pop\_tgt }\OperatorTok{=}\NormalTok{ norm\_dist\_df(qr\_pop\_align.groupby(}\StringTok{\textquotesingle{}topic\_id\textquotesingle{}}\NormalTok{).}\BuiltInTok{sum}\NormalTok{())}
\NormalTok{output.save\_table(qr\_pop\_tgt, }\SpecialStringTok{f\textquotesingle{}task1{-}}\SpecialCharTok{\{}\NormalTok{DATA\_MODE}\SpecialCharTok{\}}\SpecialStringTok{{-}pop{-}target\textquotesingle{}}\NormalTok{, parquet}\OperatorTok{=}\VariableTok{True}\NormalTok{)}
\end{Highlighting}
\end{Shaded}

\begin{verbatim}
INFO:wptrec.save:saving CSV to data\metric-tables\task1-eval-pop-target.csv.gz
INFO:wptrec.save:data\metric-tables\task1-eval-pop-target.csv.gz: 2.17 KiB
INFO:wptrec.save:saving Parquet to data\metric-tables\task1-eval-pop-target.parquet
INFO:wptrec.save:data\metric-tables\task1-eval-pop-target.parquet: 6.15 KiB
\end{verbatim}

\hypertarget{df2950fe}{}
\hypertarget{multidimensional-alignment}{%
\subsection{Multidimensional
Alignment}\label{multidimensional-alignment}}

Now, we need to set up the \emph{multidimensional} alignment. The basic
version is just to multiply the targets, but that doesn't include the
target averaging we want to do for geographic and gender targets.

Doing that averaging further requires us to very carefully handle the
unknown cases.

We are going to proceed in three steps:

\begin{enumerate}
\tightlist
\item
  Define the averaged dimensions (with their background targets) and the
  un-averaged dimensions
\item
  Demonstrate the logic by working through the alignment computations
  for a single topic
\item
  Apply step (2) to all topics
\end{enumerate}

\hypertarget{1b7ebdb0-854f-449a-a947-5e1bebdd6d69}{}
\hypertarget{dimension-definitions}{%
\subsubsection{Dimension Definitions}\label{dimension-definitions}}

Let's define background distributions for some of our dimensions:

\hypertarget{8cc734a2-eb39-4343-ad72-8ca1eeb78c04}{}
\begin{Shaded}
\begin{Highlighting}[]
\NormalTok{dim\_backgrounds }\OperatorTok{=}\NormalTok{ \{}
    \StringTok{\textquotesingle{}sub{-}geo\textquotesingle{}}\NormalTok{: world\_pop,}
    \StringTok{\textquotesingle{}src{-}geo\textquotesingle{}}\NormalTok{: world\_pop,}
    \StringTok{\textquotesingle{}gender\textquotesingle{}}\NormalTok{: gender\_tgt,}
\NormalTok{\}}
\end{Highlighting}
\end{Shaded}

\leavevmode\vadjust pre{\hypertarget{88824f07-dac0-4df0-87b3-e79373cbd63a}{}}%
Now we'll make a list of dimensions to treat with averaging:

\hypertarget{e7176ebc-2260-471f-be1f-00692e52ae26}{}
\begin{Shaded}
\begin{Highlighting}[]
\NormalTok{DR }\OperatorTok{=}\NormalTok{ namedtuple(}\StringTok{\textquotesingle{}DimRec\textquotesingle{}}\NormalTok{, [}\StringTok{\textquotesingle{}name\textquotesingle{}}\NormalTok{, }\StringTok{\textquotesingle{}align\textquotesingle{}}\NormalTok{, }\StringTok{\textquotesingle{}background\textquotesingle{}}\NormalTok{], defaults}\OperatorTok{=}\NormalTok{[}\VariableTok{None}\NormalTok{])}
\NormalTok{avg\_dims }\OperatorTok{=}\NormalTok{ [}
\NormalTok{    DR(d.name, d.page\_align\_xr, xr.DataArray(dim\_backgrounds[d.name], dims}\OperatorTok{=}\NormalTok{[d.name]))}
    \ControlFlowTok{for}\NormalTok{ d }\KeywordTok{in}\NormalTok{ dimensions}
    \ControlFlowTok{if}\NormalTok{ d.name }\KeywordTok{in}\NormalTok{ dim\_backgrounds}
\NormalTok{]}
\NormalTok{[d.name }\ControlFlowTok{for}\NormalTok{ d }\KeywordTok{in}\NormalTok{ avg\_dims]}
\end{Highlighting}
\end{Shaded}

\begin{verbatim}
['sub-geo', 'src-geo', 'gender']
\end{verbatim}

\leavevmode\vadjust pre{\hypertarget{219ad004-5223-4a9c-9585-4aebe62aa9c9}{}}%
And a list of dimensions to use as-is:

\hypertarget{ff83c5cd-437c-4bcf-98f5-dbcad1538e02}{}
\begin{Shaded}
\begin{Highlighting}[]
\NormalTok{raw\_dims }\OperatorTok{=}\NormalTok{ [}
\NormalTok{    DR(d.name, d.page\_align\_xr)}
    \ControlFlowTok{for}\NormalTok{ d }\KeywordTok{in}\NormalTok{ dimensions}
    \ControlFlowTok{if}\NormalTok{ d.name }\KeywordTok{not} \KeywordTok{in}\NormalTok{ dim\_backgrounds}
\NormalTok{]}
\NormalTok{[d.name }\ControlFlowTok{for}\NormalTok{ d }\KeywordTok{in}\NormalTok{ raw\_dims]}
\end{Highlighting}
\end{Shaded}

\begin{verbatim}
['occ', 'alpha', 'age', 'pop', 'langs']
\end{verbatim}

\leavevmode\vadjust pre{\hypertarget{cee501c6-caf3-4f8c-96aa-a0b20fc0ba23}{}}%
Now: these dimension are in the original order - \texttt{dimensions} has
the averaged dimensions before the non-averaged ones. \textbf{This is
critical for the rest of the code to work.}

\hypertarget{623411db}{}
\hypertarget{demo}{%
\subsubsection{Demo}\label{demo}}

To demonstrate how the logic works, let's first work it out in cells for
one query (1).

What are its documents?

\hypertarget{632b8a14}{}
\begin{Shaded}
\begin{Highlighting}[]
\NormalTok{qno }\OperatorTok{=}\NormalTok{ qrels[}\StringTok{\textquotesingle{}topic\_id\textquotesingle{}}\NormalTok{].iloc[}\DecValTok{0}\NormalTok{]}
\NormalTok{qdf }\OperatorTok{=}\NormalTok{ qrels[qrels[}\StringTok{\textquotesingle{}topic\_id\textquotesingle{}}\NormalTok{] }\OperatorTok{==}\NormalTok{ qno]}
\NormalTok{qdf.name }\OperatorTok{=}\NormalTok{ qno}
\NormalTok{qdf}
\end{Highlighting}
\end{Shaded}

\begin{verbatim}
       topic_id   page_id
0           187       682
1           187       954
2           187      1170
3           187      1315
4           187      1322
...         ...       ...
68641       187  69882575
68642       187  69890514
68643       187  69891122
68644       187  69891390
68645       187  69892653

[68646 rows x 2 columns]
\end{verbatim}

\leavevmode\vadjust pre{\hypertarget{797bc0f6}{}}%
We can use these page IDs to get its alignments.

\hypertarget{1c69eee3-e7cf-4b4f-ac8f-8638cca6788e}{}
\begin{Shaded}
\begin{Highlighting}[]
\NormalTok{q\_pages }\OperatorTok{=}\NormalTok{ qdf[}\StringTok{\textquotesingle{}page\_id\textquotesingle{}}\NormalTok{].values}
\end{Highlighting}
\end{Shaded}

\hypertarget{82d70197-00dc-4bde-b26f-b81cf6a2ce56}{}
\hypertarget{accumulating-initial-targets}{%
\paragraph{Accumulating Initial
Targets}\label{accumulating-initial-targets}}

\leavevmode\vadjust pre{\hypertarget{8fa295d7-530b-46bd-88ee-55df3d1edb6b}{}}%
We're now going to grab the dimensions that have targets, and create a
single xarray with all of them:

\hypertarget{6a331ce2-532f-4b8b-8ae8-c102858ad14b}{}
\begin{Shaded}
\begin{Highlighting}[]
\NormalTok{q\_xta }\OperatorTok{=} \BuiltInTok{reduce}\NormalTok{(operator.mul, [d.align.loc[q\_pages] }\ControlFlowTok{for}\NormalTok{ d }\KeywordTok{in}\NormalTok{ avg\_dims])}
\NormalTok{q\_xta}
\end{Highlighting}
\end{Shaded}

\begin{verbatim}
<xarray.DataArray (page: 68646, sub-geo: 21, src-geo: 21, gender: 4)>
array([[[[0.4       , 0.        , 0.        , 0.        ],
         [0.        , 0.        , 0.        , 0.        ],
         [0.        , 0.        , 0.        , 0.        ],
         ...,
         [0.        , 0.        , 0.        , 0.        ],
         [0.        , 0.        , 0.        , 0.        ],
         [0.05      , 0.        , 0.        , 0.        ]],

        [[0.        , 0.        , 0.        , 0.        ],
         [0.        , 0.        , 0.        , 0.        ],
         [0.        , 0.        , 0.        , 0.        ],
         ...,
         [0.        , 0.        , 0.        , 0.        ],
         [0.        , 0.        , 0.        , 0.        ],
         [0.        , 0.        , 0.        , 0.        ]],

        [[0.        , 0.        , 0.        , 0.        ],
         [0.        , 0.        , 0.        , 0.        ],
         [0.        , 0.        , 0.        , 0.        ],
         ...,
...
         ...,
         [0.        , 0.        , 0.        , 0.        ],
         [0.        , 0.        , 0.        , 0.        ],
         [0.        , 0.        , 0.        , 0.        ]],

        [[0.        , 0.        , 0.        , 0.        ],
         [0.        , 0.        , 0.        , 0.        ],
         [0.        , 0.        , 0.        , 0.        ],
         ...,
         [0.        , 0.        , 0.        , 0.        ],
         [0.        , 0.        , 0.        , 0.        ],
         [0.        , 0.        , 0.        , 0.        ]],

        [[0.        , 0.        , 0.        , 0.        ],
         [0.        , 0.        , 0.        , 0.        ],
         [0.        , 0.        , 0.        , 0.        ],
         ...,
         [0.        , 0.        , 0.        , 0.        ],
         [0.        , 0.        , 0.        , 0.        ],
         [0.        , 0.        , 0.        , 0.        ]]]])
Coordinates:
  * page     (page) int64 682 954 1170 1315 ... 69891122 69891390 69892653
  * sub-geo  (sub-geo) object '@UNKNOWN' 'Antarctica' ... 'Western Europe'
  * src-geo  (src-geo) object '@UNKNOWN' 'Antarctica' ... 'Western Europe'
  * gender   (gender) object '@UNKNOWN' 'female' 'male' 'NB'
\end{verbatim}

\leavevmode\vadjust pre{\hypertarget{877a5140-c5bd-44eb-9bd6-756e2fcfeadb}{}}%
We can similarly do this for the dimensions without targets:

\hypertarget{768a4688-8028-4d3c-b6e5-dde8b787da49}{}
\begin{Shaded}
\begin{Highlighting}[]
\NormalTok{q\_raw\_xta }\OperatorTok{=} \BuiltInTok{reduce}\NormalTok{(operator.mul, [d.align.loc[q\_pages] }\ControlFlowTok{for}\NormalTok{ d }\KeywordTok{in}\NormalTok{ raw\_dims])}
\NormalTok{q\_raw\_xta}
\end{Highlighting}
\end{Shaded}

\begin{verbatim}
<xarray.DataArray (page: 68646, occ: 33, alpha: 4, age: 4, pop: 4, langs: 3)>
array([[[[[[0., 1., 0.],
           [0., 0., 0.],
           [0., 0., 0.],
           [0., 0., 0.]],

          [[0., 0., 0.],
           [0., 0., 0.],
           [0., 0., 0.],
           [0., 0., 0.]],

          [[0., 0., 0.],
           [0., 0., 0.],
           [0., 0., 0.],
           [0., 0., 0.]],

          [[0., 0., 0.],
           [0., 0., 0.],
           [0., 0., 0.],
           [0., 0., 0.]]],

...

         [[[0., 0., 0.],
           [0., 0., 0.],
           [0., 0., 0.],
           [0., 0., 0.]],

          [[0., 0., 0.],
           [0., 0., 0.],
           [0., 0., 0.],
           [0., 0., 0.]],

          [[0., 0., 0.],
           [0., 0., 0.],
           [0., 0., 0.],
           [0., 0., 0.]],

          [[0., 0., 0.],
           [0., 0., 0.],
           [0., 0., 0.],
           [0., 0., 0.]]]]]])
Coordinates:
  * page     (page) int64 682 954 1170 1315 ... 69891122 69891390 69892653
  * occ      (occ) object '@UNKNOWN' 'activist' ... 'writer'
  * alpha    (alpha) object 'a-d' 'e-k' 'l-r' 's-'
  * age      (age) object '2001-2006' '2007-2011' '2012-2016' '2017-2022'
  * pop      (pop) object 'High' 'Low' 'Medium-High' 'Medium-Low'
  * langs    (langs) object '2-4 languages' '5+ languages' 'English only'
\end{verbatim}

\leavevmode\vadjust pre{\hypertarget{2f3156ab-7fb4-44d2-90ff-b3de1045aab7}{}}%
Now, we need to combine this with the other matrix to produce a complete
alignment matrix, which we then will collapse into a query target
matrix. However, we don't have memory to do the whole thing at one go.
Therefore, we will do it page by page.

The \texttt{mean\_outer} function does this:

\hypertarget{f303fab9-093b-475c-aa30-9f190faacf73}{}
\begin{Shaded}
\begin{Highlighting}[]
\ImportTok{from}\NormalTok{ wptrec.dimension }\ImportTok{import}\NormalTok{ mean\_outer}
\end{Highlighting}
\end{Shaded}

\hypertarget{4fd407f8-d3e4-4e97-8ef7-e95df8c70d74}{}
\begin{Shaded}
\begin{Highlighting}[]
\NormalTok{q\_tam }\OperatorTok{=}\NormalTok{ mean\_outer(q\_xta, q\_raw\_xta)}
\NormalTok{q\_tam}
\end{Highlighting}
\end{Shaded}

\begin{verbatim}
<xarray.DataArray (sub-geo: 21, src-geo: 21, gender: 4, occ: 33, alpha: 4,
                   age: 4, pop: 4, langs: 3)>
array([[[[[[[[3.90778732e-05, 9.13512756e-04, 0.00000000e+00],
             [1.07309246e-03, 1.09248385e-03, 8.45444299e-04],
             [2.37733808e-04, 9.16356065e-04, 5.09862192e-05],
             [3.91334003e-04, 3.40654865e-04, 1.97493559e-04]],

            [[8.32428068e-06, 4.32454542e-05, 2.61467791e-06],
             [4.90345161e-04, 3.17004794e-04, 7.68906390e-04],
             [7.80659667e-05, 1.96644900e-04, 1.78047115e-05],
             [2.89383842e-04, 2.64631783e-04, 3.21959246e-04]],

            [[2.03267319e-06, 7.67543443e-06, 0.00000000e+00],
             [3.34556557e-04, 1.53049653e-04, 5.17167629e-04],
             [4.49973617e-05, 2.80944473e-05, 2.21113706e-05],
             [8.65276618e-05, 7.67250106e-05, 1.43730437e-04]],

            [[2.70602185e-05, 4.39063929e-06, 7.07563858e-06],
             [2.58404302e-04, 9.37175266e-05, 7.87475979e-04],
             [9.01797073e-06, 1.61861013e-06, 4.82104549e-05],
             [1.18014878e-04, 2.01703724e-05, 9.15858957e-05]]],

...

           [[[0.00000000e+00, 0.00000000e+00, 0.00000000e+00],
             [0.00000000e+00, 0.00000000e+00, 0.00000000e+00],
             [0.00000000e+00, 0.00000000e+00, 0.00000000e+00],
             [0.00000000e+00, 0.00000000e+00, 0.00000000e+00]],

            [[0.00000000e+00, 0.00000000e+00, 0.00000000e+00],
             [0.00000000e+00, 0.00000000e+00, 0.00000000e+00],
             [0.00000000e+00, 0.00000000e+00, 0.00000000e+00],
             [0.00000000e+00, 0.00000000e+00, 0.00000000e+00]],

            [[0.00000000e+00, 0.00000000e+00, 0.00000000e+00],
             [0.00000000e+00, 0.00000000e+00, 0.00000000e+00],
             [0.00000000e+00, 0.00000000e+00, 0.00000000e+00],
             [0.00000000e+00, 0.00000000e+00, 0.00000000e+00]],

            [[0.00000000e+00, 0.00000000e+00, 0.00000000e+00],
             [0.00000000e+00, 0.00000000e+00, 0.00000000e+00],
             [0.00000000e+00, 0.00000000e+00, 0.00000000e+00],
             [0.00000000e+00, 0.00000000e+00, 0.00000000e+00]]]]]]]])
Coordinates:
  * sub-geo  (sub-geo) object '@UNKNOWN' 'Antarctica' ... 'Western Europe'
  * src-geo  (src-geo) object '@UNKNOWN' 'Antarctica' ... 'Western Europe'
  * gender   (gender) object '@UNKNOWN' 'female' 'male' 'NB'
  * occ      (occ) object '@UNKNOWN' 'activist' ... 'writer'
  * alpha    (alpha) object 'a-d' 'e-k' 'l-r' 's-'
  * age      (age) object '2001-2006' '2007-2011' '2012-2016' '2017-2022'
  * pop      (pop) object 'High' 'Low' 'Medium-High' 'Medium-Low'
  * langs    (langs) object '2-4 languages' '5+ languages' 'English only'
\end{verbatim}

\hypertarget{3263aae5-460f-4832-9131-0c12b9c5b8d3}{}
\begin{Shaded}
\begin{Highlighting}[]
\NormalTok{q\_tam}
\end{Highlighting}
\end{Shaded}

\begin{verbatim}
<xarray.DataArray (sub-geo: 21, src-geo: 21, gender: 4, occ: 33, alpha: 4,
                   age: 4, pop: 4, langs: 3)>
array([[[[[[[[3.90778732e-05, 9.13512756e-04, 0.00000000e+00],
             [1.07309246e-03, 1.09248385e-03, 8.45444299e-04],
             [2.37733808e-04, 9.16356065e-04, 5.09862192e-05],
             [3.91334003e-04, 3.40654865e-04, 1.97493559e-04]],

            [[8.32428068e-06, 4.32454542e-05, 2.61467791e-06],
             [4.90345161e-04, 3.17004794e-04, 7.68906390e-04],
             [7.80659667e-05, 1.96644900e-04, 1.78047115e-05],
             [2.89383842e-04, 2.64631783e-04, 3.21959246e-04]],

            [[2.03267319e-06, 7.67543443e-06, 0.00000000e+00],
             [3.34556557e-04, 1.53049653e-04, 5.17167629e-04],
             [4.49973617e-05, 2.80944473e-05, 2.21113706e-05],
             [8.65276618e-05, 7.67250106e-05, 1.43730437e-04]],

            [[2.70602185e-05, 4.39063929e-06, 7.07563858e-06],
             [2.58404302e-04, 9.37175266e-05, 7.87475979e-04],
             [9.01797073e-06, 1.61861013e-06, 4.82104549e-05],
             [1.18014878e-04, 2.01703724e-05, 9.15858957e-05]]],

...

           [[[0.00000000e+00, 0.00000000e+00, 0.00000000e+00],
             [0.00000000e+00, 0.00000000e+00, 0.00000000e+00],
             [0.00000000e+00, 0.00000000e+00, 0.00000000e+00],
             [0.00000000e+00, 0.00000000e+00, 0.00000000e+00]],

            [[0.00000000e+00, 0.00000000e+00, 0.00000000e+00],
             [0.00000000e+00, 0.00000000e+00, 0.00000000e+00],
             [0.00000000e+00, 0.00000000e+00, 0.00000000e+00],
             [0.00000000e+00, 0.00000000e+00, 0.00000000e+00]],

            [[0.00000000e+00, 0.00000000e+00, 0.00000000e+00],
             [0.00000000e+00, 0.00000000e+00, 0.00000000e+00],
             [0.00000000e+00, 0.00000000e+00, 0.00000000e+00],
             [0.00000000e+00, 0.00000000e+00, 0.00000000e+00]],

            [[0.00000000e+00, 0.00000000e+00, 0.00000000e+00],
             [0.00000000e+00, 0.00000000e+00, 0.00000000e+00],
             [0.00000000e+00, 0.00000000e+00, 0.00000000e+00],
             [0.00000000e+00, 0.00000000e+00, 0.00000000e+00]]]]]]]])
Coordinates:
  * sub-geo  (sub-geo) object '@UNKNOWN' 'Antarctica' ... 'Western Europe'
  * src-geo  (src-geo) object '@UNKNOWN' 'Antarctica' ... 'Western Europe'
  * gender   (gender) object '@UNKNOWN' 'female' 'male' 'NB'
  * occ      (occ) object '@UNKNOWN' 'activist' ... 'writer'
  * alpha    (alpha) object 'a-d' 'e-k' 'l-r' 's-'
  * age      (age) object '2001-2006' '2007-2011' '2012-2016' '2017-2022'
  * pop      (pop) object 'High' 'Low' 'Medium-High' 'Medium-Low'
  * langs    (langs) object '2-4 languages' '5+ languages' 'English only'
\end{verbatim}

\hypertarget{cec140b7-ed73-4def-9629-a2875d1de7d8}{}
\begin{Shaded}
\begin{Highlighting}[]
\NormalTok{q\_tam.}\BuiltInTok{sum}\NormalTok{()}
\end{Highlighting}
\end{Shaded}

\begin{verbatim}
<xarray.DataArray ()>
array(1.00001457)
\end{verbatim}

\leavevmode\vadjust pre{\hypertarget{03c5d7ae}{}}%
In 2021, we ignored fully-unknown for Task 1. However, it isn't clear
hot to properly do that with some attributes that are never fully
unknown - they still need to be counted. Therefore, we consistently
treat fully-unknown as a distinct category for both Task 1 and Task 2
metrics.

\hypertarget{3acc0e56}{}
\hypertarget{data-subsetting}{%
\paragraph{Data Subsetting}\label{data-subsetting}}

Before we average, we need to be able to select data by its
known/unknown status.

Let's start by making a list of cases - the known/unknown status of each
dimension.

\hypertarget{a53eb29c-76ce-4580-942e-ac43424ad8e4}{}
\begin{Shaded}
\begin{Highlighting}[]
\NormalTok{avg\_cases }\OperatorTok{=} \BuiltInTok{list}\NormalTok{(product(}\OperatorTok{*}\NormalTok{[[}\VariableTok{True}\NormalTok{, }\VariableTok{False}\NormalTok{] }\ControlFlowTok{for}\NormalTok{ d }\KeywordTok{in}\NormalTok{ avg\_dims]))}
\NormalTok{avg\_cases}
\end{Highlighting}
\end{Shaded}

\begin{verbatim}
[(True, True, True),
 (True, True, False),
 (True, False, True),
 (True, False, False),
 (False, True, True),
 (False, True, False),
 (False, False, True),
 (False, False, False)]
\end{verbatim}

\leavevmode\vadjust pre{\hypertarget{7ab2c9f7-0a78-449a-9c27-733a3eb81259}{}}%
The last entry is the all-unknown case - remove it:

\hypertarget{0602baa0-02d6-4555-b63c-6e0c5ef479e6}{}
\begin{Shaded}
\begin{Highlighting}[]
\NormalTok{avg\_cases.pop()}
\NormalTok{avg\_cases}
\end{Highlighting}
\end{Shaded}

\begin{verbatim}
[(True, True, True),
 (True, True, False),
 (True, False, True),
 (True, False, False),
 (False, True, True),
 (False, True, False),
 (False, False, True)]
\end{verbatim}

\leavevmode\vadjust pre{\hypertarget{1f9d1df8-47ac-42aa-9bd0-2c437b9a3956}{}}%
We now want the ability to create an indexer to look up the subset of
the alignment frame corresponding to a case. Let's write that function:

\hypertarget{ac94ffb7-9c4c-4009-b521-6e9c7ac2cfc2}{}
\begin{Shaded}
\begin{Highlighting}[]
\KeywordTok{def}\NormalTok{ case\_selector(case):}
    \KeywordTok{def}\NormalTok{ mksel(known):}
        \ControlFlowTok{if}\NormalTok{ known:}
            \CommentTok{\# select all but 1st column}
            \ControlFlowTok{return} \BuiltInTok{slice}\NormalTok{(}\DecValTok{1}\NormalTok{, }\VariableTok{None}\NormalTok{, }\VariableTok{None}\NormalTok{)}
        \ControlFlowTok{else}\NormalTok{:}
            \CommentTok{\# select 1st column}
            \ControlFlowTok{return} \DecValTok{0}
    
    \ControlFlowTok{return} \BuiltInTok{tuple}\NormalTok{(mksel(k) }\ControlFlowTok{for}\NormalTok{ k }\KeywordTok{in}\NormalTok{ case)}
\end{Highlighting}
\end{Shaded}

\leavevmode\vadjust pre{\hypertarget{a7292a11-482e-4933-815d-c925f7b27b91}{}}%
Let's test this function quick:

\hypertarget{9ab9d515-47f2-4a70-9e17-a96b62d490ed}{}
\begin{Shaded}
\begin{Highlighting}[]
\NormalTok{case\_selector(avg\_cases[}\DecValTok{0}\NormalTok{])}
\end{Highlighting}
\end{Shaded}

\begin{verbatim}
(slice(1, None, None), slice(1, None, None), slice(1, None, None))
\end{verbatim}

\hypertarget{ea928b45-9f28-4aee-b748-19f6acdcebf9}{}
\begin{Shaded}
\begin{Highlighting}[]
\NormalTok{case\_selector(avg\_cases[}\OperatorTok{{-}}\DecValTok{1}\NormalTok{])}
\end{Highlighting}
\end{Shaded}

\begin{verbatim}
(0, 0, slice(1, None, None))
\end{verbatim}

\leavevmode\vadjust pre{\hypertarget{758346b1-3fc8-4170-85e6-693cce5d21a9}{}}%
And make sure we can use it:

\hypertarget{303d14f5-9983-47bc-9e95-a10ad7ba1e15}{}
\begin{Shaded}
\begin{Highlighting}[]
\NormalTok{q\_tam[case\_selector(avg\_cases[}\DecValTok{1}\NormalTok{])]}
\end{Highlighting}
\end{Shaded}

\begin{verbatim}
<xarray.DataArray (sub-geo: 20, src-geo: 20, occ: 33, alpha: 4, age: 4, pop: 4,
                   langs: 3)>
array([[[[[[[0.00000000e+00, 0.00000000e+00, 0.00000000e+00],
            [0.00000000e+00, 0.00000000e+00, 0.00000000e+00],
            [0.00000000e+00, 0.00000000e+00, 0.00000000e+00],
            [0.00000000e+00, 0.00000000e+00, 0.00000000e+00]],

           [[0.00000000e+00, 0.00000000e+00, 0.00000000e+00],
            [0.00000000e+00, 0.00000000e+00, 0.00000000e+00],
            [0.00000000e+00, 0.00000000e+00, 0.00000000e+00],
            [0.00000000e+00, 0.00000000e+00, 0.00000000e+00]],

           [[0.00000000e+00, 0.00000000e+00, 0.00000000e+00],
            [0.00000000e+00, 0.00000000e+00, 0.00000000e+00],
            [0.00000000e+00, 0.00000000e+00, 0.00000000e+00],
            [0.00000000e+00, 0.00000000e+00, 0.00000000e+00]],

           [[0.00000000e+00, 0.00000000e+00, 0.00000000e+00],
            [0.00000000e+00, 0.00000000e+00, 0.00000000e+00],
            [0.00000000e+00, 0.00000000e+00, 0.00000000e+00],
            [0.00000000e+00, 0.00000000e+00, 0.00000000e+00]]],

...

          [[[0.00000000e+00, 0.00000000e+00, 0.00000000e+00],
            [0.00000000e+00, 0.00000000e+00, 0.00000000e+00],
            [0.00000000e+00, 0.00000000e+00, 0.00000000e+00],
            [0.00000000e+00, 0.00000000e+00, 0.00000000e+00]],

           [[0.00000000e+00, 0.00000000e+00, 0.00000000e+00],
            [0.00000000e+00, 0.00000000e+00, 0.00000000e+00],
            [0.00000000e+00, 0.00000000e+00, 0.00000000e+00],
            [0.00000000e+00, 0.00000000e+00, 0.00000000e+00]],

           [[0.00000000e+00, 0.00000000e+00, 0.00000000e+00],
            [0.00000000e+00, 0.00000000e+00, 0.00000000e+00],
            [0.00000000e+00, 0.00000000e+00, 0.00000000e+00],
            [0.00000000e+00, 0.00000000e+00, 0.00000000e+00]],

           [[0.00000000e+00, 0.00000000e+00, 0.00000000e+00],
            [0.00000000e+00, 0.00000000e+00, 0.00000000e+00],
            [0.00000000e+00, 0.00000000e+00, 0.00000000e+00],
            [0.00000000e+00, 0.00000000e+00, 0.00000000e+00]]]]]]])
Coordinates:
  * sub-geo  (sub-geo) object 'Antarctica' 'Caribbean' ... 'Western Europe'
  * src-geo  (src-geo) object 'Antarctica' 'Caribbean' ... 'Western Europe'
    gender   <U8 '@UNKNOWN'
  * occ      (occ) object '@UNKNOWN' 'activist' ... 'writer'
  * alpha    (alpha) object 'a-d' 'e-k' 'l-r' 's-'
  * age      (age) object '2001-2006' '2007-2011' '2012-2016' '2017-2022'
  * pop      (pop) object 'High' 'Low' 'Medium-High' 'Medium-Low'
  * langs    (langs) object '2-4 languages' '5+ languages' 'English only'
\end{verbatim}

\leavevmode\vadjust pre{\hypertarget{bb065aa3-8b64-4d74-89ca-23274a78ece1}{}}%
Fantastic! Given a case (known and unknown statuses), we can select the
subset of the target matrix with exactly those.

\hypertarget{15469b24-e30b-4ab5-8c8a-dd99fa5cace0}{}
\hypertarget{averaging}{%
\paragraph{Averaging}\label{averaging}}

Ok, now we have to - very carefully - average with our target modifier.
For each dimension that is not fully-unknown, we average with the
intersectional target defined over the known dimensions.

At all times, we also need to respect the fraction of the total it
represents.

We'll use the selection capabilities above to handle this.

First, let's make sure that our target matrix sums to 1 to start with:

\hypertarget{791db0d8-6b15-4351-9482-b1cc505e5da8}{}
\begin{Shaded}
\begin{Highlighting}[]
\NormalTok{q\_tam.}\BuiltInTok{sum}\NormalTok{()}
\end{Highlighting}
\end{Shaded}

\begin{verbatim}
<xarray.DataArray ()>
array(1.00001457)
\end{verbatim}

\leavevmode\vadjust pre{\hypertarget{fd0cda25-6f96-48db-afc9-e3237e03e567}{}}%
Fantastic. This means that if we sum up a subset of the data, it will
give us the fraction of the distribution that has that combination of
known/unknown status.

For each condition, we are going to proceed as follows:

\begin{enumerate}
\tightlist
\item
  Compute an appropriate intersectional background distribution (based
  on the dimensions that are "known")
\item
  Select the subset of the target matrix with this known status
\item
  Compute the sum of this subset
\item
  Re-normalize the subset to sum to 1
\item
  Compute a normalization table such that each coordinate in the
  distributions to correct sums to 1 (so multiplying this by the
  background distribution spreads the background across the other
  dimensions appropriately), and use this to spread the background
  distribution
\item
  Average with the spread background distribution
\item
  Re-normalize to preserve the original sum
\end{enumerate}

Let's define the whole process as a function:

\hypertarget{a584d651-c27f-46f5-bf16-40526babaa91}{}
\begin{Shaded}
\begin{Highlighting}[]
\KeywordTok{def}\NormalTok{ avg\_with\_bg(tm, verbose}\OperatorTok{=}\VariableTok{False}\NormalTok{):}
\NormalTok{    tm }\OperatorTok{=}\NormalTok{ tm.copy()}
    
\NormalTok{    tail\_names }\OperatorTok{=}\NormalTok{ [d.name }\ControlFlowTok{for}\NormalTok{ d }\KeywordTok{in}\NormalTok{ raw\_dims]}
    
    \CommentTok{\# compute the tail mass for each coordinate (can be done once)}
\NormalTok{    tail\_mass }\OperatorTok{=}\NormalTok{ tm.}\BuiltInTok{sum}\NormalTok{(tail\_names)}
    
    \CommentTok{\# now some things don\textquotesingle{}t have any mass, but we still need to distribute background distributions.}
    \CommentTok{\# solution: we impute the marginal tail distribution}
    \CommentTok{\# first compute it}
\NormalTok{    tail\_marg }\OperatorTok{=}\NormalTok{ tm.}\BuiltInTok{sum}\NormalTok{([d.name }\ControlFlowTok{for}\NormalTok{ d }\KeywordTok{in}\NormalTok{ avg\_dims])}
    \CommentTok{\# then impute that where we don\textquotesingle{}t have mass}
\NormalTok{    tm\_imputed }\OperatorTok{=}\NormalTok{ xr.where(tail\_mass }\OperatorTok{\textgreater{}} \DecValTok{0}\NormalTok{, tm, tail\_marg)}
    \CommentTok{\# and re{-}compute the tail mass}
\NormalTok{    tail\_mass }\OperatorTok{=}\NormalTok{ tm\_imputed.}\BuiltInTok{sum}\NormalTok{(tail\_names)}
    \CommentTok{\# and finally we compute the rescaled matrix}
\NormalTok{    tail\_scale }\OperatorTok{=}\NormalTok{ tm\_imputed }\OperatorTok{/}\NormalTok{ tail\_mass}
    \KeywordTok{del}\NormalTok{ tm\_imputed}
    
    \ControlFlowTok{for}\NormalTok{ case }\KeywordTok{in}\NormalTok{ avg\_cases:}
        \CommentTok{\# for deugging: get names}
\NormalTok{        known\_names }\OperatorTok{=}\NormalTok{ [d.name }\ControlFlowTok{for}\NormalTok{ (d, known) }\KeywordTok{in} \BuiltInTok{zip}\NormalTok{(avg\_dims, case) }\ControlFlowTok{if}\NormalTok{ known]}
        \ControlFlowTok{if}\NormalTok{ verbose:}
            \BuiltInTok{print}\NormalTok{(}\StringTok{\textquotesingle{}processing known:\textquotesingle{}}\NormalTok{, known\_names)}
        
        \CommentTok{\# Step 1: background}
\NormalTok{        bg }\OperatorTok{=} \BuiltInTok{reduce}\NormalTok{(operator.mul, [}
\NormalTok{            d.background}
            \ControlFlowTok{for}\NormalTok{ (d, known) }\KeywordTok{in} \BuiltInTok{zip}\NormalTok{(avg\_dims, case)}
            \ControlFlowTok{if}\NormalTok{ known}
\NormalTok{        ])}
        \ControlFlowTok{if} \KeywordTok{not}\NormalTok{ np.allclose(bg.}\BuiltInTok{sum}\NormalTok{(), }\FloatTok{1.0}\NormalTok{):}
\NormalTok{            warnings.warn(}\StringTok{\textquotesingle{}background distribution for }\SpecialCharTok{\{\}}\StringTok{ sums to }\SpecialCharTok{\{\}}\StringTok{, expected 1\textquotesingle{}}\NormalTok{.}\BuiltInTok{format}\NormalTok{(known\_names, bg.values.}\BuiltInTok{sum}\NormalTok{()))}
        
        \CommentTok{\# Step 2: selector}
\NormalTok{        sel }\OperatorTok{=}\NormalTok{ case\_selector(case)}
        
        \CommentTok{\# Steps 3: sum in preparation for normalization}
\NormalTok{        c\_sum }\OperatorTok{=}\NormalTok{ tm[sel].}\BuiltInTok{sum}\NormalTok{()}
        
        \CommentTok{\# Step 5: spread the background}
\NormalTok{        bg\_spread }\OperatorTok{=}\NormalTok{ bg }\OperatorTok{*}\NormalTok{ tail\_scale[sel] }\OperatorTok{*}\NormalTok{ c\_sum}
        \ControlFlowTok{if} \KeywordTok{not}\NormalTok{ np.allclose(bg\_spread.}\BuiltInTok{sum}\NormalTok{(), c\_sum):}
\NormalTok{            warnings.warn(}\StringTok{\textquotesingle{}rescaled background sums to }\SpecialCharTok{\{\}}\StringTok{, expected c\_sum\textquotesingle{}}\NormalTok{.}\BuiltInTok{format}\NormalTok{(bg\_spread.values.}\BuiltInTok{sum}\NormalTok{()))}
        
        \CommentTok{\# Step 4 \& 6: average with the background}
\NormalTok{        tm[sel] }\OperatorTok{*=} \FloatTok{0.5}
\NormalTok{        bg\_spread }\OperatorTok{*=} \FloatTok{0.5}
\NormalTok{        tm[sel] }\OperatorTok{+=}\NormalTok{ bg\_spread}
                        
        \ControlFlowTok{if} \KeywordTok{not}\NormalTok{ np.allclose(tm[sel].}\BuiltInTok{sum}\NormalTok{(), c\_sum):}
\NormalTok{            warnings.warn(}\StringTok{\textquotesingle{}target distribution for }\SpecialCharTok{\{\}}\StringTok{ sums to }\SpecialCharTok{\{\}}\StringTok{, expected }\SpecialCharTok{\{\}}\StringTok{\textquotesingle{}}\NormalTok{.}\BuiltInTok{format}\NormalTok{(known\_names, tm[sel].values.}\BuiltInTok{sum}\NormalTok{(), c\_sum))}
    
    \ControlFlowTok{return}\NormalTok{ tm}
\end{Highlighting}
\end{Shaded}

\leavevmode\vadjust pre{\hypertarget{2ee5241e}{}}%
And apply it:

\hypertarget{5641af93-1bb2-4ee7-bb0e-196f72256dd9}{}
\begin{Shaded}
\begin{Highlighting}[]
\NormalTok{q\_target }\OperatorTok{=}\NormalTok{ avg\_with\_bg(q\_tam, }\VariableTok{True}\NormalTok{)}
\NormalTok{q\_target.}\BuiltInTok{sum}\NormalTok{()}
\end{Highlighting}
\end{Shaded}

\begin{verbatim}
processing known: ['sub-geo', 'src-geo', 'gender']
processing known: ['sub-geo', 'src-geo']
processing known: ['sub-geo', 'gender']
processing known: ['sub-geo']
processing known: ['src-geo', 'gender']
processing known: ['src-geo']
processing known: ['gender']
\end{verbatim}

\begin{verbatim}
<xarray.DataArray ()>
array(1.00001457)
\end{verbatim}

\hypertarget{b896812a-2892-4f3e-94da-228094cf548c}{}
\begin{Shaded}
\begin{Highlighting}[]
\NormalTok{q\_target}
\end{Highlighting}
\end{Shaded}

\begin{verbatim}
<xarray.DataArray (sub-geo: 21, src-geo: 21, gender: 4, occ: 33, alpha: 4,
                   age: 4, pop: 4, langs: 3)>
array([[[[[[[[3.90778732e-05, 9.13512756e-04, 0.00000000e+00],
             [1.07309246e-03, 1.09248385e-03, 8.45444299e-04],
             [2.37733808e-04, 9.16356065e-04, 5.09862192e-05],
             [3.91334003e-04, 3.40654865e-04, 1.97493559e-04]],

            [[8.32428068e-06, 4.32454542e-05, 2.61467791e-06],
             [4.90345161e-04, 3.17004794e-04, 7.68906390e-04],
             [7.80659667e-05, 1.96644900e-04, 1.78047115e-05],
             [2.89383842e-04, 2.64631783e-04, 3.21959246e-04]],

            [[2.03267319e-06, 7.67543443e-06, 0.00000000e+00],
             [3.34556557e-04, 1.53049653e-04, 5.17167629e-04],
             [4.49973617e-05, 2.80944473e-05, 2.21113706e-05],
             [8.65276618e-05, 7.67250106e-05, 1.43730437e-04]],

            [[2.70602185e-05, 4.39063929e-06, 7.07563858e-06],
             [2.58404302e-04, 9.37175266e-05, 7.87475979e-04],
             [9.01797073e-06, 1.61861013e-06, 4.82104549e-05],
             [1.18014878e-04, 2.01703724e-05, 9.15858957e-05]]],

...

           [[[0.00000000e+00, 0.00000000e+00, 0.00000000e+00],
             [7.50104470e-13, 3.37547011e-12, 0.00000000e+00],
             [0.00000000e+00, 3.18794400e-12, 0.00000000e+00],
             [0.00000000e+00, 1.63415617e-12, 0.00000000e+00]],

            [[0.00000000e+00, 0.00000000e+00, 0.00000000e+00],
             [7.76358126e-12, 2.32532386e-12, 2.25031341e-12],
             [0.00000000e+00, 4.50062682e-13, 0.00000000e+00],
             [7.50104470e-13, 0.00000000e+00, 0.00000000e+00]],

            [[0.00000000e+00, 0.00000000e+00, 0.00000000e+00],
             [1.87526117e-12, 0.00000000e+00, 2.43783953e-12],
             [0.00000000e+00, 0.00000000e+00, 0.00000000e+00],
             [0.00000000e+00, 0.00000000e+00, 0.00000000e+00]],

            [[0.00000000e+00, 0.00000000e+00, 0.00000000e+00],
             [2.25031341e-12, 5.62578352e-13, 3.00041788e-12],
             [0.00000000e+00, 0.00000000e+00, 0.00000000e+00],
             [0.00000000e+00, 0.00000000e+00, 0.00000000e+00]]]]]]]])
Coordinates:
  * sub-geo  (sub-geo) object '@UNKNOWN' 'Antarctica' ... 'Western Europe'
  * src-geo  (src-geo) object '@UNKNOWN' 'Antarctica' ... 'Western Europe'
  * gender   (gender) object '@UNKNOWN' 'female' 'male' 'NB'
  * occ      (occ) object '@UNKNOWN' 'activist' ... 'writer'
  * alpha    (alpha) object 'a-d' 'e-k' 'l-r' 's-'
  * age      (age) object '2001-2006' '2007-2011' '2012-2016' '2017-2022'
  * pop      (pop) object 'High' 'Low' 'Medium-High' 'Medium-Low'
  * langs    (langs) object '2-4 languages' '5+ languages' 'English only'
\end{verbatim}

\hypertarget{af47d2db-a794-4c34-8ffe-e274262fa9cb}{}
\begin{Shaded}
\begin{Highlighting}[]
\BuiltInTok{print}\NormalTok{(number(q\_target.values.size), }\StringTok{\textquotesingle{}values taking\textquotesingle{}}\NormalTok{, binarysize(q\_target.nbytes))}
\end{Highlighting}
\end{Shaded}

\begin{verbatim}
11,176,704 values taking 89.41 MiB
\end{verbatim}

\leavevmode\vadjust pre{\hypertarget{16e7686a-1120-4fd6-9216-973ee503b194}{}}%
Is it still a distribution?

\hypertarget{bff0d1c1-d393-42c8-87b1-92e2ccff7524}{}
\begin{Shaded}
\begin{Highlighting}[]
\NormalTok{q\_target.}\BuiltInTok{sum}\NormalTok{()}
\end{Highlighting}
\end{Shaded}

\begin{verbatim}
<xarray.DataArray ()>
array(1.00001457)
\end{verbatim}

\leavevmode\vadjust pre{\hypertarget{5f0b1cdd-7892-48c6-b0fa-da0edfa80cd7}{}}%
We can unravel this value into a single-dimensional array representing
the multidimensional target:

\hypertarget{ef2789e8-ce11-43e3-85a8-0a70d9ffbd54}{}
\begin{Shaded}
\begin{Highlighting}[]
\NormalTok{q\_target.values.ravel()}
\end{Highlighting}
\end{Shaded}

\begin{verbatim}
array([3.90778732e-05, 9.13512756e-04, 0.00000000e+00, ...,
       0.00000000e+00, 0.00000000e+00, 0.00000000e+00])
\end{verbatim}

\leavevmode\vadjust pre{\hypertarget{662a0a02}{}}%
Now we have all the pieces to compute this for each of our queries.

\hypertarget{5a176fb7}{}
\hypertarget{implementing-function}{%
\subsubsection{Implementing Function}\label{implementing-function}}

To perform this combination for every query, we'll use a function that
takes a data frame for a query's relevant docs and performs all of the
above operations:

\hypertarget{984c7b0a}{}
\begin{Shaded}
\begin{Highlighting}[]
\KeywordTok{def}\NormalTok{ query\_xalign(pages):}
    \CommentTok{\# compute targets to average}
\NormalTok{    avg\_pages }\OperatorTok{=} \BuiltInTok{reduce}\NormalTok{(operator.mul, [d.align.loc[pages] }\ControlFlowTok{for}\NormalTok{ d }\KeywordTok{in}\NormalTok{ avg\_dims])}
\NormalTok{    raw\_pages }\OperatorTok{=} \BuiltInTok{reduce}\NormalTok{(operator.mul, [d.align.loc[pages] }\ControlFlowTok{for}\NormalTok{ d }\KeywordTok{in}\NormalTok{ raw\_dims])}

    \CommentTok{\# convert to query distribution}
\NormalTok{    tgt }\OperatorTok{=}\NormalTok{ mean\_outer(avg\_pages, raw\_pages)}

    \CommentTok{\# average with background distributions}
\NormalTok{    tgt }\OperatorTok{=}\NormalTok{ avg\_with\_bg(tgt)}
    
    \CommentTok{\# and return the result}
    \ControlFlowTok{return}\NormalTok{ tgt}
\end{Highlighting}
\end{Shaded}

\leavevmode\vadjust pre{\hypertarget{4ac9c20d}{}}%
Make sure it works:

\hypertarget{e35a7003}{}
\begin{Shaded}
\begin{Highlighting}[]
\NormalTok{query\_xalign(qdf.page\_id.values)}
\end{Highlighting}
\end{Shaded}

\begin{verbatim}
<xarray.DataArray (sub-geo: 21, src-geo: 21, gender: 4, occ: 33, alpha: 4,
                   age: 4, pop: 4, langs: 3)>
array([[[[[[[[3.90778732e-05, 9.13512756e-04, 0.00000000e+00],
             [1.07309246e-03, 1.09248385e-03, 8.45444299e-04],
             [2.37733808e-04, 9.16356065e-04, 5.09862192e-05],
             [3.91334003e-04, 3.40654865e-04, 1.97493559e-04]],

            [[8.32428068e-06, 4.32454542e-05, 2.61467791e-06],
             [4.90345161e-04, 3.17004794e-04, 7.68906390e-04],
             [7.80659667e-05, 1.96644900e-04, 1.78047115e-05],
             [2.89383842e-04, 2.64631783e-04, 3.21959246e-04]],

            [[2.03267319e-06, 7.67543443e-06, 0.00000000e+00],
             [3.34556557e-04, 1.53049653e-04, 5.17167629e-04],
             [4.49973617e-05, 2.80944473e-05, 2.21113706e-05],
             [8.65276618e-05, 7.67250106e-05, 1.43730437e-04]],

            [[2.70602185e-05, 4.39063929e-06, 7.07563858e-06],
             [2.58404302e-04, 9.37175266e-05, 7.87475979e-04],
             [9.01797073e-06, 1.61861013e-06, 4.82104549e-05],
             [1.18014878e-04, 2.01703724e-05, 9.15858957e-05]]],

...

           [[[0.00000000e+00, 0.00000000e+00, 0.00000000e+00],
             [7.50104470e-13, 3.37547011e-12, 0.00000000e+00],
             [0.00000000e+00, 3.18794400e-12, 0.00000000e+00],
             [0.00000000e+00, 1.63415617e-12, 0.00000000e+00]],

            [[0.00000000e+00, 0.00000000e+00, 0.00000000e+00],
             [7.76358126e-12, 2.32532386e-12, 2.25031341e-12],
             [0.00000000e+00, 4.50062682e-13, 0.00000000e+00],
             [7.50104470e-13, 0.00000000e+00, 0.00000000e+00]],

            [[0.00000000e+00, 0.00000000e+00, 0.00000000e+00],
             [1.87526117e-12, 0.00000000e+00, 2.43783953e-12],
             [0.00000000e+00, 0.00000000e+00, 0.00000000e+00],
             [0.00000000e+00, 0.00000000e+00, 0.00000000e+00]],

            [[0.00000000e+00, 0.00000000e+00, 0.00000000e+00],
             [2.25031341e-12, 5.62578352e-13, 3.00041788e-12],
             [0.00000000e+00, 0.00000000e+00, 0.00000000e+00],
             [0.00000000e+00, 0.00000000e+00, 0.00000000e+00]]]]]]]])
Coordinates:
  * sub-geo  (sub-geo) object '@UNKNOWN' 'Antarctica' ... 'Western Europe'
  * src-geo  (src-geo) object '@UNKNOWN' 'Antarctica' ... 'Western Europe'
  * gender   (gender) object '@UNKNOWN' 'female' 'male' 'NB'
  * occ      (occ) object '@UNKNOWN' 'activist' ... 'writer'
  * alpha    (alpha) object 'a-d' 'e-k' 'l-r' 's-'
  * age      (age) object '2001-2006' '2007-2011' '2012-2016' '2017-2022'
  * pop      (pop) object 'High' 'Low' 'Medium-High' 'Medium-Low'
  * langs    (langs) object '2-4 languages' '5+ languages' 'English only'
\end{verbatim}

\hypertarget{5d986450-d142-4531-9e6f-517ea33fd79d}{}
\hypertarget{computing-query-targets}{%
\subsubsection{Computing Query Targets}\label{computing-query-targets}}

\leavevmode\vadjust pre{\hypertarget{4f259d2b}{}}%
Now with that function, we can compute the alignment vector for each
query. Extract queries into a dictionary:

\hypertarget{99cdc931-8650-47f2-b5f7-77521bb7674c}{}
\begin{Shaded}
\begin{Highlighting}[]
\NormalTok{queries }\OperatorTok{=}\NormalTok{ \{}
\NormalTok{    t: df[}\StringTok{\textquotesingle{}page\_id\textquotesingle{}}\NormalTok{].values}
    \ControlFlowTok{for}\NormalTok{ (t, df) }\KeywordTok{in}\NormalTok{ qrels.groupby(}\StringTok{\textquotesingle{}topic\_id\textquotesingle{}}\NormalTok{)}
\NormalTok{\}}
\end{Highlighting}
\end{Shaded}

\leavevmode\vadjust pre{\hypertarget{bbe813d6-934a-45f4-adb5-f708db5b0656}{}}%
Make an index that we'll need later for setting up the XArray dimension:

\hypertarget{3fb22653-4512-4f20-bd15-07cd1142531d}{}
\begin{Shaded}
\begin{Highlighting}[]
\NormalTok{q\_ids }\OperatorTok{=}\NormalTok{ pd.Index(queries.keys(), name}\OperatorTok{=}\StringTok{\textquotesingle{}topic\_id\textquotesingle{}}\NormalTok{)}
\NormalTok{q\_ids}
\end{Highlighting}
\end{Shaded}

\begin{verbatim}
Int64Index([ 187,  270,  359,  365,  400,  404,  480,  517,  568,  596,  715,
             807,  834,  881,  883,  949,  951,  955,  995, 1018, 1180, 1233,
            1328, 1406, 1417, 1448, 1449, 1479, 1499, 1548, 1558, 1647, 1685,
            1806, 1821, 1877, 1884, 1890, 2000, 2028, 2106, 2153, 2160, 2229,
            2244, 2448, 2483, 2758, 2867, 2872],
           dtype='int64', name='topic_id')
\end{verbatim}

\leavevmode\vadjust pre{\hypertarget{b83a26f9-31be-449a-a85d-a928998ed182}{}}%
Now let's create targets for each of these:

\hypertarget{f4c2c70f-1caa-4cc3-a4f3-26a75b5fc5f0}{}
\begin{Shaded}
\begin{Highlighting}[]
\NormalTok{q\_tgts }\OperatorTok{=}\NormalTok{ [query\_xalign(queries[q]) }\ControlFlowTok{for}\NormalTok{ q }\KeywordTok{in}\NormalTok{ tqdm(q\_ids)]}
\end{Highlighting}
\end{Shaded}

\begin{Shaded}
\begin{Highlighting}[]
\FunctionTok{\{}\DataTypeTok{"model\_id"}\FunctionTok{:}\StringTok{"d7cf659921754083b3c99d2a487c1f52"}\FunctionTok{,}\DataTypeTok{"version\_major"}\FunctionTok{:}\DecValTok{2}\FunctionTok{,}\DataTypeTok{"version\_minor"}\FunctionTok{:}\DecValTok{0}\FunctionTok{\}}
\end{Highlighting}
\end{Shaded}

\leavevmode\vadjust pre{\hypertarget{4912e850-384c-403f-8cdb-b5cfe7229419}{}}%
Assemble a composite xarray:

\hypertarget{73a708fb-21d1-4664-8b99-f7997ac40b6c}{}
\begin{Shaded}
\begin{Highlighting}[]
\NormalTok{q\_tgts }\OperatorTok{=}\NormalTok{ xr.concat(q\_tgts, q\_ids)}
\NormalTok{q\_tgts}
\end{Highlighting}
\end{Shaded}

\begin{verbatim}
<xarray.DataArray (topic_id: 50, sub-geo: 21, src-geo: 21, gender: 4, occ: 33,
                   alpha: 4, age: 4, pop: 4, langs: 3)>
array([[[[[[[[[3.90778732e-05, 9.13512756e-04, 0.00000000e+00],
              [1.07309246e-03, 1.09248385e-03, 8.45444299e-04],
              [2.37733808e-04, 9.16356065e-04, 5.09862192e-05],
              [3.91334003e-04, 3.40654865e-04, 1.97493559e-04]],

             [[8.32428068e-06, 4.32454542e-05, 2.61467791e-06],
              [4.90345161e-04, 3.17004794e-04, 7.68906390e-04],
              [7.80659667e-05, 1.96644900e-04, 1.78047115e-05],
              [2.89383842e-04, 2.64631783e-04, 3.21959246e-04]],

             [[2.03267319e-06, 7.67543443e-06, 0.00000000e+00],
              [3.34556557e-04, 1.53049653e-04, 5.17167629e-04],
              [4.49973617e-05, 2.80944473e-05, 2.21113706e-05],
              [8.65276618e-05, 7.67250106e-05, 1.43730437e-04]],

             [[2.70602185e-05, 4.39063929e-06, 7.07563858e-06],
              [2.58404302e-04, 9.37175266e-05, 7.87475979e-04],
              [9.01797073e-06, 1.61861013e-06, 4.82104549e-05],
              [1.18014878e-04, 2.01703724e-05, 9.15858957e-05]]],

...

            [[[0.00000000e+00, 0.00000000e+00, 0.00000000e+00],
              [0.00000000e+00, 0.00000000e+00, 0.00000000e+00],
              [0.00000000e+00, 0.00000000e+00, 0.00000000e+00],
              [0.00000000e+00, 0.00000000e+00, 0.00000000e+00]],

             [[0.00000000e+00, 0.00000000e+00, 0.00000000e+00],
              [0.00000000e+00, 0.00000000e+00, 0.00000000e+00],
              [0.00000000e+00, 0.00000000e+00, 0.00000000e+00],
              [0.00000000e+00, 0.00000000e+00, 0.00000000e+00]],

             [[0.00000000e+00, 0.00000000e+00, 0.00000000e+00],
              [0.00000000e+00, 0.00000000e+00, 0.00000000e+00],
              [0.00000000e+00, 0.00000000e+00, 0.00000000e+00],
              [0.00000000e+00, 0.00000000e+00, 0.00000000e+00]],

             [[0.00000000e+00, 0.00000000e+00, 0.00000000e+00],
              [0.00000000e+00, 0.00000000e+00, 0.00000000e+00],
              [0.00000000e+00, 0.00000000e+00, 0.00000000e+00],
              [0.00000000e+00, 0.00000000e+00, 0.00000000e+00]]]]]]]]])
Coordinates:
  * sub-geo   (sub-geo) object '@UNKNOWN' 'Antarctica' ... 'Western Europe'
  * src-geo   (src-geo) object '@UNKNOWN' 'Antarctica' ... 'Western Europe'
  * gender    (gender) object '@UNKNOWN' 'female' 'male' 'NB'
  * occ       (occ) object '@UNKNOWN' 'activist' ... 'writer'
  * alpha     (alpha) object 'a-d' 'e-k' 'l-r' 's-'
  * age       (age) object '2001-2006' '2007-2011' '2012-2016' '2017-2022'
  * pop       (pop) object 'High' 'Low' 'Medium-High' 'Medium-Low'
  * langs     (langs) object '2-4 languages' '5+ languages' 'English only'
  * topic_id  (topic_id) int64 187 270 359 365 400 ... 2448 2483 2758 2867 2872
\end{verbatim}

\leavevmode\vadjust pre{\hypertarget{0636d848-8b10-4647-9f10-f147e30c5211}{}}%
Save this to NetCDF (xarray's recommended format):

\hypertarget{26dbf0d4-4ba9-444c-aefc-fb053c2b6ac9}{}
\begin{Shaded}
\begin{Highlighting}[]
\NormalTok{output.save\_xarray(q\_tgts, }\SpecialStringTok{f\textquotesingle{}task1{-}}\SpecialCharTok{\{}\NormalTok{DATA\_MODE}\SpecialCharTok{\}}\SpecialStringTok{{-}int{-}targets\textquotesingle{}}\NormalTok{)}
\end{Highlighting}
\end{Shaded}

\begin{verbatim}
INFO:wptrec.save:saving NetCDF to data\metric-tables\task1-eval-int-targets.nc
\end{verbatim}

\hypertarget{7298042a-d3bb-440a-96c4-7a859a8c1d8d}{}
\begin{Shaded}
\begin{Highlighting}[]

\end{Highlighting}
\end{Shaded}

\hypertarget{4a1f8258}{}
\hypertarget{task-2-alignment}{%
\section{Task 2 Alignment}\label{task-2-alignment}}

This notebook computes the target distributions and retrieved page
alignments for \textbf{Task 2}. It depends on the output of the
PageAlignments notebook, as imported by MetricInputs.

\leavevmode\vadjust pre{\hypertarget{708161fe}{}}%
This notebook can be run in two modes: 'train', to process the training
topics, and 'eval' for the eval topics.

\hypertarget{d24320b4}{}
\begin{Shaded}
\begin{Highlighting}[]
\NormalTok{DATA\_MODE }\OperatorTok{=} \StringTok{\textquotesingle{}eval\textquotesingle{}}
\end{Highlighting}
\end{Shaded}

\hypertarget{c7f2d588}{}
\hypertarget{setup}{%
\subsection{Setup}\label{setup}}

We begin by loading necessary libraries:

\hypertarget{b979c0a3}{}
\begin{Shaded}
\begin{Highlighting}[]
\ImportTok{import}\NormalTok{ sys}
\ImportTok{import}\NormalTok{ operator}
\ImportTok{from}\NormalTok{ functools }\ImportTok{import} \BuiltInTok{reduce}
\ImportTok{from}\NormalTok{ itertools }\ImportTok{import}\NormalTok{ product}
\ImportTok{from}\NormalTok{ collections }\ImportTok{import}\NormalTok{ namedtuple}
\ImportTok{from}\NormalTok{ pathlib }\ImportTok{import}\NormalTok{ Path}
\ImportTok{import}\NormalTok{ pandas }\ImportTok{as}\NormalTok{ pd}
\ImportTok{import}\NormalTok{ xarray }\ImportTok{as}\NormalTok{ xr}
\ImportTok{import}\NormalTok{ numpy }\ImportTok{as}\NormalTok{ np}
\ImportTok{import}\NormalTok{ matplotlib.pyplot }\ImportTok{as}\NormalTok{ plt}
\ImportTok{import}\NormalTok{ seaborn }\ImportTok{as}\NormalTok{ sns}
\ImportTok{import}\NormalTok{ gzip}
\ImportTok{import}\NormalTok{ json}
\ImportTok{from}\NormalTok{ natural.size }\ImportTok{import}\NormalTok{ binarysize}
\end{Highlighting}
\end{Shaded}

\leavevmode\vadjust pre{\hypertarget{d4ec1474}{}}%
Set up progress bar and logging support:

\hypertarget{639655ac}{}
\begin{Shaded}
\begin{Highlighting}[]
\ImportTok{from}\NormalTok{ tqdm.auto }\ImportTok{import}\NormalTok{ tqdm}
\NormalTok{tqdm.pandas(leave}\OperatorTok{=}\VariableTok{False}\NormalTok{)}
\end{Highlighting}
\end{Shaded}

\hypertarget{32212ded}{}
\begin{Shaded}
\begin{Highlighting}[]
\ImportTok{import}\NormalTok{ sys, logging}
\NormalTok{logging.basicConfig(level}\OperatorTok{=}\NormalTok{logging.INFO, stream}\OperatorTok{=}\NormalTok{sys.stderr)}
\NormalTok{log }\OperatorTok{=}\NormalTok{ logging.getLogger(}\StringTok{\textquotesingle{}Task2Alignment\textquotesingle{}}\NormalTok{)}
\end{Highlighting}
\end{Shaded}

\leavevmode\vadjust pre{\hypertarget{369de555}{}}%
And set up an output directory:

\hypertarget{02788756}{}
\begin{Shaded}
\begin{Highlighting}[]
\ImportTok{from}\NormalTok{ wptrec.save }\ImportTok{import}\NormalTok{ OutRepo}
\NormalTok{output }\OperatorTok{=}\NormalTok{ OutRepo(}\StringTok{\textquotesingle{}data/metric{-}tables\textquotesingle{}}\NormalTok{)}
\end{Highlighting}
\end{Shaded}

\hypertarget{ea23c240-c161-4b8e-a200-c9b830b8e67b}{}
\begin{Shaded}
\begin{Highlighting}[]
\ImportTok{from}\NormalTok{ wptrec }\ImportTok{import}\NormalTok{ metrics}
\ImportTok{from}\NormalTok{ wptrec.dimension }\ImportTok{import}\NormalTok{ sum\_outer}
\end{Highlighting}
\end{Shaded}

\hypertarget{1332f75e-bd68-4c83-b888-67a44010d0f7}{}
\hypertarget{data-and-helpers}{%
\subsection{Data and Helpers}\label{data-and-helpers}}

Most data loading is outsourced to \texttt{MetricInputs}. First we save
the data mode where metric inputs can find it:

\hypertarget{9725d318-608b-4ae6-9ff1-be23af6eebb8}{}
\begin{Shaded}
\begin{Highlighting}[]
\ImportTok{import}\NormalTok{ wptrec}
\NormalTok{wptrec.DATA\_MODE }\OperatorTok{=}\NormalTok{ DATA\_MODE}
\end{Highlighting}
\end{Shaded}

\hypertarget{410189f5-cc87-4e73-bb4e-7d917af62b78}{}
\begin{Shaded}
\begin{Highlighting}[]
\ImportTok{from}\NormalTok{ MetricInputs }\ImportTok{import} \OperatorTok{*}
\end{Highlighting}
\end{Shaded}

\begin{verbatim}
INFO:MetricInputs:reading data\metric-tables\page-sub-geo-align.parquet
INFO:MetricInputs:reading data\metric-tables\page-src-geo-align.parquet
INFO:MetricInputs:reading data\metric-tables\page-gender-align.parquet
INFO:MetricInputs:reading data\metric-tables\page-occ-align.parquet
INFO:MetricInputs:reading data\metric-tables\page-alpha-align.parquet
INFO:MetricInputs:reading data\metric-tables\page-age-align.parquet
INFO:MetricInputs:reading data\metric-tables\page-pop-align.parquet
INFO:MetricInputs:reading data\metric-tables\page-langs-align.parquet
\end{verbatim}

\hypertarget{d1fa1114-b729-4ae6-8c5c-efead0b10a7b}{}
\begin{Shaded}
\begin{Highlighting}[]
\NormalTok{dimensions}
\end{Highlighting}
\end{Shaded}

\begin{verbatim}
[<dimension "sub-geo": 21 levels>,
 <dimension "src-geo": 21 levels>,
 <dimension "gender": 4 levels>,
 <dimension "occ": 33 levels>,
 <dimension "alpha": 4 levels>,
 <dimension "age": 4 levels>,
 <dimension "pop": 4 levels>,
 <dimension "langs": 3 levels>]
\end{verbatim}

\hypertarget{9aa6a9d5-efbe-4ef9-8dc4-212e5018b8ee}{}
\hypertarget{qrel-join}{%
\subsubsection{qrel join}\label{qrel-join}}

We want a function to join alignments with qrels:

\hypertarget{f03e4261-a9ca-427f-901e-a4571b844141}{}
\begin{Shaded}
\begin{Highlighting}[]
\KeywordTok{def}\NormalTok{ qr\_join(align):}
    \ControlFlowTok{return}\NormalTok{ qrels.join(align, on}\OperatorTok{=}\StringTok{\textquotesingle{}page\_id\textquotesingle{}}\NormalTok{).set\_index([}\StringTok{\textquotesingle{}topic\_id\textquotesingle{}}\NormalTok{, }\StringTok{\textquotesingle{}page\_id\textquotesingle{}}\NormalTok{])}
\end{Highlighting}
\end{Shaded}

\hypertarget{9c0a9401-546b-4d10-b9de-830a778946bd}{}
\hypertarget{norm_dist}{%
\subsubsection{norm\_dist}\label{norm_dist}}

And a function to normalize to a distribution:

\hypertarget{fe8aca82-64b3-4d22-bc6b-431ff1e67342}{}
\begin{Shaded}
\begin{Highlighting}[]
\KeywordTok{def}\NormalTok{ norm\_dist\_df(mat):}
\NormalTok{    sums }\OperatorTok{=}\NormalTok{ mat.}\BuiltInTok{sum}\NormalTok{(}\StringTok{\textquotesingle{}columns\textquotesingle{}}\NormalTok{)}
    \ControlFlowTok{return}\NormalTok{ mat.divide(sums, }\StringTok{\textquotesingle{}rows\textquotesingle{}}\NormalTok{)}
\end{Highlighting}
\end{Shaded}

\hypertarget{c541f3eb}{}
\hypertarget{work-and-target-exposure}{%
\subsection{Work and Target Exposure}\label{work-and-target-exposure}}

The first thing we need to do to prepare the metric is to compute the
work-needed for each topic's pages, and use that to compute the target
exposure for each (relevant) page in the topic.

This is because an ideal ranking orders relevant documents in decreasing
order of work needed, followed by irrelevant documents. All relevant
documents at a given work level should receive the same expected
exposure.

First, look up the work for each query page ('query page work', or qpw):

\hypertarget{504fd5f3}{}
\begin{Shaded}
\begin{Highlighting}[]
\NormalTok{qpw }\OperatorTok{=}\NormalTok{ qrels.join(page\_quality, on}\OperatorTok{=}\StringTok{\textquotesingle{}page\_id\textquotesingle{}}\NormalTok{)}
\NormalTok{qpw}
\end{Highlighting}
\end{Shaded}

\begin{verbatim}
         topic_id   page_id quality
0             187       682       B
1             187       954       C
2             187      1170       C
3             187      1315       B
4             187      1322       B
...           ...       ...     ...
2737607      2872  69877511    Stub
2737608      2872  69878912       C
2737609      2872  69879322   Start
2737610      2872  69881345    Stub
2737611      2872  69883661   Start

[2737612 rows x 3 columns]
\end{verbatim}

\leavevmode\vadjust pre{\hypertarget{92356276}{}}%
And now use that to compute the number of documents at each work level:

\hypertarget{bafe6d00}{}
\begin{Shaded}
\begin{Highlighting}[]
\NormalTok{qwork }\OperatorTok{=}\NormalTok{ qpw.groupby([}\StringTok{\textquotesingle{}topic\_id\textquotesingle{}}\NormalTok{, }\StringTok{\textquotesingle{}quality\textquotesingle{}}\NormalTok{])[}\StringTok{\textquotesingle{}page\_id\textquotesingle{}}\NormalTok{].count()}
\NormalTok{qwork}
\end{Highlighting}
\end{Shaded}

\begin{verbatim}
topic_id  quality
187       Stub       31076
          Start      20015
          C          11853
          B           4146
          GA          1479
                     ...  
2872      Start      21769
          C           9480
          B           2627
          GA           806
          FA            69
Name: page_id, Length: 300, dtype: int64
\end{verbatim}

\leavevmode\vadjust pre{\hypertarget{eba51fd9}{}}%
Now we need to convert this into target exposure levels. This function
will, given a series of counts for each work level, compute the expected
exposure a page at that work level should receive.

\hypertarget{72e22ea5}{}
\begin{Shaded}
\begin{Highlighting}[]
\KeywordTok{def}\NormalTok{ qw\_tgt\_exposure(qw\_counts: pd.Series) }\OperatorTok{{-}\textgreater{}}\NormalTok{ pd.Series:}
    \ControlFlowTok{if} \StringTok{\textquotesingle{}topic\_id\textquotesingle{}} \OperatorTok{==}\NormalTok{ qw\_counts.index.names[}\DecValTok{0}\NormalTok{]:}
\NormalTok{        qw\_counts }\OperatorTok{=}\NormalTok{ qw\_counts.reset\_index(level}\OperatorTok{=}\StringTok{\textquotesingle{}topic\_id\textquotesingle{}}\NormalTok{, drop}\OperatorTok{=}\VariableTok{True}\NormalTok{)}
\NormalTok{    qwc }\OperatorTok{=}\NormalTok{ qw\_counts.reindex(work\_order, fill\_value}\OperatorTok{=}\DecValTok{0}\NormalTok{).astype(}\StringTok{\textquotesingle{}i4\textquotesingle{}}\NormalTok{)}
\NormalTok{    tot }\OperatorTok{=} \BuiltInTok{int}\NormalTok{(qwc.}\BuiltInTok{sum}\NormalTok{())}
\NormalTok{    da }\OperatorTok{=}\NormalTok{ metrics.discount(tot)}
\NormalTok{    qwp }\OperatorTok{=}\NormalTok{ qwc.shift(}\DecValTok{1}\NormalTok{, fill\_value}\OperatorTok{=}\DecValTok{0}\NormalTok{)}
\NormalTok{    qwc\_s }\OperatorTok{=}\NormalTok{ qwc.cumsum()}
\NormalTok{    qwp\_s }\OperatorTok{=}\NormalTok{ qwp.cumsum()}
\NormalTok{    res }\OperatorTok{=}\NormalTok{ pd.Series(}
\NormalTok{        [np.mean(da[s:e]) }\ControlFlowTok{for}\NormalTok{ (s, e) }\KeywordTok{in} \BuiltInTok{zip}\NormalTok{(qwp\_s, qwc\_s)],}
\NormalTok{        index}\OperatorTok{=}\NormalTok{qwc.index}
\NormalTok{    )}
    \ControlFlowTok{return}\NormalTok{ res}
\end{Highlighting}
\end{Shaded}

\leavevmode\vadjust pre{\hypertarget{c3c56c2c}{}}%
We'll then apply this to each topic, to determine the per-topic target
exposures:

\hypertarget{d96e7dc7}{}
\begin{Shaded}
\begin{Highlighting}[]
\NormalTok{qw\_pp\_target }\OperatorTok{=}\NormalTok{ qwork.groupby(}\StringTok{\textquotesingle{}topic\_id\textquotesingle{}}\NormalTok{).}\BuiltInTok{apply}\NormalTok{(qw\_tgt\_exposure)}
\NormalTok{qw\_pp\_target.name }\OperatorTok{=} \StringTok{\textquotesingle{}tgt\_exposure\textquotesingle{}}
\NormalTok{qw\_pp\_target}
\end{Highlighting}
\end{Shaded}

\begin{verbatim}
C:\Users\michaelekstrand\scoop\apps\mambaforge\current\envs\wptrec\lib\site-packages\numpy\core\fromnumeric.py:3474: RuntimeWarning: Mean of empty slice.
  return _methods._mean(a, axis=axis, dtype=dtype,
C:\Users\michaelekstrand\scoop\apps\mambaforge\current\envs\wptrec\lib\site-packages\numpy\core\_methods.py:189: RuntimeWarning: invalid value encountered in true_divide
  ret = ret.dtype.type(ret / rcount)
\end{verbatim}

\begin{verbatim}
topic_id  quality
187       Stub       0.075443
          Start      0.065321
          C          0.063307
          B          0.062546
          GA         0.062307
                       ...   
2872      Start      0.062570
          C          0.061352
          B          0.060958
          GA         0.060853
          FA         0.060827
Name: tgt_exposure, Length: 300, dtype: float32
\end{verbatim}

\leavevmode\vadjust pre{\hypertarget{8aba5221}{}}%
We can now merge the relevant document work categories with this
exposure, to compute the target exposure for each relevant document:

\hypertarget{7af3d40c}{}
\begin{Shaded}
\begin{Highlighting}[]
\NormalTok{qp\_exp }\OperatorTok{=}\NormalTok{ qpw.join(qw\_pp\_target, on}\OperatorTok{=}\NormalTok{[}\StringTok{\textquotesingle{}topic\_id\textquotesingle{}}\NormalTok{, }\StringTok{\textquotesingle{}quality\textquotesingle{}}\NormalTok{])}
\NormalTok{qp\_exp }\OperatorTok{=}\NormalTok{ qp\_exp.set\_index([}\StringTok{\textquotesingle{}topic\_id\textquotesingle{}}\NormalTok{, }\StringTok{\textquotesingle{}page\_id\textquotesingle{}}\NormalTok{])[}\StringTok{\textquotesingle{}tgt\_exposure\textquotesingle{}}\NormalTok{]}
\NormalTok{qp\_exp}
\end{Highlighting}
\end{Shaded}

\begin{verbatim}
topic_id  page_id 
187       682         0.062546
          954         0.063307
          1170        0.063307
          1315        0.062546
          1322        0.062546
                        ...   
2872      69877511    0.071035
          69878912    0.061352
          69879322    0.062570
          69881345    0.071035
          69883661    0.062570
Name: tgt_exposure, Length: 2737612, dtype: float32
\end{verbatim}

\hypertarget{8e975036-deef-4240-acf8-7b050b1cb485}{}
\hypertarget{subject-geography}{%
\subsection{Subject Geography}\label{subject-geography}}

Subject geography targets the average of the relevant set alignments and
the world population.

\hypertarget{d08efb3c-50ba-4027-8b62-bf8d88268476}{}
\begin{Shaded}
\begin{Highlighting}[]
\NormalTok{qr\_sub\_geo\_align }\OperatorTok{=}\NormalTok{ qr\_join(sub\_geo\_align)}
\NormalTok{qr\_sub\_geo\_align}
\end{Highlighting}
\end{Shaded}

\begin{verbatim}
                   @UNKNOWN  Antarctica  Caribbean  Central America  \
topic_id page_id                                                      
187      682            1.0         0.0        0.0              0.0   
         954            0.0         0.0        0.0              0.0   
         1170           1.0         0.0        0.0              0.0   
         1315           1.0         0.0        0.0              0.0   
         1322           0.0         0.0        0.0              0.0   
...                     ...         ...        ...              ...   
2872     69877511       1.0         0.0        0.0              0.0   
         69878912       1.0         0.0        0.0              0.0   
         69879322       1.0         0.0        0.0              0.0   
         69881345       0.0         0.0        0.0              0.0   
         69883661       1.0         0.0        0.0              0.0   

                   Central Asia  Eastern Africa  Eastern Asia  Eastern Europe  \
topic_id page_id                                                                
187      682                0.0             0.0           0.0             0.0   
         954                0.0             0.0           0.0             0.0   
         1170               0.0             0.0           0.0             0.0   
         1315               0.0             0.0           0.0             0.0   
         1322               0.0             0.0           0.0             0.0   
...                         ...             ...           ...             ...   
2872     69877511           0.0             0.0           0.0             0.0   
         69878912           0.0             0.0           0.0             0.0   
         69879322           0.0             0.0           0.0             0.0   
         69881345           0.0             0.0           0.0             0.0   
         69883661           0.0             0.0           0.0             0.0   

                   Middle Africa  Northern Africa  ...  Northern Europe  \
topic_id page_id                                   ...                    
187      682                 0.0              0.0  ...              0.0   
         954                 0.0              0.0  ...              0.0   
         1170                0.0              0.0  ...              0.0   
         1315                0.0              0.0  ...              0.0   
         1322                0.0              0.0  ...              0.0   
...                          ...              ...  ...              ...   
2872     69877511            0.0              0.0  ...              0.0   
         69878912            0.0              0.0  ...              0.0   
         69879322            0.0              0.0  ...              0.0   
         69881345            0.0              0.0  ...              0.0   
         69883661            0.0              0.0  ...              0.0   

                   Oceania  South America  South-eastern Asia  \
topic_id page_id                                                
187      682           0.0            0.0                 0.0   
         954           0.0            0.0                 0.0   
         1170          0.0            0.0                 0.0   
         1315          0.0            0.0                 0.0   
         1322          0.0            0.0                 0.0   
...                    ...            ...                 ...   
2872     69877511      0.0            0.0                 0.0   
         69878912      0.0            0.0                 0.0   
         69879322      0.0            0.0                 0.0   
         69881345      0.0            0.0                 1.0   
         69883661      0.0            0.0                 0.0   

                   Southern Africa  Southern Asia  Southern Europe  \
topic_id page_id                                                     
187      682                   0.0            0.0              0.0   
         954                   0.0            0.0              0.0   
         1170                  0.0            0.0              0.0   
         1315                  0.0            0.0              0.0   
         1322                  0.0            0.0              1.0   
...                            ...            ...              ...   
2872     69877511              0.0            0.0              0.0   
         69878912              0.0            0.0              0.0   
         69879322              0.0            0.0              0.0   
         69881345              0.0            0.0              0.0   
         69883661              0.0            0.0              0.0   

                   Western Africa  Western Asia  Western Europe  
topic_id page_id                                                 
187      682                  0.0           0.0             0.0  
         954                  0.0           0.0             1.0  
         1170                 0.0           0.0             0.0  
         1315                 0.0           0.0             0.0  
         1322                 0.0           0.0             0.0  
...                           ...           ...             ...  
2872     69877511             0.0           0.0             0.0  
         69878912             0.0           0.0             0.0  
         69879322             0.0           0.0             0.0  
         69881345             0.0           0.0             0.0  
         69883661             0.0           0.0             0.0  

[2737612 rows x 21 columns]
\end{verbatim}

\leavevmode\vadjust pre{\hypertarget{4a20ac65-a373-4df4-a4d0-15010530f253}{}}%
Compute a raw target, factoring in weights:

\hypertarget{5f07bc9b}{}
\begin{Shaded}
\begin{Highlighting}[]
\NormalTok{qr\_sub\_geo\_tgt }\OperatorTok{=}\NormalTok{ qr\_sub\_geo\_align.multiply(qp\_exp, axis}\OperatorTok{=}\StringTok{\textquotesingle{}rows\textquotesingle{}}\NormalTok{).groupby(}\StringTok{\textquotesingle{}topic\_id\textquotesingle{}}\NormalTok{).}\BuiltInTok{sum}\NormalTok{()}
\end{Highlighting}
\end{Shaded}

\leavevmode\vadjust pre{\hypertarget{8fc3d349-d5f3-43e9-9d70-5c723210b82f}{}}%
And now we need to average the known-geo with the background.

\hypertarget{cfc0f2bf-9823-4f7c-9f47-2d60321ec98d}{}
\begin{Shaded}
\begin{Highlighting}[]
\NormalTok{qr\_sub\_geo\_fk }\OperatorTok{=}\NormalTok{ qr\_sub\_geo\_tgt.iloc[:, }\DecValTok{1}\NormalTok{:].}\BuiltInTok{sum}\NormalTok{(}\StringTok{\textquotesingle{}columns\textquotesingle{}}\NormalTok{)}
\NormalTok{qr\_sub\_geo\_tgt.iloc[:, }\DecValTok{1}\NormalTok{:] }\OperatorTok{*=} \FloatTok{0.5}
\NormalTok{qr\_sub\_geo\_tgt.iloc[:, }\DecValTok{1}\NormalTok{:] }\OperatorTok{+=}\NormalTok{ qr\_sub\_geo\_fk.}\BuiltInTok{apply}\NormalTok{(}\KeywordTok{lambda}\NormalTok{ k: world\_pop }\OperatorTok{*}\NormalTok{ k }\OperatorTok{*} \FloatTok{0.5}\NormalTok{)}
\NormalTok{qr\_sub\_geo\_tgt.head()}
\end{Highlighting}
\end{Shaded}

\begin{verbatim}
             @UNKNOWN  Antarctica  Caribbean  Central America  Central Asia  \
topic_id                                                                      
187        758.390795    0.000309  19.328449        59.318134     21.122305   
270        967.129024    0.000233  69.231601        59.217295     23.466741   
359        641.628435    0.000220  59.452325        50.316697     12.890808   
365        481.821710    0.000148  18.649567        31.101461      9.853304   
400       2137.392223    0.000465  39.636681       106.665337     28.905838   

          Eastern Africa  Eastern Asia  Eastern Europe  Middle Africa  \
topic_id                                                                
187           109.070472    538.041050      160.059755      39.543253   
270           147.703296    423.954414      216.488611      39.756533   
359            74.364014    418.392361       59.464259      27.060083   
365            53.778891    251.973232       88.152130      28.231344   
400           168.787290    825.397926      224.909498      61.032641   

          Northern Africa  ...  Northern Europe     Oceania  South America  \
topic_id                   ...                                               
187             70.002691  ...       629.703684   93.268184     144.134466   
270             68.587236  ...       234.564466   82.918686     151.424207   
359             41.801031  ...        23.286746   19.606259     102.636117   
365             36.627649  ...        70.381866   38.938076      88.112552   
400            101.411871  ...       623.047574  189.240030     250.638365   

          South-eastern Asia  Southern Africa  Southern Asia  Southern Europe  \
topic_id                                                                        
187               206.460548        22.555736     554.706743       289.109630   
270               154.497647        35.594573     402.112346       174.454966   
359               124.679452        12.871306     348.961634        36.449902   
365               128.244847         9.595809     242.239786       158.985835   
400               297.782968        44.412480     820.697054       224.451157   

          Western Africa  Western Asia  Western Europe  
topic_id                                                
187            97.058510    112.506820      279.935254  
270           106.516209     91.341129      224.075828  
359            66.136114     49.743391       42.552754  
365            58.330269     70.821535       78.130276  
400           154.170802    152.891150      466.963698  

[5 rows x 21 columns]
\end{verbatim}

\leavevmode\vadjust pre{\hypertarget{11f2920b-849d-4398-a9ff-7f9a9b00c317}{}}%
These are \textbf{not} distributions, let's fix that!

\hypertarget{051e7233-a9bf-4c16-9ef5-922d7d542ee3}{}
\begin{Shaded}
\begin{Highlighting}[]
\NormalTok{qr\_sub\_geo\_tgt }\OperatorTok{=}\NormalTok{ norm\_dist\_df(qr\_sub\_geo\_tgt)}
\end{Highlighting}
\end{Shaded}

\hypertarget{3b0cc18d-f55d-4eb3-b775-b205c10efb4f}{}
\begin{Shaded}
\begin{Highlighting}[]
\NormalTok{output.save\_table(qr\_sub\_geo\_tgt, }\SpecialStringTok{f\textquotesingle{}task2{-}}\SpecialCharTok{\{}\NormalTok{DATA\_MODE}\SpecialCharTok{\}}\SpecialStringTok{{-}sub{-}geo{-}target\textquotesingle{}}\NormalTok{, parquet}\OperatorTok{=}\VariableTok{True}\NormalTok{)}
\end{Highlighting}
\end{Shaded}

\begin{verbatim}
INFO:wptrec.save:saving CSV to data\metric-tables\task2-eval-sub-geo-target.csv.gz
INFO:wptrec.save:data\metric-tables\task2-eval-sub-geo-target.csv.gz: 10.67 KiB
INFO:wptrec.save:saving Parquet to data\metric-tables\task2-eval-sub-geo-target.parquet
INFO:wptrec.save:data\metric-tables\task2-eval-sub-geo-target.parquet: 25.97 KiB
\end{verbatim}

\hypertarget{9e6939dd-7bf2-4d91-b596-391b06b6a51e}{}
\hypertarget{source-geography}{%
\subsection{Source Geography}\label{source-geography}}

Source geography works the same way.

\hypertarget{6d102833}{}
\begin{Shaded}
\begin{Highlighting}[]
\NormalTok{qr\_src\_geo\_align }\OperatorTok{=}\NormalTok{ qr\_join(src\_geo\_align)}
\NormalTok{qr\_src\_geo\_align}
\end{Highlighting}
\end{Shaded}

\begin{verbatim}
                   @UNKNOWN  Antarctica  Caribbean  Central America  \
topic_id page_id                                                      
187      682       0.400000         0.0        0.0              0.0   
         954       0.257143         0.0        0.0              0.0   
         1170      0.368421         0.0        0.0              0.0   
         1315      0.375000         0.0        0.0              0.0   
         1322      0.428571         0.0        0.0              0.0   
...                     ...         ...        ...              ...   
2872     69877511  1.000000         0.0        0.0              0.0   
         69878912  0.366667         0.0        0.0              0.0   
         69879322  0.200000         0.0        0.0              0.0   
         69881345  0.500000         0.0        0.0              0.0   
         69883661  0.000000         0.0        0.0              0.0   

                   Central Asia  Eastern Africa  Eastern Asia  Eastern Europe  \
topic_id page_id                                                                
187      682                0.0             0.0           0.0             0.0   
         954                0.0             0.0           0.0             0.0   
         1170               0.0             0.0           0.0             0.0   
         1315               0.0             0.0           0.0             0.0   
         1322               0.0             0.0           0.0             0.0   
...                         ...             ...           ...             ...   
2872     69877511           0.0             0.0           0.0             0.0   
         69878912           0.0             0.0           0.0             0.0   
         69879322           0.0             0.0           0.0             0.0   
         69881345           0.0             0.0           0.0             0.0   
         69883661           0.0             0.0           0.0             0.0   

                   Middle Africa  Northern Africa  ...  Northern Europe  \
topic_id page_id                                   ...                    
187      682                 0.0              0.0  ...         0.150000   
         954                 0.0              0.0  ...         0.285714   
         1170                0.0              0.0  ...         0.052632   
         1315                0.0              0.0  ...         0.000000   
         1322                0.0              0.0  ...         0.000000   
...                          ...              ...  ...              ...   
2872     69877511            0.0              0.0  ...         0.000000   
         69878912            0.0              0.0  ...         0.000000   
         69879322            0.0              0.0  ...         0.000000   
         69881345            0.0              0.0  ...         0.000000   
         69883661            0.0              0.0  ...         0.000000   

                    Oceania  South America  South-eastern Asia  \
topic_id page_id                                                 
187      682       0.000000            0.0                 0.0   
         954       0.000000            0.0                 0.0   
         1170      0.052632            0.0                 0.0   
         1315      0.000000            0.0                 0.0   
         1322      0.000000            0.0                 0.0   
...                     ...            ...                 ...   
2872     69877511  0.000000            0.0                 0.0   
         69878912  0.000000            0.0                 0.1   
         69879322  0.000000            0.0                 0.0   
         69881345  0.000000            0.0                 0.5   
         69883661  0.000000            0.0                 0.0   

                   Southern Africa  Southern Asia  Southern Europe  \
topic_id page_id                                                     
187      682                   0.0            0.0         0.000000   
         954                   0.0            0.0         0.000000   
         1170                  0.0            0.0         0.000000   
         1315                  0.0            0.0         0.000000   
         1322                  0.0            0.0         0.571429   
...                            ...            ...              ...   
2872     69877511              0.0            0.0         0.000000   
         69878912              0.0            0.0         0.000000   
         69879322              0.0            0.0         0.000000   
         69881345              0.0            0.0         0.000000   
         69883661              0.0            0.0         0.000000   

                   Western Africa  Western Asia  Western Europe  
topic_id page_id                                                 
187      682                  0.0         0.000        0.050000  
         954                  0.0         0.000        0.171429  
         1170                 0.0         0.000        0.000000  
         1315                 0.0         0.125        0.000000  
         1322                 0.0         0.000        0.000000  
...                           ...           ...             ...  
2872     69877511             0.0         0.000        0.000000  
         69878912             0.0         0.000        0.000000  
         69879322             0.0         0.600        0.000000  
         69881345             0.0         0.000        0.000000  
         69883661             0.0         0.000        0.000000  

[2737612 rows x 21 columns]
\end{verbatim}

\leavevmode\vadjust pre{\hypertarget{47cec80b-f2a9-42b2-bdf6-d2acb1bf6ee6}{}}%
And now we repeat these computations!

\hypertarget{0ca0cb12-f79a-4e0b-80be-fafb74ca9ca1}{}
\begin{Shaded}
\begin{Highlighting}[]
\NormalTok{qr\_src\_geo\_tgt }\OperatorTok{=}\NormalTok{ qr\_src\_geo\_align.multiply(qp\_exp, axis}\OperatorTok{=}\StringTok{\textquotesingle{}rows\textquotesingle{}}\NormalTok{).groupby(}\StringTok{\textquotesingle{}topic\_id\textquotesingle{}}\NormalTok{).}\BuiltInTok{sum}\NormalTok{()}
\end{Highlighting}
\end{Shaded}

\hypertarget{cdbc0bf3-cfc4-4be2-aef7-b17f2ff33a38}{}
\begin{Shaded}
\begin{Highlighting}[]
\NormalTok{qr\_src\_geo\_fk }\OperatorTok{=}\NormalTok{ qr\_src\_geo\_tgt.iloc[:, }\DecValTok{1}\NormalTok{:].}\BuiltInTok{sum}\NormalTok{(}\StringTok{\textquotesingle{}columns\textquotesingle{}}\NormalTok{)}
\NormalTok{qr\_src\_geo\_tgt.iloc[:, }\DecValTok{1}\NormalTok{:] }\OperatorTok{*=} \FloatTok{0.5}
\NormalTok{qr\_src\_geo\_tgt.iloc[:, }\DecValTok{1}\NormalTok{:] }\OperatorTok{+=}\NormalTok{ qr\_src\_geo\_fk.}\BuiltInTok{apply}\NormalTok{(}\KeywordTok{lambda}\NormalTok{ k: world\_pop }\OperatorTok{*}\NormalTok{ k }\OperatorTok{*} \FloatTok{0.5}\NormalTok{)}
\NormalTok{qr\_src\_geo\_tgt.head()}
\end{Highlighting}
\end{Shaded}

\begin{verbatim}
             @UNKNOWN  Antarctica  Caribbean  Central America  Central Asia  \
topic_id                                                                      
187       1892.369467    0.000221  10.629244        38.085204     13.580445   
270       1682.383393    0.000177  14.119208        32.195247     10.694027   
359       1349.305462    0.000166  10.812019        28.257371      9.637102   
365        899.571884    0.000116  24.578317        20.163280      7.058418   
400       3510.441727    0.002120  20.067844        67.028356     21.829953   

          Eastern Africa  Eastern Asia  Eastern Europe  Middle Africa  \
topic_id                                                                
187            76.133518    368.324160       91.512896      27.292950   
270            62.403539    291.840082       76.966103      21.889625   
359            55.966959    288.368964       44.738826      20.314799   
365            40.687112    195.637555       43.383442      18.290350   
400           124.063329    603.794372      146.012407      44.342043   

          Northern Africa  ...  Northern Europe     Oceania  South America  \
topic_id                   ...                                               
187             43.321614  ...       518.410807   53.513703      92.059446   
270             34.559267  ...       175.354880   40.800385      81.370613   
359             31.459874  ...        27.303292   12.455743      62.663487   
365             22.941681  ...        51.127647   29.052902      48.157030   
400             71.306778  ...       564.339012  159.334304     158.261887   

          South-eastern Asia  Southern Africa  Southern Asia  Southern Europe  \
topic_id                                                                        
187               137.914296        14.561724     383.654770        94.527498   
270               104.008024        13.928225     291.129055        74.411532   
359                93.284401         9.119594     262.223930        24.364869   
365                91.720399         6.467410     185.621213        93.737043   
400               220.074822        27.370868     631.930727       129.760455   

          Western Africa  Western Asia  Western Europe  
topic_id                                                
187            67.295433     64.363892      158.997834  
270            54.628367     45.613038      244.158227  
359            49.612454     37.221849       34.950278  
365            35.679642     47.403614       70.055359  
400           111.504376    103.379965      207.421067  

[5 rows x 21 columns]
\end{verbatim}

\leavevmode\vadjust pre{\hypertarget{8e995c9e-469c-40d2-9300-028daa23c7f7}{}}%
Make sure the rows are distributions:

\hypertarget{73010603}{}
\begin{Shaded}
\begin{Highlighting}[]
\NormalTok{qr\_src\_geo\_tgt }\OperatorTok{=}\NormalTok{ norm\_dist\_df(qr\_src\_geo\_tgt)}
\end{Highlighting}
\end{Shaded}

\hypertarget{2f6de863}{}
\begin{Shaded}
\begin{Highlighting}[]
\NormalTok{output.save\_table(qr\_src\_geo\_tgt, }\SpecialStringTok{f\textquotesingle{}task2{-}}\SpecialCharTok{\{}\NormalTok{DATA\_MODE}\SpecialCharTok{\}}\SpecialStringTok{{-}src{-}geo{-}target\textquotesingle{}}\NormalTok{, parquet}\OperatorTok{=}\VariableTok{True}\NormalTok{)}
\end{Highlighting}
\end{Shaded}

\begin{verbatim}
INFO:wptrec.save:saving CSV to data\metric-tables\task2-eval-src-geo-target.csv.gz
INFO:wptrec.save:data\metric-tables\task2-eval-src-geo-target.csv.gz: 10.62 KiB
INFO:wptrec.save:saving Parquet to data\metric-tables\task2-eval-src-geo-target.parquet
INFO:wptrec.save:data\metric-tables\task2-eval-src-geo-target.parquet: 25.97 KiB
\end{verbatim}

\hypertarget{611ad56d-c102-474d-8538-e65618934a72}{}
\hypertarget{gender}{%
\subsection{Gender}\label{gender}}

Now we're going to grab the gender alignments. Works the same way.

\hypertarget{326b87c0-91e6-4b65-8020-cadc23d21a9c}{}
\begin{Shaded}
\begin{Highlighting}[]
\NormalTok{qr\_gender\_align }\OperatorTok{=}\NormalTok{ qr\_join(gender\_align)}
\NormalTok{qr\_gender\_align.head()}
\end{Highlighting}
\end{Shaded}

\begin{verbatim}
                  @UNKNOWN  female  male   NB
topic_id page_id                             
187      682           1.0     0.0   0.0  0.0
         954           0.0     0.0   1.0  0.0
         1170          1.0     0.0   0.0  0.0
         1315          1.0     0.0   0.0  0.0
         1322          1.0     0.0   0.0  0.0
\end{verbatim}

\hypertarget{8efb59ad-5fb6-4a00-a2ef-1f1bad3258b8}{}
\begin{Shaded}
\begin{Highlighting}[]
\NormalTok{qr\_gender\_tgt }\OperatorTok{=}\NormalTok{ qr\_gender\_align.multiply(qp\_exp, axis}\OperatorTok{=}\StringTok{\textquotesingle{}rows\textquotesingle{}}\NormalTok{).groupby(}\StringTok{\textquotesingle{}topic\_id\textquotesingle{}}\NormalTok{).}\BuiltInTok{sum}\NormalTok{()}
\end{Highlighting}
\end{Shaded}

\hypertarget{6dc2fdc7-04b4-4161-92c6-ca996923b820}{}
\begin{Shaded}
\begin{Highlighting}[]
\NormalTok{qr\_gender\_fk }\OperatorTok{=}\NormalTok{ qr\_gender\_tgt.iloc[:, }\DecValTok{1}\NormalTok{:].}\BuiltInTok{sum}\NormalTok{(}\StringTok{\textquotesingle{}columns\textquotesingle{}}\NormalTok{)}
\NormalTok{qr\_gender\_tgt.iloc[:, }\DecValTok{1}\NormalTok{:] }\OperatorTok{*=} \FloatTok{0.5}
\NormalTok{qr\_gender\_tgt.iloc[:, }\DecValTok{1}\NormalTok{:] }\OperatorTok{+=}\NormalTok{ qr\_gender\_fk.}\BuiltInTok{apply}\NormalTok{(}\KeywordTok{lambda}\NormalTok{ k: gender\_tgt }\OperatorTok{*}\NormalTok{ k }\OperatorTok{*} \FloatTok{0.5}\NormalTok{)}
\NormalTok{qr\_gender\_tgt.head()}
\end{Highlighting}
\end{Shaded}

\begin{verbatim}
             @UNKNOWN       female         male         NB
topic_id                                                  
187       4231.726279   159.708759   364.436851   2.704633
270       1461.677295  1029.567013  1476.707985  12.917147
359       1164.868940   601.468537  1714.967051  11.640380
365       1012.069178   445.784544   938.953553   6.958483
400         94.885554  3323.222661  4707.223206  42.888097
\end{verbatim}

\hypertarget{28cb07b4-c4e0-40d6-9d9d-28c75dc09606}{}
\begin{Shaded}
\begin{Highlighting}[]
\NormalTok{qr\_gender\_tgt }\OperatorTok{=}\NormalTok{ norm\_dist\_df(qr\_gender\_tgt)}
\end{Highlighting}
\end{Shaded}

\hypertarget{b74608a4-ec64-48c9-8702-060723194a62}{}
\begin{Shaded}
\begin{Highlighting}[]
\NormalTok{output.save\_table(qr\_gender\_tgt, }\SpecialStringTok{f\textquotesingle{}task2{-}}\SpecialCharTok{\{}\NormalTok{DATA\_MODE}\SpecialCharTok{\}}\SpecialStringTok{{-}gender{-}target\textquotesingle{}}\NormalTok{, parquet}\OperatorTok{=}\VariableTok{True}\NormalTok{)}
\end{Highlighting}
\end{Shaded}

\begin{verbatim}
INFO:wptrec.save:saving CSV to data\metric-tables\task2-eval-gender-target.csv.gz
INFO:wptrec.save:data\metric-tables\task2-eval-gender-target.csv.gz: 2.24 KiB
INFO:wptrec.save:saving Parquet to data\metric-tables\task2-eval-gender-target.parquet
INFO:wptrec.save:data\metric-tables\task2-eval-gender-target.parquet: 6.90 KiB
\end{verbatim}

\hypertarget{b4a5905b-11ab-426e-ab57-faae3722d5ad}{}
\hypertarget{occupation}{%
\subsection{Occupation}\label{occupation}}

Occupation is more straightforward, since we don't have a global target
to average with. We do need to drop unknown.

\hypertarget{03193ea0-0d82-47f2-af6c-faf524c9c632}{}
\begin{Shaded}
\begin{Highlighting}[]
\NormalTok{qr\_occ\_align }\OperatorTok{=}\NormalTok{ qr\_join(occ\_align).multiply(qp\_exp, axis}\OperatorTok{=}\StringTok{\textquotesingle{}rows\textquotesingle{}}\NormalTok{)}
\NormalTok{qr\_occ\_tgt }\OperatorTok{=}\NormalTok{ qr\_occ\_align.iloc[:, }\DecValTok{1}\NormalTok{:].groupby(}\StringTok{\textquotesingle{}topic\_id\textquotesingle{}}\NormalTok{).}\BuiltInTok{sum}\NormalTok{()}
\NormalTok{qr\_occ\_tgt }\OperatorTok{=}\NormalTok{ norm\_dist\_df(qr\_occ\_tgt)}
\NormalTok{qr\_occ\_tgt.head()}
\end{Highlighting}
\end{Shaded}

\begin{verbatim}
          activist  agricultural worker    artist   athlete  biologist  \
topic_id                                                                 
187       0.001742             0.000423  0.046779  0.003448   0.001719   
270       0.000220             0.000236  0.000887  0.963747   0.000225   
359       0.000308             0.000073  0.000855  0.913443   0.000066   
365       0.000134             0.000030  0.000319  0.874820   0.000039   
400       0.004460             0.000402  0.331925  0.003775   0.001594   

          businessperson   chemist  civil servant  clergyperson  \
topic_id                                                          
187             0.025363  0.000046       0.001743      0.000760   
270             0.001724  0.000190       0.001045      0.000168   
359             0.007293  0.000094       0.000495      0.000071   
365             0.003116  0.000089       0.000365      0.000151   
400             0.020481  0.000277       0.002385      0.001815   

          computer scientist  ...  military personnel  musician  \
topic_id                      ...                                 
187                 0.000127  ...            0.003020  0.001195   
270                 0.000015  ...            0.001331  0.000621   
359                 0.000000  ...            0.002207  0.001371   
365                 0.000024  ...            0.001284  0.000451   
400                 0.000278  ...            0.002132  0.011309   

          performing artist  physicist  politician  scientist  \
topic_id                                                        
187                0.000999   0.000467    0.009501   0.010534   
270                0.001608   0.000035    0.003671   0.000428   
359                0.004169   0.000014    0.002663   0.000054   
365                0.002861   0.000000    0.001718   0.000100   
400                0.133634   0.000404    0.007652   0.003079   

          social scientist  sportsperson (non-athlete)  \
topic_id                                                 
187               0.004196                    0.000352   
270               0.000431                    0.013288   
359               0.000069                    0.047321   
365               0.000131                    0.106472   
400               0.003482                    0.001700   

          transportation occupation    writer  
topic_id                                       
187                        0.000268  0.012910  
270                        0.000434  0.001255  
359                        0.000085  0.002104  
365                        0.000160  0.001281  
400                        0.000531  0.262259  

[5 rows x 32 columns]
\end{verbatim}

\hypertarget{bdf9e942-4da6-4a15-9e9b-3401be54d10e}{}
\begin{Shaded}
\begin{Highlighting}[]
\NormalTok{output.save\_table(qr\_occ\_tgt, }\SpecialStringTok{f\textquotesingle{}task2{-}}\SpecialCharTok{\{}\NormalTok{DATA\_MODE}\SpecialCharTok{\}}\SpecialStringTok{{-}occ{-}target\textquotesingle{}}\NormalTok{, parquet}\OperatorTok{=}\VariableTok{True}\NormalTok{)}
\end{Highlighting}
\end{Shaded}

\begin{verbatim}
INFO:wptrec.save:saving CSV to data\metric-tables\task2-eval-occ-target.csv.gz
INFO:wptrec.save:data\metric-tables\task2-eval-occ-target.csv.gz: 14.69 KiB
INFO:wptrec.save:saving Parquet to data\metric-tables\task2-eval-occ-target.parquet
INFO:wptrec.save:data\metric-tables\task2-eval-occ-target.parquet: 37.48 KiB
\end{verbatim}

\hypertarget{1cac193c-9c72-47fa-abc7-574db0922eed}{}
\hypertarget{remaining-attributes}{%
\subsection{Remaining Attributes}\label{remaining-attributes}}

The remaining attributes don't need any further processing, as they are
completely known.

\hypertarget{733266c6-f7b7-48c2-988a-4edca70f62d8}{}
\begin{Shaded}
\begin{Highlighting}[]
\NormalTok{qr\_age\_align }\OperatorTok{=}\NormalTok{ qr\_join(age\_align).multiply(qp\_exp, axis}\OperatorTok{=}\StringTok{\textquotesingle{}rows\textquotesingle{}}\NormalTok{)}
\NormalTok{qr\_age\_tgt }\OperatorTok{=}\NormalTok{ norm\_dist\_df(qr\_age\_align.groupby(}\StringTok{\textquotesingle{}topic\_id\textquotesingle{}}\NormalTok{).}\BuiltInTok{sum}\NormalTok{())}
\NormalTok{output.save\_table(qr\_age\_tgt, }\SpecialStringTok{f\textquotesingle{}task2{-}}\SpecialCharTok{\{}\NormalTok{DATA\_MODE}\SpecialCharTok{\}}\SpecialStringTok{{-}age{-}target\textquotesingle{}}\NormalTok{, parquet}\OperatorTok{=}\VariableTok{True}\NormalTok{)}
\end{Highlighting}
\end{Shaded}

\begin{verbatim}
INFO:wptrec.save:saving CSV to data\metric-tables\task2-eval-age-target.csv.gz
INFO:wptrec.save:data\metric-tables\task2-eval-age-target.csv.gz: 1.20 KiB
INFO:wptrec.save:saving Parquet to data\metric-tables\task2-eval-age-target.parquet
INFO:wptrec.save:data\metric-tables\task2-eval-age-target.parquet: 5.24 KiB
\end{verbatim}

\hypertarget{b897d166-b0c0-4f7a-9e08-17c2c8208412}{}
\begin{Shaded}
\begin{Highlighting}[]
\NormalTok{qr\_alpha\_align }\OperatorTok{=}\NormalTok{ qr\_join(alpha\_align).multiply(qp\_exp, axis}\OperatorTok{=}\StringTok{\textquotesingle{}rows\textquotesingle{}}\NormalTok{)}
\NormalTok{qr\_alpha\_tgt }\OperatorTok{=}\NormalTok{ norm\_dist\_df(qr\_alpha\_align.groupby(}\StringTok{\textquotesingle{}topic\_id\textquotesingle{}}\NormalTok{).}\BuiltInTok{sum}\NormalTok{())}
\NormalTok{output.save\_table(qr\_alpha\_tgt, }\SpecialStringTok{f\textquotesingle{}task2{-}}\SpecialCharTok{\{}\NormalTok{DATA\_MODE}\SpecialCharTok{\}}\SpecialStringTok{{-}alpha{-}target\textquotesingle{}}\NormalTok{, parquet}\OperatorTok{=}\VariableTok{True}\NormalTok{)}
\end{Highlighting}
\end{Shaded}

\begin{verbatim}
INFO:wptrec.save:saving CSV to data\metric-tables\task2-eval-alpha-target.csv.gz
INFO:wptrec.save:data\metric-tables\task2-eval-alpha-target.csv.gz: 1.16 KiB
INFO:wptrec.save:saving Parquet to data\metric-tables\task2-eval-alpha-target.parquet
INFO:wptrec.save:data\metric-tables\task2-eval-alpha-target.parquet: 5.00 KiB
\end{verbatim}

\hypertarget{a4705eb1-02b6-479d-b833-c9de2decd730}{}
\begin{Shaded}
\begin{Highlighting}[]
\NormalTok{qr\_langs\_align }\OperatorTok{=}\NormalTok{ qr\_join(langs\_align).multiply(qp\_exp, axis}\OperatorTok{=}\StringTok{\textquotesingle{}rows\textquotesingle{}}\NormalTok{)}
\NormalTok{qr\_langs\_tgt }\OperatorTok{=}\NormalTok{ norm\_dist\_df(qr\_langs\_align.groupby(}\StringTok{\textquotesingle{}topic\_id\textquotesingle{}}\NormalTok{).}\BuiltInTok{sum}\NormalTok{())}
\NormalTok{output.save\_table(qr\_langs\_tgt, }\SpecialStringTok{f\textquotesingle{}task2{-}}\SpecialCharTok{\{}\NormalTok{DATA\_MODE}\SpecialCharTok{\}}\SpecialStringTok{{-}langs{-}target\textquotesingle{}}\NormalTok{, parquet}\OperatorTok{=}\VariableTok{True}\NormalTok{)}
\end{Highlighting}
\end{Shaded}

\begin{verbatim}
INFO:wptrec.save:saving CSV to data\metric-tables\task2-eval-langs-target.csv.gz
INFO:wptrec.save:data\metric-tables\task2-eval-langs-target.csv.gz: 978.00 iB
INFO:wptrec.save:saving Parquet to data\metric-tables\task2-eval-langs-target.parquet
INFO:wptrec.save:data\metric-tables\task2-eval-langs-target.parquet: 4.46 KiB
\end{verbatim}

\hypertarget{f5f336dc-cc67-44dd-98ae-3673c6de4ef7}{}
\begin{Shaded}
\begin{Highlighting}[]
\NormalTok{qr\_pop\_align }\OperatorTok{=}\NormalTok{ qr\_join(pop\_align).multiply(qp\_exp, axis}\OperatorTok{=}\StringTok{\textquotesingle{}rows\textquotesingle{}}\NormalTok{)}
\NormalTok{qr\_pop\_tgt }\OperatorTok{=}\NormalTok{ norm\_dist\_df(qr\_pop\_align.groupby(}\StringTok{\textquotesingle{}topic\_id\textquotesingle{}}\NormalTok{).}\BuiltInTok{sum}\NormalTok{())}
\NormalTok{output.save\_table(qr\_pop\_tgt, }\SpecialStringTok{f\textquotesingle{}task2{-}}\SpecialCharTok{\{}\NormalTok{DATA\_MODE}\SpecialCharTok{\}}\SpecialStringTok{{-}pop{-}target\textquotesingle{}}\NormalTok{, parquet}\OperatorTok{=}\VariableTok{True}\NormalTok{)}
\end{Highlighting}
\end{Shaded}

\begin{verbatim}
INFO:wptrec.save:saving CSV to data\metric-tables\task2-eval-pop-target.csv.gz
INFO:wptrec.save:data\metric-tables\task2-eval-pop-target.csv.gz: 1.24 KiB
INFO:wptrec.save:saving Parquet to data\metric-tables\task2-eval-pop-target.parquet
INFO:wptrec.save:data\metric-tables\task2-eval-pop-target.parquet: 5.15 KiB
\end{verbatim}

\hypertarget{6c43bd63-e991-4fef-9c01-fd14fa4d4fad}{}
\hypertarget{multidimensional-alignment}{%
\subsection{Multidimensional
Alignment}\label{multidimensional-alignment}}

Now let's dive into the multidmensional alignment. This is going to
proceed a lot like the Task 1 alignment.

\hypertarget{f017ee11-a257-477a-ab87-eeea0f98f4af}{}
\hypertarget{dimension-definitions}{%
\subsubsection{Dimension Definitions}\label{dimension-definitions}}

Let's define background distributions for some of our dimensions:

\hypertarget{8cc734a2-eb39-4343-ad72-8ca1eeb78c04}{}
\begin{Shaded}
\begin{Highlighting}[]
\NormalTok{dim\_backgrounds }\OperatorTok{=}\NormalTok{ \{}
    \StringTok{\textquotesingle{}sub{-}geo\textquotesingle{}}\NormalTok{: world\_pop,}
    \StringTok{\textquotesingle{}src{-}geo\textquotesingle{}}\NormalTok{: world\_pop,}
    \StringTok{\textquotesingle{}gender\textquotesingle{}}\NormalTok{: gender\_tgt,}
\NormalTok{\}}
\end{Highlighting}
\end{Shaded}

\leavevmode\vadjust pre{\hypertarget{311d0e0d-413b-4ede-bf9a-754510eb4dba}{}}%
Now we'll make a list of dimensions to treat with averaging:

\hypertarget{e7176ebc-2260-471f-be1f-00692e52ae26}{}
\begin{Shaded}
\begin{Highlighting}[]
\NormalTok{DR }\OperatorTok{=}\NormalTok{ namedtuple(}\StringTok{\textquotesingle{}DimRec\textquotesingle{}}\NormalTok{, [}\StringTok{\textquotesingle{}name\textquotesingle{}}\NormalTok{, }\StringTok{\textquotesingle{}align\textquotesingle{}}\NormalTok{, }\StringTok{\textquotesingle{}background\textquotesingle{}}\NormalTok{], defaults}\OperatorTok{=}\NormalTok{[}\VariableTok{None}\NormalTok{])}
\NormalTok{avg\_dims }\OperatorTok{=}\NormalTok{ [}
\NormalTok{    DR(d.name, d.page\_align\_xr, xr.DataArray(dim\_backgrounds[d.name], dims}\OperatorTok{=}\NormalTok{[d.name]))}
    \ControlFlowTok{for}\NormalTok{ d }\KeywordTok{in}\NormalTok{ dimensions}
    \ControlFlowTok{if}\NormalTok{ d.name }\KeywordTok{in}\NormalTok{ dim\_backgrounds}
\NormalTok{]}
\NormalTok{[d.name }\ControlFlowTok{for}\NormalTok{ d }\KeywordTok{in}\NormalTok{ avg\_dims]}
\end{Highlighting}
\end{Shaded}

\begin{verbatim}
['sub-geo', 'src-geo', 'gender']
\end{verbatim}

\leavevmode\vadjust pre{\hypertarget{e1494101-4748-4238-add9-62deb4d35e83}{}}%
And a list of dimensions to use as-is:

\hypertarget{ff83c5cd-437c-4bcf-98f5-dbcad1538e02}{}
\begin{Shaded}
\begin{Highlighting}[]
\NormalTok{raw\_dims }\OperatorTok{=}\NormalTok{ [}
\NormalTok{    DR(d.name, d.page\_align\_xr)}
    \ControlFlowTok{for}\NormalTok{ d }\KeywordTok{in}\NormalTok{ dimensions}
    \ControlFlowTok{if}\NormalTok{ d.name }\KeywordTok{not} \KeywordTok{in}\NormalTok{ dim\_backgrounds}
\NormalTok{]}
\NormalTok{[d.name }\ControlFlowTok{for}\NormalTok{ d }\KeywordTok{in}\NormalTok{ raw\_dims]}
\end{Highlighting}
\end{Shaded}

\begin{verbatim}
['occ', 'alpha', 'age', 'pop', 'langs']
\end{verbatim}

\leavevmode\vadjust pre{\hypertarget{d8950a73-24fa-447f-8612-f8bc3cf2368c}{}}%
Now: these dimension are in the original order - \texttt{dimensions} has
the averaged dimensions before the non-averaged ones. \textbf{This is
critical for the rest of the code to work.}

\hypertarget{a242988c-5f4b-49d3-a04d-6a15f1116f53}{}
\hypertarget{data-subsetting}{%
\subsubsection{Data Subsetting}\label{data-subsetting}}

Also from Task 1.

\hypertarget{a53eb29c-76ce-4580-942e-ac43424ad8e4}{}
\begin{Shaded}
\begin{Highlighting}[]
\NormalTok{avg\_cases }\OperatorTok{=} \BuiltInTok{list}\NormalTok{(product(}\OperatorTok{*}\NormalTok{[[}\VariableTok{True}\NormalTok{, }\VariableTok{False}\NormalTok{] }\ControlFlowTok{for}\NormalTok{ d }\KeywordTok{in}\NormalTok{ avg\_dims]))}
\NormalTok{avg\_cases.pop()}
\NormalTok{avg\_cases}
\end{Highlighting}
\end{Shaded}

\begin{verbatim}
[(True, True, True),
 (True, True, False),
 (True, False, True),
 (True, False, False),
 (False, True, True),
 (False, True, False),
 (False, False, True)]
\end{verbatim}

\hypertarget{ac94ffb7-9c4c-4009-b521-6e9c7ac2cfc2}{}
\begin{Shaded}
\begin{Highlighting}[]
\KeywordTok{def}\NormalTok{ case\_selector(case):}
    \KeywordTok{def}\NormalTok{ mksel(known):}
        \ControlFlowTok{if}\NormalTok{ known:}
            \CommentTok{\# select all but 1st column}
            \ControlFlowTok{return} \BuiltInTok{slice}\NormalTok{(}\DecValTok{1}\NormalTok{, }\VariableTok{None}\NormalTok{, }\VariableTok{None}\NormalTok{)}
        \ControlFlowTok{else}\NormalTok{:}
            \CommentTok{\# select 1st column}
            \ControlFlowTok{return} \DecValTok{0}
    
    \ControlFlowTok{return} \BuiltInTok{tuple}\NormalTok{(mksel(k) }\ControlFlowTok{for}\NormalTok{ k }\KeywordTok{in}\NormalTok{ case)}
\end{Highlighting}
\end{Shaded}

\hypertarget{6c8a225c-ba1c-4b37-9c1d-e0a14cfe75ae}{}
\hypertarget{background-averaging}{%
\subsubsection{Background Averaging}\label{background-averaging}}

We're now going to define our background-averaging function; this is
reused from the Task 1 alignment code.

For each condition, we are going to proceed as follows:

\begin{enumerate}
\tightlist
\item
  Compute an appropriate intersectional background distribution (based
  on the dimensions that are "known")
\item
  Select the subset of the target matrix with this known status
\item
  Compute the sum of this subset
\item
  Re-normalize the subset to sum to 1
\item
  Compute a normalization table such that each coordinate in the
  distributions to correct sums to 1 (so multiplying this by the
  background distribution spreads the background across the other
  dimensions appropriately), and use this to spread the background
  distribution
\item
  Average with the spread background distribution
\item
  Re-normalize to preserve the original sum
\end{enumerate}

Let's define the whole process as a function:

\hypertarget{a584d651-c27f-46f5-bf16-40526babaa91}{}
\begin{Shaded}
\begin{Highlighting}[]
\KeywordTok{def}\NormalTok{ avg\_with\_bg(tm, verbose}\OperatorTok{=}\VariableTok{False}\NormalTok{):}
\NormalTok{    tm }\OperatorTok{=}\NormalTok{ tm.copy()}
    
\NormalTok{    tail\_names }\OperatorTok{=}\NormalTok{ [d.name }\ControlFlowTok{for}\NormalTok{ d }\KeywordTok{in}\NormalTok{ raw\_dims]}
    
    \CommentTok{\# compute the tail mass for each coordinate (can be done once)}
\NormalTok{    tail\_mass }\OperatorTok{=}\NormalTok{ tm.}\BuiltInTok{sum}\NormalTok{(tail\_names)}
    
    \CommentTok{\# now some things don\textquotesingle{}t have any mass, but we still need to distribute background distributions.}
    \CommentTok{\# solution: we impute the marginal tail distribution}
    \CommentTok{\# first compute it}
\NormalTok{    tail\_marg }\OperatorTok{=}\NormalTok{ tm.}\BuiltInTok{sum}\NormalTok{([d.name }\ControlFlowTok{for}\NormalTok{ d }\KeywordTok{in}\NormalTok{ avg\_dims])}
    \CommentTok{\# then impute that where we don\textquotesingle{}t have mass}
\NormalTok{    tm\_imputed }\OperatorTok{=}\NormalTok{ xr.where(tail\_mass }\OperatorTok{\textgreater{}} \DecValTok{0}\NormalTok{, tm, tail\_marg)}
    \CommentTok{\# and re{-}compute the tail mass}
\NormalTok{    tail\_mass }\OperatorTok{=}\NormalTok{ tm\_imputed.}\BuiltInTok{sum}\NormalTok{(tail\_names)}
    \CommentTok{\# and finally we compute the rescaled matrix}
\NormalTok{    tail\_scale }\OperatorTok{=}\NormalTok{ tm\_imputed }\OperatorTok{/}\NormalTok{ tail\_mass}
    \KeywordTok{del}\NormalTok{ tm\_imputed}
    
    \ControlFlowTok{for}\NormalTok{ case }\KeywordTok{in}\NormalTok{ avg\_cases:}
        \CommentTok{\# for deugging: get names}
\NormalTok{        known\_names }\OperatorTok{=}\NormalTok{ [d.name }\ControlFlowTok{for}\NormalTok{ (d, known) }\KeywordTok{in} \BuiltInTok{zip}\NormalTok{(avg\_dims, case) }\ControlFlowTok{if}\NormalTok{ known]}
        \ControlFlowTok{if}\NormalTok{ verbose:}
            \BuiltInTok{print}\NormalTok{(}\StringTok{\textquotesingle{}processing known:\textquotesingle{}}\NormalTok{, known\_names)}
        
        \CommentTok{\# Step 1: background}
\NormalTok{        bg }\OperatorTok{=} \BuiltInTok{reduce}\NormalTok{(operator.mul, [}
\NormalTok{            d.background}
            \ControlFlowTok{for}\NormalTok{ (d, known) }\KeywordTok{in} \BuiltInTok{zip}\NormalTok{(avg\_dims, case)}
            \ControlFlowTok{if}\NormalTok{ known}
\NormalTok{        ])}
        \ControlFlowTok{if} \KeywordTok{not}\NormalTok{ np.allclose(bg.}\BuiltInTok{sum}\NormalTok{(), }\FloatTok{1.0}\NormalTok{):}
\NormalTok{            warnings.warn(}\StringTok{\textquotesingle{}background distribution for }\SpecialCharTok{\{\}}\StringTok{ sums to }\SpecialCharTok{\{\}}\StringTok{, expected 1\textquotesingle{}}\NormalTok{.}\BuiltInTok{format}\NormalTok{(known\_names, bg.values.}\BuiltInTok{sum}\NormalTok{()))}
        
        \CommentTok{\# Step 2: selector}
\NormalTok{        sel }\OperatorTok{=}\NormalTok{ case\_selector(case)}
        
        \CommentTok{\# Steps 3: sum in preparation for normalization}
\NormalTok{        c\_sum }\OperatorTok{=}\NormalTok{ tm[sel].}\BuiltInTok{sum}\NormalTok{()}
        
        \CommentTok{\# Step 5: spread the background}
\NormalTok{        bg\_spread }\OperatorTok{=}\NormalTok{ bg }\OperatorTok{*}\NormalTok{ tail\_scale[sel] }\OperatorTok{*}\NormalTok{ c\_sum}
        \ControlFlowTok{if} \KeywordTok{not}\NormalTok{ np.allclose(bg\_spread.}\BuiltInTok{sum}\NormalTok{(), c\_sum):}
\NormalTok{            warnings.warn(}\StringTok{\textquotesingle{}rescaled background sums to }\SpecialCharTok{\{\}}\StringTok{, expected c\_sum\textquotesingle{}}\NormalTok{.}\BuiltInTok{format}\NormalTok{(bg\_spread.values.}\BuiltInTok{sum}\NormalTok{()))}
        
        \CommentTok{\# Step 4 \& 6: average with the background}
\NormalTok{        tm[sel] }\OperatorTok{*=} \FloatTok{0.5}
\NormalTok{        bg\_spread }\OperatorTok{*=} \FloatTok{0.5}
\NormalTok{        tm[sel] }\OperatorTok{+=}\NormalTok{ bg\_spread}
                        
        \ControlFlowTok{if} \KeywordTok{not}\NormalTok{ np.allclose(tm[sel].}\BuiltInTok{sum}\NormalTok{(), c\_sum):}
\NormalTok{            warnings.warn(}\StringTok{\textquotesingle{}target distribution for }\SpecialCharTok{\{\}}\StringTok{ sums to }\SpecialCharTok{\{\}}\StringTok{, expected }\SpecialCharTok{\{\}}\StringTok{\textquotesingle{}}\NormalTok{.}\BuiltInTok{format}\NormalTok{(known\_names, tm[sel].values.}\BuiltInTok{sum}\NormalTok{(), c\_sum))}
    
    \ControlFlowTok{return}\NormalTok{ tm}
\end{Highlighting}
\end{Shaded}

\hypertarget{15272953-adc2-4cfd-89a4-10a9cc5e473c}{}
\hypertarget{computing-targets}{%
\subsubsection{Computing Targets}\label{computing-targets}}

We're now ready to compute a multidimensional target. This works like
the Task 1, with the difference that we are propagating work needed into
the targets as well; the input will be series whose \emph{index} is page
IDs and values are the work levels.

\hypertarget{921722fd-684c-43b0-8b2a-07c3a8f873ac}{}
\begin{Shaded}
\begin{Highlighting}[]
\KeywordTok{def}\NormalTok{ query\_xalign(pages):}
    \CommentTok{\# compute targets to average}
\NormalTok{    avg\_pages }\OperatorTok{=} \BuiltInTok{reduce}\NormalTok{(operator.mul, [d.align.loc[pages.index] }\ControlFlowTok{for}\NormalTok{ d }\KeywordTok{in}\NormalTok{ avg\_dims])}
\NormalTok{    raw\_pages }\OperatorTok{=} \BuiltInTok{reduce}\NormalTok{(operator.mul, [d.align.loc[pages.index] }\ControlFlowTok{for}\NormalTok{ d }\KeywordTok{in}\NormalTok{ raw\_dims])}
    
    \CommentTok{\# weight the left pages}
\NormalTok{    pages.index.name }\OperatorTok{=} \StringTok{\textquotesingle{}page\textquotesingle{}}
\NormalTok{    qpw }\OperatorTok{=}\NormalTok{ xr.DataArray.from\_series(pages)}
\NormalTok{    avg\_pages }\OperatorTok{=}\NormalTok{ avg\_pages }\OperatorTok{*}\NormalTok{ qpw}

    \CommentTok{\# convert to query distribution}
\NormalTok{    tgt }\OperatorTok{=}\NormalTok{ sum\_outer(avg\_pages, raw\_pages)}
\NormalTok{    tgt }\OperatorTok{/=}\NormalTok{ qpw.}\BuiltInTok{sum}\NormalTok{()}

    \CommentTok{\# average with background distributions}
\NormalTok{    tgt }\OperatorTok{=}\NormalTok{ avg\_with\_bg(tgt)}
    
    \CommentTok{\# and return the result}
    \ControlFlowTok{return}\NormalTok{ tgt}
\end{Highlighting}
\end{Shaded}

\hypertarget{e51433a1-0163-4fa0-b05a-908fd3a71d5d}{}
\hypertarget{applying-computations}{%
\subsubsection{Applying Computations}\label{applying-computations}}

Now let's run this thing - compute all the target distributions:

\hypertarget{6fe51dde-1c99-4913-9718-2eff63d94954}{}
\begin{Shaded}
\begin{Highlighting}[]
\NormalTok{q\_ids }\OperatorTok{=}\NormalTok{ qp\_exp.index.levels[}\DecValTok{0}\NormalTok{].copy()}
\NormalTok{q\_ids}
\end{Highlighting}
\end{Shaded}

\begin{verbatim}
Int64Index([ 187,  270,  359,  365,  400,  404,  480,  517,  568,  596,  715,
             807,  834,  881,  883,  949,  951,  955,  995, 1018, 1180, 1233,
            1328, 1406, 1417, 1448, 1449, 1479, 1499, 1548, 1558, 1647, 1685,
            1806, 1821, 1877, 1884, 1890, 2000, 2028, 2106, 2153, 2160, 2229,
            2244, 2448, 2483, 2758, 2867, 2872],
           dtype='int64', name='topic_id')
\end{verbatim}

\hypertarget{f6091b56-9652-45d4-9f8b-c14395dc0975}{}
\begin{Shaded}
\begin{Highlighting}[]
\NormalTok{q\_tgts }\OperatorTok{=}\NormalTok{ [query\_xalign(qp\_exp.loc[q]) }\ControlFlowTok{for}\NormalTok{ q }\KeywordTok{in}\NormalTok{ tqdm(q\_ids)]}
\end{Highlighting}
\end{Shaded}

\begin{Shaded}
\begin{Highlighting}[]
\FunctionTok{\{}\DataTypeTok{"model\_id"}\FunctionTok{:}\StringTok{"825dff5cd101402e8910af2cb8a4abf7"}\FunctionTok{,}\DataTypeTok{"version\_major"}\FunctionTok{:}\DecValTok{2}\FunctionTok{,}\DataTypeTok{"version\_minor"}\FunctionTok{:}\DecValTok{0}\FunctionTok{\}}
\end{Highlighting}
\end{Shaded}

\hypertarget{7c153039-f250-4c9e-b0a0-26ca2a50bda3}{}
\begin{Shaded}
\begin{Highlighting}[]
\NormalTok{q\_tgts }\OperatorTok{=}\NormalTok{ xr.concat(q\_tgts, q\_ids)}
\NormalTok{q\_tgts}
\end{Highlighting}
\end{Shaded}

\begin{verbatim}
<xarray.DataArray (topic_id: 50, sub-geo: 21, src-geo: 21, gender: 4, occ: 33,
                   alpha: 4, age: 4, pop: 4, langs: 3)>
array([[[[[[[[[5.32222994e-10, 1.22700201e-08, 0.00000000e+00],
              [1.62526437e-08, 1.59783339e-08, 1.29447847e-08],
              [3.42041043e-09, 1.26698684e-08, 7.92859105e-10],
              [5.93547427e-09, 5.09430447e-09, 2.95050019e-09]],

             [[1.14269430e-10, 6.14178489e-10, 3.43674300e-11],
              [7.37725527e-09, 4.95719395e-09, 1.18193819e-08],
              [1.13716762e-09, 2.68869541e-09, 2.74025688e-10],
              [4.25413257e-09, 3.79956877e-09, 4.77083603e-09]],

             [[2.66154076e-11, 1.00500571e-10, 0.00000000e+00],
              [5.05224537e-09, 2.29518724e-09, 7.76356299e-09],
              [6.48676034e-10, 3.73767879e-10, 2.92756686e-10],
              [1.25282632e-09, 1.10224187e-09, 2.08692884e-09]],

             [[3.86682637e-10, 5.77107369e-11, 9.71290153e-11],
              [3.89256821e-09, 1.41817503e-09, 1.16465576e-08],
              [1.19974875e-10, 2.22190558e-11, 6.51717199e-10],
              [1.69738191e-09, 2.74512105e-10, 1.28763261e-09]]],

...

            [[[0.00000000e+00, 0.00000000e+00, 0.00000000e+00],
              [0.00000000e+00, 0.00000000e+00, 0.00000000e+00],
              [0.00000000e+00, 0.00000000e+00, 0.00000000e+00],
              [0.00000000e+00, 0.00000000e+00, 0.00000000e+00]],

             [[0.00000000e+00, 0.00000000e+00, 0.00000000e+00],
              [0.00000000e+00, 0.00000000e+00, 0.00000000e+00],
              [0.00000000e+00, 0.00000000e+00, 0.00000000e+00],
              [0.00000000e+00, 0.00000000e+00, 0.00000000e+00]],

             [[0.00000000e+00, 0.00000000e+00, 0.00000000e+00],
              [0.00000000e+00, 0.00000000e+00, 0.00000000e+00],
              [0.00000000e+00, 0.00000000e+00, 0.00000000e+00],
              [0.00000000e+00, 0.00000000e+00, 0.00000000e+00]],

             [[0.00000000e+00, 0.00000000e+00, 0.00000000e+00],
              [0.00000000e+00, 0.00000000e+00, 0.00000000e+00],
              [0.00000000e+00, 0.00000000e+00, 0.00000000e+00],
              [0.00000000e+00, 0.00000000e+00, 0.00000000e+00]]]]]]]]])
Coordinates:
  * sub-geo   (sub-geo) object '@UNKNOWN' 'Antarctica' ... 'Western Europe'
  * src-geo   (src-geo) object '@UNKNOWN' 'Antarctica' ... 'Western Europe'
  * gender    (gender) object '@UNKNOWN' 'female' 'male' 'NB'
  * occ       (occ) object '@UNKNOWN' 'activist' ... 'writer'
  * alpha     (alpha) object 'a-d' 'e-k' 'l-r' 's-'
  * age       (age) object '2001-2006' '2007-2011' '2012-2016' '2017-2022'
  * pop       (pop) object 'High' 'Low' 'Medium-High' 'Medium-Low'
  * langs     (langs) object '2-4 languages' '5+ languages' 'English only'
  * topic_id  (topic_id) int64 187 270 359 365 400 ... 2448 2483 2758 2867 2872
\end{verbatim}

\leavevmode\vadjust pre{\hypertarget{63a6026f-17a5-4eee-9da1-b9e5dd3ca87d}{}}%
Save this to NetCDF (xarray's recommended format):

\hypertarget{26dbf0d4-4ba9-444c-aefc-fb053c2b6ac9}{}
\begin{Shaded}
\begin{Highlighting}[]
\NormalTok{output.save\_xarray(q\_tgts, }\SpecialStringTok{f\textquotesingle{}task2{-}}\SpecialCharTok{\{}\NormalTok{DATA\_MODE}\SpecialCharTok{\}}\SpecialStringTok{{-}int{-}targets\textquotesingle{}}\NormalTok{)}
\end{Highlighting}
\end{Shaded}

\begin{verbatim}
INFO:wptrec.save:saving NetCDF to data\metric-tables\task2-eval-int-targets.nc
\end{verbatim}

\hypertarget{b7e904ba}{}
\hypertarget{task-2b---equity-of-underexposure---not-yet-done}{%
\subsection{Task 2B - Equity of Underexposure - NOT YET
DONE}\label{task-2b---equity-of-underexposure---not-yet-done}}

For 2022, we are using a diffrent version of the metric. \textbf{Equity
of Underexposure} looks at each page's underexposure (system exposure is
less than target exposure), and looks for underexposure to be equitably
distributed between groups.

On its own, this isn't too difficult; averaging with background
distributions, however, gets rather subtle. Background distributions are
at the roup level, but we need to propgagate that into the page level,
so we can compute the difference between system and target exposure at
the page level, and then aggregate the underexposure within each group.

The idea of equity of underexposure is that we
\(\epsilon = \operatorname{E}_\pi [\eta]\) and
\(\epsilon^* = \operatorname{E}_\tau [\eta]\). We then compute
\(u = min(\epsilon^* - \epsilon, 0)\), and restrict it to be negative,
and aggregate it by group; if \(A\) is our page alignment matrix and
\(\vec{u}\), we compute the group underexposure by \(A^T \vec{u}\).

That's the key idea. However, we want to use \(\epsilon^\dagger\) that
has the equivalent of averaging group-aggregated \(\epsilon^*\) with
global target distributions \(w_g\). We can do this in a few stages.
First, we compute the total attention of each group, and use that to
compute the fraction of group global weight that should go to each unit
of alignment:

\textbackslash begin\{align*\} s\_g \& = \textbackslash sum\_d a\_\{dg\}
\textbackslash{} \textbackslash hat\{w\}\_g \& =
\textbackslash frac\{w\_g\}\{s\_g\} \textbackslash end\{align*\}

We can then average:

\textbackslash begin\{align\emph{\}
\textbackslash epsilon\^{}\textbackslash dagger\_d \& =
\textbackslash frac\{1\}\{2\}\textbackslash left(\textbackslash epsilon\^{}}\_d
+ \textbackslash sum\_g a\_\{dg\} \textbackslash hat\{w\}\_g
\textbackslash epsilon\^{}*\_\{\textbackslash mathrm\{total\}\}
\textbackslash right) \textbackslash{} \textbackslash end\{align*\}

This is all on a per-topic basis.

\hypertarget{60a609dd}{}
\hypertarget{demo-topic}{%
\subsubsection{Demo Topic}\label{demo-topic}}

We're going to reuse demo topic data from before:

\hypertarget{775fbe1a}{}
\begin{Shaded}
\begin{Highlighting}[]
\NormalTok{q\_xa}
\end{Highlighting}
\end{Shaded}

\leavevmode\vadjust pre{\hypertarget{c5a28bd0}{}}%
Compute the total for each attribute:

\hypertarget{bfe6124f}{}
\begin{Shaded}
\begin{Highlighting}[]
\NormalTok{s\_xg }\OperatorTok{=}\NormalTok{ q\_xa.}\BuiltInTok{sum}\NormalTok{(axis}\OperatorTok{=}\DecValTok{0}\NormalTok{) }\OperatorTok{+} \FloatTok{1e{-}10}
\NormalTok{s\_xg}
\end{Highlighting}
\end{Shaded}

\leavevmode\vadjust pre{\hypertarget{77bbb8cf}{}}%
Let's get some fractions out of that:

\hypertarget{3e682cb2}{}
\begin{Shaded}
\begin{Highlighting}[]
\NormalTok{s\_xgf }\OperatorTok{=}\NormalTok{ s\_xg }\OperatorTok{/}\NormalTok{ s\_xg.}\BuiltInTok{sum}\NormalTok{()}
\NormalTok{s\_xgf}
\end{Highlighting}
\end{Shaded}

\leavevmode\vadjust pre{\hypertarget{d43e7c1d}{}}%
Now, let's make a copy, and start building up a world target matrix that
properly accounts for missing values:

\hypertarget{2955a69f}{}
\begin{Shaded}
\begin{Highlighting}[]
\NormalTok{W }\OperatorTok{=}\NormalTok{ s\_xgf.copy()}
\end{Highlighting}
\end{Shaded}

\leavevmode\vadjust pre{\hypertarget{5f8f895d}{}}%
Now, let's put in the known intersectional targets:

\hypertarget{950146ab}{}
\begin{Shaded}
\begin{Highlighting}[]
\NormalTok{W[}\DecValTok{1}\NormalTok{:, }\DecValTok{1}\NormalTok{:] }\OperatorTok{=}\NormalTok{ int\_tgt }\OperatorTok{*}\NormalTok{ W[}\DecValTok{1}\NormalTok{:, }\DecValTok{1}\NormalTok{:].}\BuiltInTok{sum}\NormalTok{()}
\end{Highlighting}
\end{Shaded}

\leavevmode\vadjust pre{\hypertarget{14b699f0}{}}%
Now we need the known-gender / unknown-geo targets:

\hypertarget{a7c580d2}{}
\begin{Shaded}
\begin{Highlighting}[]
\NormalTok{W[}\DecValTok{0}\NormalTok{, }\DecValTok{1}\NormalTok{:] }\OperatorTok{=}\NormalTok{ int\_tgt.}\BuiltInTok{sum}\NormalTok{(axis}\OperatorTok{=}\DecValTok{0}\NormalTok{) }\OperatorTok{*}\NormalTok{ W[}\DecValTok{0}\NormalTok{, }\DecValTok{1}\NormalTok{:].}\BuiltInTok{sum}\NormalTok{()}
\end{Highlighting}
\end{Shaded}

\leavevmode\vadjust pre{\hypertarget{32ba4e94}{}}%
And the known-geo / unknown-gender targets:

\hypertarget{81f85ee7}{}
\begin{Shaded}
\begin{Highlighting}[]
\NormalTok{W[}\DecValTok{1}\NormalTok{:, }\DecValTok{0}\NormalTok{] }\OperatorTok{=}\NormalTok{ int\_tgt.}\BuiltInTok{sum}\NormalTok{(axis}\OperatorTok{=}\DecValTok{1}\NormalTok{) }\OperatorTok{*}\NormalTok{ W[}\DecValTok{1}\NormalTok{:, }\DecValTok{0}\NormalTok{].}\BuiltInTok{sum}\NormalTok{()}
\end{Highlighting}
\end{Shaded}

\leavevmode\vadjust pre{\hypertarget{e4b41485}{}}%
Let's see what we have:

\hypertarget{e73e084f}{}
\begin{Shaded}
\begin{Highlighting}[]
\NormalTok{W}
\end{Highlighting}
\end{Shaded}

\leavevmode\vadjust pre{\hypertarget{6748521a}{}}%
Now we normalize it by \(s_g\):

\hypertarget{603aed01}{}
\begin{Shaded}
\begin{Highlighting}[]
\NormalTok{Wh }\OperatorTok{=}\NormalTok{ W }\OperatorTok{/}\NormalTok{ s\_xg}
\NormalTok{Wh}
\end{Highlighting}
\end{Shaded}

\leavevmode\vadjust pre{\hypertarget{daee30d7}{}}%
The massive values are only where we have no relevant items, so they'll
never actually be used.

We can now compute the query-aligned target matrix.

\hypertarget{46e22a3f}{}
\begin{Shaded}
\begin{Highlighting}[]
\NormalTok{qp\_gt }\OperatorTok{=}\NormalTok{ (q\_xa }\OperatorTok{*}\NormalTok{ (Wh }\OperatorTok{*}\NormalTok{ qp\_exp[}\DecValTok{1}\NormalTok{].}\BuiltInTok{sum}\NormalTok{())).}\BuiltInTok{sum}\NormalTok{(axis}\OperatorTok{=}\NormalTok{(}\DecValTok{1}\NormalTok{,}\DecValTok{2}\NormalTok{)).to\_series()}
\NormalTok{qp\_gt.index.name }\OperatorTok{=} \StringTok{\textquotesingle{}page\_id\textquotesingle{}}
\NormalTok{qp\_gt}
\end{Highlighting}
\end{Shaded}

\hypertarget{a3ce00cf}{}
\begin{Shaded}
\begin{Highlighting}[]
\NormalTok{qp\_exp[}\DecValTok{1}\NormalTok{]}
\end{Highlighting}
\end{Shaded}

\hypertarget{34a60814}{}
\begin{Shaded}
\begin{Highlighting}[]
\NormalTok{qp\_tgt }\OperatorTok{=} \FloatTok{0.5} \OperatorTok{*}\NormalTok{ (qp\_exp[}\DecValTok{1}\NormalTok{] }\OperatorTok{+}\NormalTok{ qp\_gt)}
\NormalTok{qp\_tgt}
\end{Highlighting}
\end{Shaded}

\hypertarget{72799ddc}{}
\hypertarget{setting-up-matrix}{%
\subsubsection{Setting Up Matrix}\label{setting-up-matrix}}

Now that we have the math worked out, we can create actual global target
frames for each query.

\hypertarget{aac8aa39}{}
\begin{Shaded}
\begin{Highlighting}[]
\KeywordTok{def}\NormalTok{ topic\_page\_tgt(qdf):}
\NormalTok{    pages }\OperatorTok{=}\NormalTok{ qdf[}\StringTok{\textquotesingle{}page\_id\textquotesingle{}}\NormalTok{]}
\NormalTok{    pages }\OperatorTok{=}\NormalTok{ pages[pages.isin(page\_xalign.indexes[}\StringTok{\textquotesingle{}page\textquotesingle{}}\NormalTok{])]}
\NormalTok{    q\_xa }\OperatorTok{=}\NormalTok{ page\_xalign.loc[pages.values, :, :]}
    
    \CommentTok{\# now we need to get the exposure for the pages}
\NormalTok{    p\_exp }\OperatorTok{=}\NormalTok{ qp\_exp.loc[qdf.name]}
    \ControlFlowTok{assert}\NormalTok{ p\_exp.index.is\_unique}
    
    \CommentTok{\# need our sums}
\NormalTok{    s\_xg }\OperatorTok{=}\NormalTok{ q\_xa.}\BuiltInTok{sum}\NormalTok{(axis}\OperatorTok{=}\DecValTok{0}\NormalTok{) }\OperatorTok{+} \FloatTok{1e{-}10}
    
    \CommentTok{\# set up the global target}
\NormalTok{    W }\OperatorTok{=}\NormalTok{ s\_xg }\OperatorTok{/}\NormalTok{ s\_xg.}\BuiltInTok{sum}\NormalTok{()}
\NormalTok{    W[}\DecValTok{1}\NormalTok{:, }\DecValTok{1}\NormalTok{:] }\OperatorTok{=}\NormalTok{ int\_tgt }\OperatorTok{*}\NormalTok{ W[}\DecValTok{1}\NormalTok{:, }\DecValTok{1}\NormalTok{:].}\BuiltInTok{sum}\NormalTok{()}
\NormalTok{    W[}\DecValTok{0}\NormalTok{, }\DecValTok{1}\NormalTok{:] }\OperatorTok{=}\NormalTok{ int\_tgt.}\BuiltInTok{sum}\NormalTok{(axis}\OperatorTok{=}\DecValTok{0}\NormalTok{) }\OperatorTok{*}\NormalTok{ W[}\DecValTok{0}\NormalTok{, }\DecValTok{1}\NormalTok{:].}\BuiltInTok{sum}\NormalTok{()}
\NormalTok{    W[}\DecValTok{1}\NormalTok{:, }\DecValTok{0}\NormalTok{] }\OperatorTok{=}\NormalTok{ int\_tgt.}\BuiltInTok{sum}\NormalTok{(axis}\OperatorTok{=}\DecValTok{1}\NormalTok{) }\OperatorTok{*}\NormalTok{ W[}\DecValTok{1}\NormalTok{:, }\DecValTok{0}\NormalTok{].}\BuiltInTok{sum}\NormalTok{()}
    
    \CommentTok{\# per{-}unit global weights, de{-}normalized by total exposure}
\NormalTok{    Wh }\OperatorTok{=}\NormalTok{ W }\OperatorTok{/}\NormalTok{ s\_xg}
\NormalTok{    Wh }\OperatorTok{*=}\NormalTok{ p\_exp.}\BuiltInTok{sum}\NormalTok{()}
    
    \CommentTok{\# compute global target}
\NormalTok{    gtgt }\OperatorTok{=}\NormalTok{ q\_xa }\OperatorTok{*}\NormalTok{ Wh}
\NormalTok{    gtgt }\OperatorTok{=}\NormalTok{ gtgt.}\BuiltInTok{sum}\NormalTok{(axis}\OperatorTok{=}\NormalTok{(}\DecValTok{1}\NormalTok{,}\DecValTok{2}\NormalTok{)).to\_series()}
    
    \CommentTok{\# compute average target}
\NormalTok{    avg\_tgt }\OperatorTok{=} \FloatTok{0.5} \OperatorTok{*}\NormalTok{ (p\_exp }\OperatorTok{+}\NormalTok{ gtgt)}
\NormalTok{    avg\_tgt.index.name }\OperatorTok{=} \StringTok{\textquotesingle{}page\textquotesingle{}}
    
    \ControlFlowTok{return}\NormalTok{ avg\_tgt}
\end{Highlighting}
\end{Shaded}

\leavevmode\vadjust pre{\hypertarget{ca57adc4}{}}%
Test it quick:

\hypertarget{80e2e264}{}
\begin{Shaded}
\begin{Highlighting}[]
\NormalTok{topic\_page\_tgt(qdf)}
\end{Highlighting}
\end{Shaded}

\leavevmode\vadjust pre{\hypertarget{f7fb3560}{}}%
And create our targets:

\hypertarget{a0cd5399}{}
\begin{Shaded}
\begin{Highlighting}[]
\NormalTok{qp\_tgt }\OperatorTok{=}\NormalTok{ qrels.groupby(}\StringTok{\textquotesingle{}id\textquotesingle{}}\NormalTok{).progress\_apply(topic\_page\_tgt)}
\NormalTok{qp\_tgt}
\end{Highlighting}
\end{Shaded}

\hypertarget{482d7fe8}{}
\begin{Shaded}
\begin{Highlighting}[]
\NormalTok{save\_table(qp\_tgt.to\_frame(}\StringTok{\textquotesingle{}target\textquotesingle{}}\NormalTok{), }\StringTok{\textquotesingle{}task2{-}all{-}page{-}targets\textquotesingle{}}\NormalTok{)}
\end{Highlighting}
\end{Shaded}

\hypertarget{30cf3298}{}
\begin{Shaded}
\begin{Highlighting}[]
\NormalTok{train\_qptgt }\OperatorTok{=}\NormalTok{ qp\_tgt.loc[train\_topics[}\StringTok{\textquotesingle{}id\textquotesingle{}}\NormalTok{]].to\_frame(}\StringTok{\textquotesingle{}target\textquotesingle{}}\NormalTok{)}
\NormalTok{eval\_qptgt }\OperatorTok{=}\NormalTok{ qp\_tgt.loc[eval\_topics[}\StringTok{\textquotesingle{}id\textquotesingle{}}\NormalTok{]].to\_frame(}\StringTok{\textquotesingle{}target\textquotesingle{}}\NormalTok{)}
\end{Highlighting}
\end{Shaded}

\hypertarget{a213228f}{}
\begin{Shaded}
\begin{Highlighting}[]
\NormalTok{save\_table(train\_qptgt, }\StringTok{\textquotesingle{}task2{-}train{-}page{-}targets\textquotesingle{}}\NormalTok{)}
\NormalTok{save\_table(eval\_qptgt, }\StringTok{\textquotesingle{}task2{-}eval{-}page{-}targets\textquotesingle{}}\NormalTok{)}
\end{Highlighting}
\end{Shaded}

\end{document}